\documentclass[
  journal=largetwo,
  manuscript=article-type,
  year=2024,
  volume=37,
]{cup-journal}

\usepackage{amsmath}
\usepackage[nopatch]{microtype}
\usepackage{booktabs}
\usepackage{graphicx}
\usepackage{subcaption}
\usepackage{multicol}
\usepackage{cuted}
\usepackage{aas_macros}
\usepackage{hyperref}
\usepackage{amssymb}
\usepackage{soul}
\DeclareUnicodeCharacter{0301}{\'{e}}

\title{Constraining the Physical Parameters of Blazars Using the Seed Factor Approach}

\author{Chang-Bin Deng}
\affiliation{Department of Physics, Zhejiang Normal University, Jinhua 321004, People's Republic of China}
\alsoaffiliation{These authors contributed equally to this work.}

\author{Yong-You Shi}
\affiliation{Department of Physics, Yunnan Normal University, Kunming 650500, People's Republic of China}
\alsoaffiliation{These authors contributed equally to this work.}

\author{Yu-Jie Song}
\affiliation{Department of Physics, Yunnan Normal University, Kunming 650500, People's Republic of China}

\author{Rui Xue}
\affiliation{Department of Physics, Zhejiang Normal University, Jinhua 321004, People's Republic of China}

\author{Lei-Ming Du}
\affiliation{Department of Physics, Yunnan Normal University, Kunming 650500, People's Republic of China}

\author{Ze-Rui Wang}
\affiliation{College of Physics and Electronic Engineering, Qilu Normal University, Jinan 250200, People's Republic of China}

\author{Zhao-Hua Xie}
\affiliation{Department of Physics, Yunnan Normal University, Kunming 650500, People's Republic of China}
\email[Rui Xue, Lei-Ming Du, Ze-Rui Wang]{ruixue@zjnu.edu.cn, duleiming@ynnu.edu.cn, zerui\_wang62@163.com}


\begin{document}

\begin{abstract}
The discovery that blazars dominate the extra-galactic $\gamma$-ray sky is a triumph in the {\it{Fermi}} era. However, the exact location of $\gamma$-ray emission region still remains in debate. Low-synchrotron-peaked blazars (LSPs) are estimated to produce high-energy radiation through the external Compton process, thus their emission regions are closely related to the external photon fields. We employed {\it{the seed factor approach}} proposed by Georganopoulos et al. It directly matches the observed seed factor of each LSP with the characteristic seed factors of external photon fields to locate the $\gamma$-ray emission region. A sample of 1138 LSPs with peak frequencies and peak luminosities was adopted to plot a histogram distribution of observed seed factors. We also collected some spectral energy distributions (SEDs) of historical flare states to investigate the variation of $\gamma$-ray emission region. Those SEDs were fitted by both quadratic and cubic functions using the Markov-chain Monte Carlo method. Furthermore, we derived some physical parameters of blazars and compared them with the constraint of internal $\gamma\gamma$-absorption. We find that dusty torus dominates the soft photon fields of LSPs and most $\gamma$-ray emission regions of LSPs are located at 1-10 pc. The soft photon fields could also transition from dusty torus to broad line region and cosmic microwave background in different flare states. Our results suggest that the cubic function is better than the quadratic function to fit the SEDs.
\end{abstract}

\section{Introduction}

Blazars are the most extreme subclass of active galactic nuclei (AGN) with a relativistic jet pointing at the Earth \citep{Urry_Padovani95}. Due to the beaming effect, they have high luminosity, fast variability and variable polarization \citep{Urry_Padovani95}. According to the equivalent width (EW) of the emission lines, blazars are divided into flat spectrum radio quasars (FSRQs) with EW $\geq$ 5 \AA \space and BL Lacertae objects (BL Lacs) with EW < 5 \AA, respectively \citep{Urry_Padovani95}. In the $\log\nu$-$\log\nu F(\nu)$ diagram, the non-thermal emission from jets dominates blazars' spectral energy distribution (SED), which usually shows a structure of double humps \citep{Macomb95}. Generally speaking, the low-energy hump is caused by synchrotron radiation of relativistic electrons moving in the magnetic field \citep{Marscher_Gear85}. Based on the peak frequency of the low-energy hump, blazars are divided into low-synchrotron-peaked (LSP; i.e., $\nu^{\rm{S}}_{\rm{p}} < 10^{14}$ Hz), intermediate-synchrotron-peaked (ISP; i.e., $10^{14} < \nu^{\rm{S}}_{\rm{p}} < 10^{15}$ Hz), high-synchrotron-peaked (HSP; i.e., $\nu^{\rm{S}}_{\rm{p}} > 10^{15}$ Hz), and extreme high-synchrotron-peaked (EHSP; i.e., $\nu^{\rm{S}}_{\rm{p}} > 10^{17}$ Hz) blazars \parencite{Padovani95,Costamante01,Abdo10d}. In the leptonic model, the high-energy hump is attributed to the inverse Compton scattering (IC) from the same population of relativistic electrons that emit the synchrotron emission. The seed photons for the IC process could be from the synchrotron radiation \citep[synchrotron self-Compton, SSC; e.g., ][]{Maraschi92,Tavecchio98} or from external photon fields \citep[external-Compton, EC; e.g., ][]{Dermer93,Sikora94,Blazejowski00}. In addition, some hadronic models have been proposed to explain the high-energy hump \citep{Aharonian00,bottcher13,Xue22Hadronuclear}.

Since the launch of the {\it{Fermi}}-Large Area Telescope ({\it{Fermi}}-LAT) in 2008, high-energy astrophysics has undergone a transformative {\it{Fermi}} era marked by profound discoveries \citep{Abdo10a,Abdo10b,Abdo10c}. Though nearly 20\% LSPs were out of detection, it was found that the diffuse extra-galactic $\gamma$-ray background is significantly dominated by emission from blazars \citep{Ajello15,Ackermann16,Arsioli18b}. However, the exact location of $\gamma$-ray emission region is still on debate \parencite{Agudo12,Hu17,Arsioli18,Tan20}. Generally speaking, the $\gamma$-ray emission of FSRQs is interpreted by the EC process, since strong ambient photon fields are detected \citep{Madejski16,Huang22}. The LSP BL Lacs (LBLs) have similar SEDs to those of FSRQs and occasionally show weak emission lines, therefore, the $\gamma$-ray emission of LBLs can also be interpreted by the EC process \citep{Madejski16, Hu24}. For LSPs whose high-energy emission originates from the EC process, investigating the dominant soft photon fields could help to locate the $\gamma$-ray emission region. If the $\gamma$-ray emission region resides at the base of the jet, the soft photons should be dominated by the accretion disc and the hot corona \citep{Dermer93,Dermer09,Xue21}. When the $\gamma$-ray emission region is positioned at sub-pc from the central supermassive black hole (SMBH), the soft photons predominantly originate from the broad line region \citep[BLR; $R_{\rm{BLR}} \approx 0.1$ pc;][]{Sikora94,Kaspi07, Bentz09,Nalewajko14}. On the other hand, if the dissipation of the $\gamma$-ray emission occurs at about 1-10 pc, the dominant soft photon source becomes the dusty torus \citep[DT; $R_{\rm{DT}} \approx 2.5$ pc;][]{Sikora08, Zhang24}. In cases where the $\gamma$-ray emission region is located in the extended jet, additional external photon fields, such as the cosmic microwave background (CMB) and starlight, play a significant role \citep{Bottcher08cmb,Potter13,Potter13b,Potter13c}.

To pinpoint $\gamma$-ray emission regions of blazars, many methods have been proposed: (i) {\it{variability}}: \cite{Tavecchio10_variability} studied the light curves of 3C 454.3 and PKS 1510-089, and found significant short variabilities, which indicates that the dissipation occurs in a very compact region located in the BLR. \cite{Dotson12} proposed that the variability timescale of flares would not change in different bands within the BLR, but should manifest faster variabilities at higher energies within the DT. Applying this method to PKS 1510-089, they analyzed four prominent $\gamma$-ray flares detected by {\it{Fermi}} in 2009 and concluded that $\gamma$-ray emission regions are distributed over an extensive range of locations beyond the BLR \citep{Dotson15}. (ii) {\it{radio core-shift}}: Based on radio core-shift measurements, \cite{Yan18} suggested that the distance between the SMBH and the $\gamma$-ray emission region is less than 3.5 pc for PKS 1510-089 and less than 0.02 pc for BL Lacertae in the framework of leptonic models. \cite{Wu18} determined the distance to be about ten times the typical size of the BLR for 23 LSPs. \cite{Jiang20} used the time lags to derive the core size and inferred that the $\gamma$-ray emission region of PMN J2345-1555 is probably inside the BLR. (iii) {\it{model}}: \cite{Cao13} reproduced the quasi-simultaneous SEDs of 21 FSRQs using the one-zone leptonic model and inferred that the locations of the $\gamma$-ray emission regions are inside the BLR for 5 FSRQs and beyond the BLR for 16 FSRQs. \cite{Tan20} fitted the quasi-simultaneous SEDs of 60 FSRQs with the same model and got similar results. Based on SED fitting, \cite{Arsioli18} analysed the electron Lorentz factor and magnetic field strength for 104 LSPs, then found they are consistent with an EC model dominated by the DT. However, SED fitting results are not always reliable due to coupling of physical parameters, underscoring the importance of constraining some of them through direct observations \citep{Yamada20,Deng21}.

In addition to the above three methods, \cite{Georganopoulos12} proposed {\it{the seed factor approach}} to study if the $\gamma$-ray emission region of blazars is located near the BLR or DT. This method provides a convenient approach by utilizing the peak frequencies and luminosities, which can be extracted from the SEDs easily. \cite{Harvey20} further applied this approach to a dataset consisting of 62 FSRQs and demonstrated that the $\gamma$-ray emission regions predominantly reside within the DT. This finding was subsequently confirmed by \cite{Huang22}, who extended their analysis to a larger sample, including 619 FSRQs.

Recently, the SEDs of blazars in the Fourth {\it{Fermi}}-LAT 12-year Source catalog (4FGL-DR3) have been fitted with the quadratic function by \cite{Yang22,Yang23}. We apply {\it{the seed factor approach}} to this latest and largest sample of $\gamma$-ray LSPs to study their dissipation region positions. Furthermore, considering that blazars are highly variable objects \citep{Dotson15,Arsioli18b}, some flare states of various epochs are collected to investigate alterations in their physical properties. This paper is organized as follows. In Section \ref{Methods}, we present the methods, including {\it{the seed factor approach}}, SED fitting, and comprehensive parameter analysis of the $\gamma$-ray emission regions. The applications are presented in Section \ref{Application}. In the end, we draw a conclusion in Section \ref{Conclusion}. The cosmological parameters $H_0 = 69.6 \rm{km\space  s^{-1}\space Mpc^{-1}}$,
$\Omega_0 = 0.29$, and $\Omega_{\Lambda} = 0.71$ are adopted in this
work \citep{Bennett14}.

\section{Methods}
\label{Methods}

\subsection{Derivation of the Seed Factor}
\label{Derivation of the Seed Factor}

In this work, we adopt {\it{the seed factor approach}} to distinguish the location of $\gamma$-ray emission region. Following \cite{Georganopoulos12}, we have peak energies of synchrotron radiation and EC scattering in the observer's frame,
\begin{equation}
    \epsilon^{\rm{obs}}_{\rm{syn}} = \frac{B}{B_{\rm{cr}}}\gamma^2_{\rm{b}}\delta/(1+z),
    \label{eq1}
\end{equation}
\begin{equation}
    \epsilon^{\rm{obs}}_{\rm{EC}} = \frac{4}{3}\epsilon_{\rm{0,ext}}\gamma^2_{\rm{b}}\delta^2/(1+z),
    \label{eq2}
\end{equation}
respectively \citep{Coppi_Blandford90,Tavecchio98,Ghisellini_Tavecchio08variability}, where $B$ is the magnetic field strength in units of Gauss; $\gamma_{\rm{b}}$ is the break Lorentz factor of relativistic electrons responsible for the SED peaks; $\epsilon_{\rm{0,ext}}$ is the dimensionless energy of ambient soft photons in the AGN frame; $B_{\rm{cr}} = m_e^2c^3/e\hbar$ is the critical magnetic field strength; $\delta = [\Gamma(1-\beta \cos \theta_{\rm{obs}})]^{-1}$ is the Doppler factor, where $\Gamma$ is the bulk Lorentz factor, $\beta c$ is the jet speed and $\theta_{\rm{obs}}$ is the viewing angle. In this paper, by assuming $\theta_{\rm{obs}} \lesssim 1 / \Gamma$, we have $\delta \approx \Gamma$. It is worth noting that equation (\ref{eq2}) is only applicable within the Thomson regime.

Dividing equation (\ref{eq1}) by equation (\ref{eq2}), we obtain
\begin{equation}
    \frac{B}{\delta} = \frac{4B_{\rm{cr}} \epsilon_{\rm{0,ext}} \epsilon^{\rm{obs}}_{\rm{syn}}}{3\epsilon^{\rm{obs}}_{\rm{EC}}}.
    \label{ratio1}
\end{equation}

And the peak luminosities of synchrotron radiation and EC scattering in the observer's frame can be written as 
\begin{equation}
    L^{\rm{obs}}_{\rm{syn,p}} = \frac{4}{3}\sigma_{\rm{T}}c\beta \gamma_{\rm{b}}^2n(\gamma_{\rm{b}})U_{\rm{B}}\delta^4,
    \label{eq3}
\end{equation}
\begin{equation}
    L^{\rm{obs}}_{\rm{EC,p}} = \frac{4}{3}\sigma_{\rm{T}}c\beta \gamma_{\rm{b}}^2n(\gamma_{\rm{b}})U_{\rm{ext}}\delta^4,
    \label{eq4}
\end{equation}
respectively \citep{Blumenthal_Gould70,Rybicki81}, where $\sigma_{\rm{T}}$ is the Thomson cross section; $n(\gamma_{\rm{b}})$ is the electron density distribution at $\gamma_{\rm{b}}$; $U_{\rm{B}} = B^2/8\pi$ is the energy density of the magnetic field. Here, the energy density of the ambient photon fields in the comoving frame can be calculated as 
\begin{equation}
    U_{\rm{ext}} = \frac{17}{12}U_{\rm{0,ext}}\Gamma^2, 
    \label{eqU}
\end{equation}
where $U_{\rm{0,ext}}$ is the energy density in the AGN frame \citep{Ghisellini_Madau96,Ghisellini_Tavecchio08sequence}. 

Take the ratio of equation (\ref{eq3}) and equation (\ref{eq4}), we then get
\begin{equation}
    \frac{B^2}{\delta^2} = \frac{34 \pi U_{\rm{0,ext}}}{3\it{CD}},
    \label{ratio2}
\end{equation}
where $\it{CD} = \it{L}^{\rm{obs}}_{\rm{EC,p}}/\it{L}^{\rm{obs}}_{\rm{syn,p}}$ is the Compton dominance.

Combining equation (\ref{ratio1}) and equation (\ref{ratio2}), we ultimately derive the seed factor as
\begin{equation}
    {\it{SF}} = \log(\frac{\sqrt{\it{U}_{\rm{0,ext}}}}{\epsilon _{\rm{0,ext}}}) \approx \log(9863 \times \frac{\nu ^{\rm{obs}}_{\rm{syn,13}}}{\nu ^{\rm{obs}}_{\rm{EC,22}}}\sqrt{\it{CD}}).
    \label{Seed Factor}
\end{equation}
Here, $\nu^{\rm{obs}}_{\rm{syn,13}}$ is the peak frequency of synchrotron radiation in units of $\rm{10^{13} Hz}$ and $\nu^{\rm{obs}}_{\rm{EC,22}}$ is the peak frequency of EC scattering in units of $\rm{10^{22} Hz}$.

\subsection{Characteristic Values of the Seed Factor}

As the ambient photon fields, the BLR and DT were discussed in the former {\it{seed factor approach}} \citep{Georganopoulos12, Harvey20, Huang22}. In addition, some studies revealed the significance of CMB and starlight \citep{Bottcher08cmb, Potter13,Potter13b,Potter13c}. In this work, we comprehensively consider the seed factors of BLR, DT, CMB, and starlight. The accretion disc is out of consideration, because it is not suitable to this method.

To calculate the characteristic seed factor of BLR, the energy density $U_{\rm{0,ext}}$ and the dimensionless energy $\epsilon_{\rm{0,ext}}$ of the soft photons need to be determined. The typical size of BLR is $R_{\rm{BLR}} \approx 1 \times 10^{17} L_{\rm{d,45}}^{1/2} \ \rm{cm}$, where $L_{\rm{d, 45}}$ is the luminosity of the accretion disc in units of $10^{45} \rm{erg \ s^{-1}}$ \citep{Kaspi07,Bentz09}. The covering factor of BLR (the fractions of the disk
luminosity $L_{\rm{d}}$ reprocessed into the BLR radiation) is $\xi_{\rm{BLR}} = 0.1$ \citep{Ghisellini_Tavecchio09}. We then obtain the energy density $U_{\rm{0, BLR}} = \xi_{\rm{BLR}}L_{\rm{d}}/4\pi R_{\rm{BLR}}^2c = 2.65\times10^{-2}\rm{erg \  cm^{-3}}$ within the characteristic distance. The BLR can be regarded as a grey body with peak frequency of $1.5 {\nu}_{{\rm{Ly}}_{\alpha} }$, then the dimensionless photon energy is $\epsilon_{\rm{0,BLR}} = 3\times10^{-5}$ \citep{Tavecchio_Ghisellini081.5Lya}. Finally, with a $5\%$ uncertainty, we derive the characteristic seed factor of BLR as ${\it{SF}}_{\rm{BLR}} = 3.74 \pm 0.19$.

The typical size of DT is found to be $R_{\rm{DT}} \approx 2.5\times10^{18} L_{\rm{d,45}}^{1/2} \ \rm{cm}$ \citep{Ghisellini_Tavecchio09,Pei22}. In this work, we set the covering factor of DT as $\xi_{\rm{DT}} = 0.5$ \citep{Ghisellini_Tavecchio09}. Then the energy density of DT within the characteristic distance is $U_{\rm{0, DT}} = \xi_{\rm{DT}} L_{\rm{d}}/4\pi R_{\rm{DT}}^2 c = 2.12 \times 10^{-4} \rm{erg\ cm^{-3}}$. In the studies, the DT, which can also be approximated by a grey body, is endowed with three different temperatures, e.g., 80K \citep{Lopez18_80K}, 370K \citep{Ghisellini_Tavecchio09}, 1500K \citep{Almeyda17_1500K,Lyu18_1500K}. Then we obtain three dimensionless photon energies for the DT, which are $\epsilon_{\rm{0,DT}}^{\rm{80K}} = 5.30 \times 10^{-8},\ \epsilon_{\rm{0,DT}}^{\rm{370K}} = 2.45\times 10^{-7}, \ \epsilon_{\rm{0,DT}}^{\rm{1500K}} = 9.94\times 10^{-7}$. Considering the $5\%$ uncertainty, three distinct characteristic seed factors of DT can be described as follows: ${\it{SF}}_{\rm{DT}}^{\rm{80K}} = 5.44 \pm 0.27 ,\ {\it{SF}}_{\rm{DT}}^{\rm{370K}} = 4.77\pm 0.24,\ {\it{SF}}_{\rm{DT}}^{\rm{1500K}} = 4.17 \pm 0.21$.

For CMB, the energy density is $U_{\rm{CMB}} = 4.02\times10^{-13}\rm{erg\ cm^{-3}}$ and the typical temperature is $T_{\rm{CMB}} = 2.72{\rm{K}}$ in the observer's frame, respectively \citep{Bottcher08cmb}. In this case, the characteristic seed factor of CMB with 5\% uncertainty is ${\it{SF}}_{\rm{CMB}} = 2.55\pm0.13$. The energy density for starlight is $U_{\rm{0,SL}} = 8.01 \times 10^{-13} \rm{erg\ cm^{-3}}$ and the typical temperature is $T_{\rm{SL}} = 30{\rm{K}}$. Then we get the characteristic seed factor of starlight with 5\% uncertainty ${\it{SF}}_{\rm{SL}} = 1.65\pm0.08$ \citep{HESS17_starlight}. 

\begin{figure}
	\hspace{-0.2in}
    \includegraphics[width=\textwidth ]{ 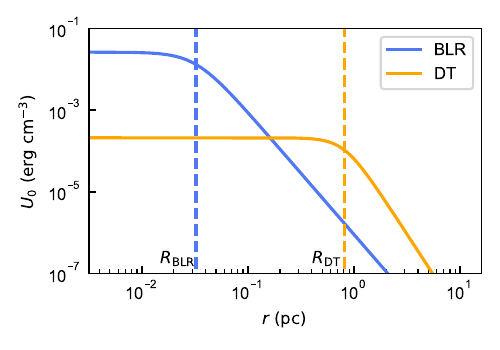}
    \caption{Energy density distribution of the broad line region (BLR) and the dusty torus (DT). $L_{\rm{d}} = 1\times 10^{45} {\rm{erg\ s^{-1}}}$ is adopted.}
    \label{energy density distribution}
\end{figure}

When applying {\it{the seed factor approach}}, there are several caveats that should be kept in mind. Firstly, the preceding derivation of the seed factor is within the Thomson regime. As a result of $\gamma \epsilon<1/4$ \citep{Moderski05}, the corresponding peak frequency of the EC radiation belonging to BLR, DT, CMB and starlight must be less than $1.03\times 10^{25}[\epsilon_0(1+z)/10^{-6}]^{-1} \rm{Hz}$. Since high-energy component of LSP usually peaks around 1 GeV, the EC process associated to BLR occurs in the Klein-Nishina regime, while others are cooling in the Thomson regime. Then the available energy density of BLR could reduce and the actual characteristic seed factor of BLR would be smaller than the above derived one. Secondly, the energy density of the BLR and DT could be smaller at farther site as proposed by \cite{Hayashida12}, i.e.,

\begin{equation}
    U_{\rm{0,BLR}}(r) = \frac{\xi_{\rm{BLR}} L_{\rm{d}}}{4\pi R^2_{\rm{BLR}}c[1+(r/R_{\rm{BLR}})^3]},
    \label{U r}
\end{equation}
and
\begin{equation}
    U_{\rm{0,DT}}(r) = \frac{\xi_{\rm{DT}} L_{\rm{d}}}{4\pi R^2_{\rm{DT}}c[1+(r/R_{\rm{DT}})^4]},
\end{equation}
where $r$ is the distance between the dissipation region and the central black hole, both energy densities of the BLR and DT have been transformed into the AGN frame (see also Figure \ref{energy density distribution}). Then the actual characteristic seed factor could also be smaller if the emission region is beyond the typical distance. Since the above derived characteristic seed factor of DT is the largest one among these four photon fields, the actual seed factor of DT covers that of the others. For example, the actual seed factor of DT could decrease to about 3.5 and equal to the actual seed factor of BLR. Therefore, the above derived seed factor of DT but not of BLR, CMB, or starlight is effective. If the observed seed factor is approximated to the derived $\it{SF}_{\rm{DT}}$, it can be ascertained that the DT dominates the soft photon fields. 

\subsection{The Fitting of Spectral Energy Distribution}

\begin{figure}
    \includegraphics[width=\textwidth ]{ 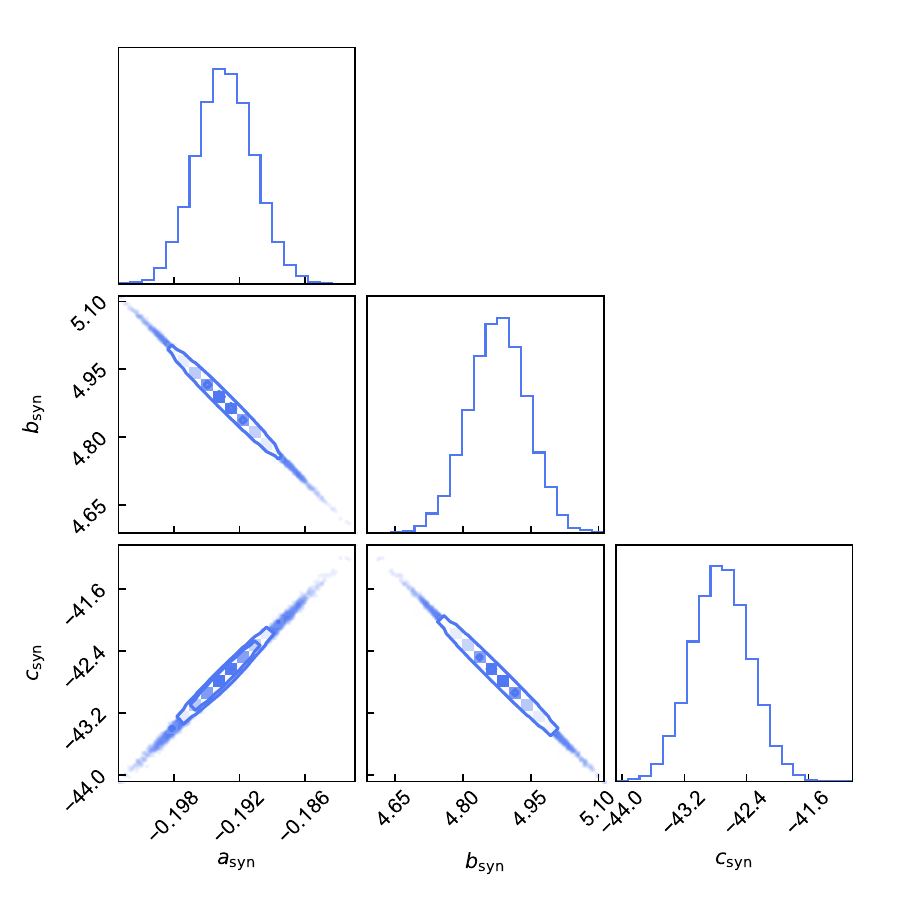}
    \caption{Posterior distribution from the Markov-chain Monte Carlo analysis of OD 166's low-energy (synchrotron) hump with quadratic function. We present only one corner plot here, those of other SEDs are available in machine-readable form.}
    \label{corner}
\end{figure}

As shown in equation (\ref{Seed Factor}), the observed seed factor can be determined by extracting the peak frequencies and luminosities associated with the two humps. Then we fit each SED by the quadratic and cubic functions, respectively. Namely,
\begin{equation}
	\left\{
	\begin{aligned}
	&\log(\nu F _{\nu}) = a_2(\log \nu)^2 + b_2\log \nu + c_2,\\
	&\log(\nu F _{\nu}) = a_3(\log \nu)^3 + b_3(\log \nu)^2 + c_3\log \nu + d_3.
	\end{aligned}\right.
\end{equation}
We use two kinds of functions because some SEDs possess high symmetry but others do not, which causes difference on the parameters \citep{Xue16_Curvature}. Markov-chain Monte Carlo (MCMC) analysis is employed since it returns robust uncertainties on the parameters \citep{Speagle19}. It works by randomly sampling from the posterior distribution, which are derived from the product of prior distribution and likelihood function. We use an uninformative uniform prior distribution because we have little knowledge about the parameters of quadratic and cubic functions in advance. Conservatively, it is expressed by 
\begin{equation}
p(m) = \left\{ 
	\begin{aligned}
	&\frac{1}{1000}, &{\rm{if}} m_0-500 < m < m_0+500\\
	&0, &\rm{otherwise}
	\end{aligned}\right.
\end{equation}
where $m$ denotes the paremeters in both quadratic and cubic functions, such as $a_2, b_2 ...$ And $m_0$ is the preprocessed $m$ derived by \texttt{numpy.polyfit}\footnote{https://numpy.org/doc/stable/reference/generated/numpy.polyfit.html}. The likelihood function is written as \citep{Yamada20} 
\begin{equation}
	{\it{L}} = \prod_{i=1}^n \frac{1}{\sqrt{2 \pi {\sigma_i}^2}} \exp{\left(-\frac{ ({\nu} F _{\nu,i} - \nu F _{\nu}(\nu_i))^2}{2\sigma^2_i} \right)}.
\end{equation}
Here, $\sigma_i$ is the Gaussian error of data point $i$ and $n$ is the number of data points in each energy hump. The \texttt{emcee} Python package\footnote{https://emcee.readthedocs.io/en/stable/} \citep[version 3.1.2,][]{Foreman13_emcee} is utilized to perform the MCMC algorithm. This package employs an affine invariant MCMC ensemble sampler with interdependent chains \citep{Goodman_Weare10}. While there is no fixed number of samples needed to make the convergence, we evaluate the convergence by inspecting the corner plot of parameters. The autocorrelation time, also the time that the chain "forgets" where it started, range from 35 to 250 in our Python program. Conservatively, we adopt 32 walkers (chains) initialized by the above preprocessed values with a $10^{-10}$ Gaussian error, run 17000 steps, burn 2000 steps and thin by 25. The results of posterior distribution are presented in Figure \ref{corner}.

While fitting the SEDs, we also compare the goodness of the two different function models with modified Akaike Information Criterion \citep[AICc;][]{Akaike74,Burnham02}, written as
\begin{equation}
	{\rm{AICc}} = -2\ln(\hat{L}) + 2k + \frac{2k^2+2k} {n-k-1}.
\end{equation}
In this formula, $\hat{L}$ is the maximum likelihood, which corresponds to the maximum posterior since we used an uniform prior, and $k$ is the number of free parameters. This criterion is chosen because the sample size of data points is small. AICc evaluates the loss of information during the fitting. The smaller AICc means the better model. Actually, the additional parameters in a model should improve the fitting to the data points, but we should also consider the increase in model complexity which causes overfitting. In this case, the AICc includes a penalty.

\subsection{Parameters of the Emission Region} 
\label{parameter analysis}

By fitting the SEDs of blazars, the observed peak frequencies and peak luminosities are determined. Then we select the blazars whose seed factors fall within the $\it{SF}_{\rm{DT}}$. With the gathered information on variability timescales and Doppler factors, we can deduce the other physical parameters of the $\gamma$-ray emission region. For blazars, we have such a formula \citep{Tavecchio98,Nalewajko14}:
\begin{equation}
    \frac{L^{\rm{obs}}_{\rm{syn}}}{U_{\rm{B}}} = \frac{L^{\rm{obs}}_{\rm{EC}}}{U_{\rm{ext}}} = \frac{L^{\rm{obs}}_{\rm{SSC}}}{U_{\rm{syn}}}.
\end{equation}
Here, $L^{\rm{obs}}_{\rm{syn}}$, $L^{\rm{obs}}_{\rm{EC}}$, $L^{\rm{obs}}_{\rm{SSC}}$ are the luminosities of the synchrotron, EC, and SSC radiations, respectively. $L^{\rm{obs}}_{\rm{syn}}$ and $L^{\rm{obs}}_{\rm{EC}}$ are determined by the integral of the low-energy and high-energy humps, respectively. For simplicity, we boldly assume that $L^{\rm{obs}}_{\rm{SSC}} = 10 \times L^{\rm{obs}}_{\rm{X}}$, where $L^{\rm{obs}}_{\rm{X}}$ is the maximum luminosity in X-ray band. $U_{\rm{syn}} = L^{\rm{obs}}_{\rm{syn}}/( 4\pi R^2c\delta^4 )$ is the energy density of the synchrotron radiation. The radius of the $\gamma$-ray emission region in the comoving frame can be estimated as
\begin{equation}
    R \approx ct_{\rm{var}}\frac{\delta}{1+z},
\end{equation}
where $t_{\rm{var}}$ is the variablility timescale. 

If we have measured the variability timescale, we can derive the following parameters by combining the above formulas:

\begin{footnotesize}
\begin{equation}
\left\{
\begin{aligned}
&\delta = 6.39 \left( \frac{L^{\rm{obs}}_{\rm{syn}}  L^{\rm{obs}}_{\rm{EC}}}{L^{\rm{obs}}_{\rm{SSC}} \cdot 10^{45} \rm{erg \ s^{-1}}}   \frac{\rm{10^{-4}erg\ cm^{-3}}}{U_{\rm{0,ext}}}\right)^{1/8} \left(\frac{t_{\rm{var}}}{1\rm{d}(1+z)} \right)^{-1/4},\\
& R = 1.66\times 10^{16} \left( \frac{L^{\rm{obs}}_{\rm{syn}}  L^{\rm{obs}}_{\rm{EC}}}{L^{\rm{obs}}_{\rm{SSC}} \cdot 10^{45} \rm{erg \  s^{-1}} }  \frac{\rm{10^{-4}erg\ cm^{-3}}}{U_{\rm{0,ext}}} \right)^{1/8} \left( \frac{t_{\rm{var}}}{1 \rm{d}(1+z)} \right)^{3/4},\\
& B = 0.382 \left( \frac{L^{\rm{obs} \  5}_{\rm{syn}}}{L^{\rm{obs} \ 3}_{\rm{EC}}  L^{\rm{obs}}_{\rm{SSC}} \cdot 10^{45}\rm{erg\ s^{-1}}} \right)^{1/8} \left( \frac{U_{\rm{0,ext}}}{\rm{10^{-4}erg\ cm^{-3}}}\right) ^{3/8}  \left( \frac{t_{\rm{var}}}{\rm{1d}(1+z)} \right)^{-1/4}.
\end{aligned}\right.
\end{equation}
\end{footnotesize}
When the Doppler factor is measured, we obtain:
\begin{footnotesize}
\begin{equation}
\left\{
\begin{aligned}
&t_{\rm{var}} = 1.44\times10^{4} \left( \frac{L^{\rm{obs}}_{\rm{syn}}  L^{\rm{obs}}_{\rm{EC}}}{L^{\rm{obs}}_{\rm{SSC}} \cdot 10^{45} \rm{erg\ s^{-s}}} \frac{\rm{10^{-4}erg\ cm^{-3}}}{U_{\rm{0,ext}}} \right)^{1/2} \left( \frac{\delta}{10} \right)^{-4} (1+z),\\
&R = 4.33 \times 10^{15} \left( \frac{L^{\rm{obs}}_{\rm{syn}}  L^{\rm{obs}}_{\rm{EC}}}{L^{\rm{obs}}_{\rm{SSC}} \cdot \rm{10^{45}erg\ s^{-1}}}  \frac{\rm{10^{-4}erg\ cm^{-3}}}{U_{\rm{{0,ext}}}}\right) ^{1/2} \left( \frac{\delta}{10} \right) ^{-3},\\
&B = 0.597 \left( \frac{L^{\rm{obs}}_{\rm{syn}}}{L^{\rm{obs}}_{\rm{EC}}}  \frac{U_{\rm{0,ext}}}{\rm{10^{-4}erg\ cm^{-3}}}\right)^{1/2} \frac{\delta}{10}.
\end{aligned}\right.
\end{equation}
\end{footnotesize}
Besides, $\gamma_{\rm{b}}$ can be calculated by \citep{Tavecchio98}
\begin{equation}
\gamma_{\rm{b}} = 5.2\times10^{-4} \left( \frac{\nu^{\rm{obs}}_{\rm{syn}}(1+z)}{B\delta} \right)^{1/2}.
\end{equation}

\subsection{Constraint of the Internal $\gamma \gamma$-Absorption}
\label{Constraint of the Internal gamma gamma absorption}

To make a comprison with the above derivation of physical parameters, we further make a constraint on $\delta$ through $\gamma\gamma$-absorption \citep[see also][]{Dondi_Ghisellini95}. Due to the $\gamma\gamma$ annihilation, electron-positron pairs are generated. Applying Delta-approximation, the corresponding optical depth can be calculated as \citep{Foffano22}: 
\begin{equation}
    \tau_{\gamma\gamma} = \sigma_{\gamma\gamma}n_{\rm{soft}}R,
\end{equation}
where $n_{\rm{soft}} = U_{\rm{soft}}/h\nu_{\rm{soft}}$ is the number density of the soft photons and $\sigma_{\gamma\gamma} = 1.68 \times 10^{-25} \rm{cm ^2}$ is the $\gamma \gamma$-absorption cross section, which is assumed to be a constant when such a condition is fulfilled:
\begin{equation}
\epsilon_{\rm{soft}}\epsilon_{\gamma} =2,
\label{eq delta}
\end{equation}
where $\epsilon_{\rm{soft}}$ and $\epsilon_{\gamma}$ are the dimensionless energies of the soft photons and $\gamma$-ray photons (comoving frame), respectively \citep{Dermer_Menon09}. 

Since the $\gamma$-ray is detected, optical depth must be less than 1. In order to solve the optical depth, energy density need to be determined. For internal soft photon fields such as synchrotron and IC radiation, we employ 
\begin{equation}
    U_{\rm{soft}} = \frac{\nu L^{\rm{obs}}_{\nu, {\rm{soft}}}}{4\pi R^2c\delta^4},
    \label{U1}
\end{equation}
where $\nu L^{\rm{obs}} _{\nu, {\rm{soft}}} = 4\pi d_{\rm{L}}^2 \nu F^{\rm{obs}}_{\nu,{\rm{soft}}}$ is the observed luminosity of the soft photons, $d_{\rm{L}}$ is the luminosity distance, and $\nu F^{\rm{obs}}_{\nu,{\rm{soft}}}$ is the observed flux. Then we derive
\begin{equation}
    \tau_{\gamma\gamma} = \frac{\sigma_{\gamma\gamma}d^2_{\rm{L}}\nu F^{\rm{obs}}_{\nu,{\rm{soft}}(1+z)}}{hc^2t_{\rm{var}}\delta^5\nu_{\rm{soft}}} < 1.
\end{equation}
Here, $\nu_{\rm{soft}}$ could be derived from equation (\ref{eq delta}) and expressed by
\begin{equation}
    \nu_{\rm{soft}} = \frac{2(m_{\rm{e}}c^2)^2}{h^2\nu_{\gamma}},
\end{equation}
where $\nu_{\gamma} = \nu_{\gamma}^{\rm{obs}}(1+z)/\delta$ is the frequency of $\gamma$-ray in the comoving frame.
Then we obtain the lower limit of $\delta$:
\begin{equation}
    \delta > \left( \frac{h\sigma_{\gamma\gamma}d_{\rm{L}}^2\nu F_{\nu,{\rm{soft}}}^{\rm{obs}}\nu_{\gamma}^{\rm{obs}}(1+z)^2}{2m_{\rm{e}}^2c^6t_{\rm{var}}} \right)^{1/6}.
    \label{delta lower limit}
\end{equation}

Not only do internal photon fields constrain the physical parameter, but external photon fields also give an additional constraint on $r$. The absorption of DT could be omitted according to equation (\ref{eq delta}), since the corresponding $\gamma$-ray ($\nu_{\gamma}^{\rm{obs}} = 10^{27}(\nu_{\rm{DT}}/10^{13} \rm{Hz})^{-1}\rm{Hz}$) is beyond detection in our collected SEDs. However, the $\gamma$-ray up to $10^{25}(\nu_{\rm{BLR}}/10^{15} \rm{Hz})^{-1}\rm{Hz}$ that is detectable could be absorbed by BLR. Therefore, we inspect the $\gamma\gamma$-absorption of BLR by unfolding its frequency spectrum. Given that the BLR is a grey body, we have
\begin{equation}
    \frac{\rm{d}\it{U}}{{\rm{d}} \nu} = \frac{8\pi h \nu^3}{c^3}(e^{h\nu / k_{\rm{B}}T}-1)^{-1},
    \label{U nu}
\end{equation}
where $T = h\nu_{\rm{BLR}}/3.93k_{\rm{B}}$ is the characteristic temperature of BLR and $k_{\rm{B}}$ is the Boltzmann constant \citep{Ghisellini_Tavecchio09}. From the combination of equation (\ref{U r}) and (\ref{U nu}), we derive the energy density of BLR as a function of both $r$ and $\nu$: 
\begin{equation}
    U_{\rm{BLR}}(r,\nu) = \Gamma^2 U_{\rm{0,BLR}}(r)\nu \frac{\rm{d} \it{U}}{{\rm{d}} \nu} / \int \frac{\rm{d} \it{U}}{{\rm{d}} \nu}{\rm{d}}\nu.
\end{equation}
With the same condition about $\tau < 1$ and lower limit of $\delta$ derived from equation (\ref{delta lower limit}), we could get the lower limit of $r$:
\begin{footnotesize}
\begin{equation}
    r > R_{\rm{BLR}} \left(   \frac{\sigma_{\gamma\gamma}R}{h} \frac{\xi_{\rm{BLR}} \Gamma^2L_{\rm{d}}}{3\pi R^2_{\rm{BLR}}c} \frac{\rm{d} \it{U}}{{\rm{d}} \nu} \bigg|_{\nu = \nu_{\rm{soft}}}/ \int\frac{\rm{d} \it{U}}{{\rm{d}} \nu}\rm{d}\nu -1 \right)^{1/3}.
\end{equation}
\end{footnotesize}

\section{Application}
\label{Application}

\subsection{Low-Synchrotron-Peaked Blazars}

\begin{footnotesize}
\begin{table*}
\begin{threeparttable}
\caption{Observed quantities and seed factors of 1138 low-synchrotron-peaked blazars.}
\centering
\begin{tabular}{*9c}
\hline\hline
Fermi Name       & Classification & $z$ & $\log \nu_{\rm{syn}}^{\rm{obs}}$ & $\log L_{\rm{syn}}^{\rm{obs}}$ & $\log \nu_{\rm{IC}}^{\rm{obs}}$ & $\log L^{\rm{obs}}_{\rm{IC}}$ & $\it{SF}$   & $\delta$  \\
(1) & (2) & (3) & (4) & (5) & (6) & (7) & (8) & (9) \\
\hline
4FGL J0001.5+2113 & FSRQ & 1.106 & 13.2    & 46.17  & 20.6   & 47.21 & 6.10 &             \\
4FGL J0003.3-1928 & BCU  & 2.000 & 13.3    & 46.21  & 22.5   & 46.42 & 3.90 &             \\
4FGL J0003.9-1149 & BLL  & 0.860 & 13.2    & 45.9   & 23.1   & 45.28 & 2.81 &             \\
4FGL J0004.3+4614 & FSRQ & 1.810 & 13.1    & 45.93  & 21.2   & 46.84 & 5.35 & 7.75        \\
4FGL J0004.4-4737 & FSRQ & 0.880 & 13      & 45.96  & 21.7   & 45.89 & 4.30 &             \\
··· & ··· & ··· & ··· & ··· & ··· & ··· & ··· & ···  \\
\hline
\label{table1}
\end{tabular}
\begin{tablenotes}
    \item NOTE: Column (1) gives the Fermi name. Column (2) represents the spectral classification. Column (3) gives the redshift. Column (4) and (5) are the synchrotron peak frequencies and luminosities from \cite{Yang22}, respectively. Column (6) and (7) are the IC peak frequencies and luminosities from \cite{Yang23}, respectively. Column (8) gives the observed seed factors. Column (9) gives the Doppler factors of 383 blazars from \cite{Liodakis18}. We present only 5 items here, full table is available in machine-readable form.
\end{tablenotes}
\end{threeparttable}
\end{table*}
\end{footnotesize}

\begin{figure}
	\hspace{-0.2in}
    \includegraphics[width=\textwidth]{ 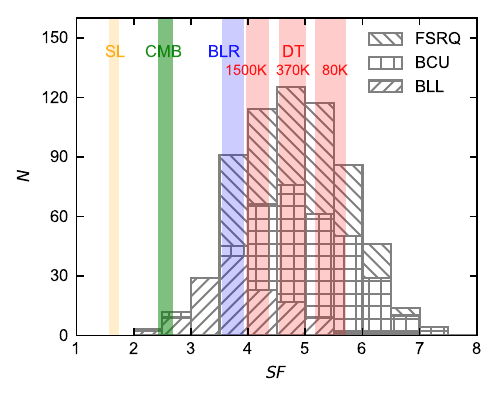}
    \caption{Histogram of observed seed factors. FSRQs, BL Lacs and blazar candidates of uncertain type (BCU) in our sample are distinguished by three kinds of grids. Three red areas represent the scales of characteristic seed factors belonging to the dusty torus (DT) in three different temperatures. Blue, green and yellow areas represent the scales of characteristic seed factors belonging to the broad line region (BLR), cosmic microwave background (CMB) and starlight (SL).}
    \label{Histogram of observed seed factors}
\end{figure}

Locating the $\gamma$-ray emission region of blazars has been a significant issue in the {\it{Fermi}} era. Debates continuously happen because of the limited accuracy of instruments and complicated radiation mechanism of blazars. In this work, we adopted {\it{the seed factor approach}} proposed by \cite{Georganopoulos12}. Based on the derivation in Section \ref{Derivation of the Seed Factor}, some observed quantities need to be determined. We collected a sample of 1138 LSPs with synchrotron peak frequencies and luminosities from \cite{Yang22}, and with IC peak frequencies and luminosities from \cite{Yang23}. There are 630 FSRQs, 132 BL Lacs and 376 blazar candidates of uncertain type (BCUs) in this sample (see also Table \ref{table1}). The Doppler factors of 383 blazars are also recorded from \cite{Liodakis18} for further calculation. Figure \ref{Histogram of observed seed factors} displays the histogram of observed seed factors belonging to the LSPs, which is obtained using equation (\ref{Seed Factor}). The observed seed factors of FSRQs and BCUs converge around the areas of DT, which demonstrates that DT dominates the soft photon fields of FSRQs and BCUs. This can be attributed to the strong radiation from DT, which is reproduced by the strong radiation from accretion disc of FSRQs and BCUs \citep{Madejski16,Huang22}. The observed seed factors of BL Lacs converge around the area of BLR. This suggests that the soft photons either originate from BLR or from DT, as the actual seed factor of DT could be smaller than the one depicted in Figure \ref{Histogram of observed seed factors}. The areas of CMB and starlight appear on the left edge of the histogram, indicating that their contributions to the soft photons in the EC process are relatively small. In general, observed seed factors of 552 in 1138 LSPs directly fall into the areas of DT, which locates the $\gamma$-ray emission region at 1-10 pc. 

This result is consistent with the former analysis using {\it{the seed factor approach}}. \cite{Harvey20} calculated the seed factors of 62 FSRQs and found the distribution peaking at a value corresponding to DT. \cite{Huang22} used a sample of 619 sources and also found the distribution is located at DT. In our work, rather than setting the temperature of DT to 1200K, we considered three different temperatures of DT because it is a relatively thick gas cloud with its inner temperature varying from the outer one \citep{Lyu18_1500K}. Figure \ref{Histogram of observed seed factors} shows 370K dominates the distribution of observed seed factors, indicating most $\gamma$-ray emission regions are located inside the DT.

\subsection{LSPs Dominated by the Dusty Torus}

\begin{figure}
\begin{subfigure}{\textwidth}
	\hspace{-0.17in}
    \includegraphics[width = \linewidth]{ 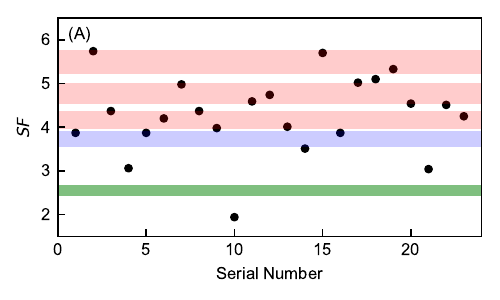}
\end{subfigure}
\begin{subfigure}{\textwidth}
	\hspace{-0.17in}
    \includegraphics[width = \linewidth]{ 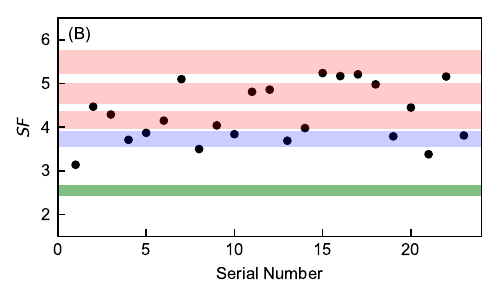}
\end{subfigure}
\caption{Scatterplots of observed seed factors belonging to low-synchrotron-peaked blazars dominated by the dusty torus. These data correspond to the first 23 flare states in Table \ref{table2}, ordered from top to bottom. The upper and lower panels are derived from SEDs fitted by quadratic and cubic functions, respectively. Red, blue and green areas depict the characteristic seed factors of dusty torus, broad line region and cosmic microwave background, respectively.}
\label{Scatter1}
\end{figure}

Although our result suggests that DT dominates the soft photon fields of LSPs, previous study demonstrated that blazars might have various $\gamma$-ray emission regions in different flare epochs \citep{Dotson15}. We investigated this property by collecting some historical flare states. LSPs whose observed seed factors directly fall into the red areas in Figure \ref{Histogram of observed seed factors} (i.e., LSPs dominated by the DT) were selected, since only their soft photon fields had been determined effectively. The SEDs that we collected fulfilled these conditions: Possessing multi-wavelength quasi-simultaneous data except for the radio band. The observation times of each band intersect within two months. In total, we collected 23 SEDs. The related references are given in Table \ref{table2}. These SEDs were fitted by both quadratic and cubic functions using the MCMC method (see also Figures \ref{SED DT Doppler} \& \ref{SED DT var}). Then we extracted the peak frequencies and luminosities of two humps, and calculated the observed seed factors. The results are displayed as scatterplots in Figure \ref{Scatter1}. The distributions of data scatters vary under two different function fits. This demonstrates that the values of seed factors are significantly influenced by the choice of the function. On the other hand, both fitting results of two different functions show that some scatters, i.e., 7 scatters of quadratic function and 9 scatters of cubic function, move outside DT areas. Although the actual seed factor of DT could be smaller and cover them, they have already moved to the areas of BLR or CMB. This indicates that the location of $\gamma$-ray emission region changed in historical flare states, and the soft photon fields could transition from DT to BLR and CMB. 

\subsection{Four Typical Low-Synchrotron-Peaked Blazars}

\begin{figure}
\begin{subfigure}{.5\textwidth}
	\hspace{-0.1in}
    \includegraphics[width = \linewidth]{ 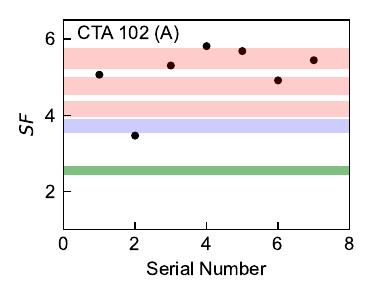}
\end{subfigure}%
\begin{subfigure}{.5\textwidth}
	\hspace{-0.1in}
    \includegraphics[width = \linewidth]{ 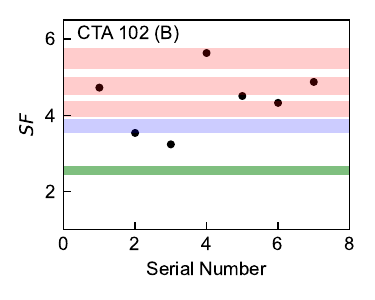}
\end{subfigure}
\begin{subfigure}{.5\textwidth}
	\hspace{-0.1in}
    \includegraphics[width = \linewidth]{ 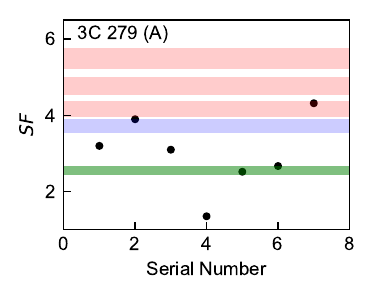}
\end{subfigure}%
\begin{subfigure}{.5\textwidth}
	\hspace{-0.1in}
    \includegraphics[width = \linewidth]{ 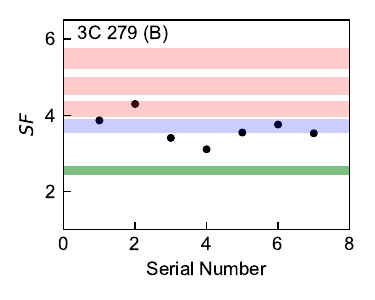}
\end{subfigure}
\begin{subfigure}{.5\textwidth}
	\hspace{-0.1in}
    \includegraphics[width = \linewidth]{ 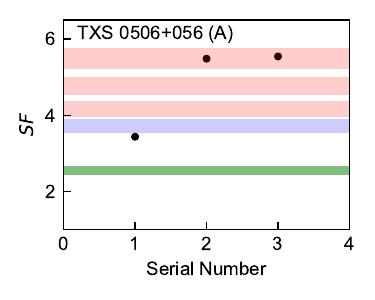}
\end{subfigure}%
\begin{subfigure}{.5\textwidth}
	\hspace{-0.1in}
    \includegraphics[width = \linewidth]{ 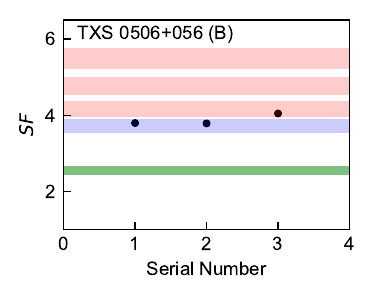}
\end{subfigure}
\begin{subfigure}{.5\textwidth}
	\hspace{-0.1in}
    \includegraphics[width = \linewidth]{ 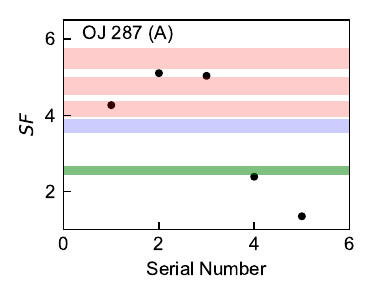}
\end{subfigure}%
\begin{subfigure}{.5\textwidth}
	\hspace{-0.1in}
    \includegraphics[width = \linewidth]{ 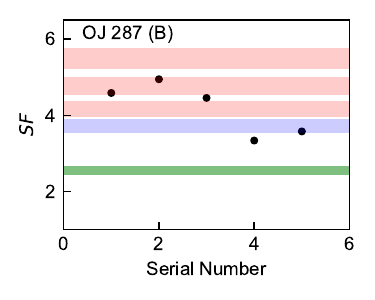}
\end{subfigure}
\caption{From top to bottom, the scatterplots represent the observed seed factor distributions of historical states for CTA 102, 3C 279, TXS 0506+056, and OJ 287, respectively. The left and right panels are derived from SEDs fitted by quadratic and cubic functions, respectively. Red, blue and green areas depict the characteristic seed factors of dusty torus, broad line region and cosmic microwave background, respectively.}
\label{Scatter2}
\end{figure}

In order to further verify the alteration of $\gamma$-ray emission region in different flare epochs, we collected SEDs of four typical LSPs, including 7 SEDs of CTA 102, 7 SEDs of 3C 279, 3 SEDs of TXS 0506+056 and 5 SEDs of OJ 287. The references of these SEDs are presented in Table \ref{table2}. These SEDs fulfilled the same conditions as above. Similarly, we fitted these SEDs with both quadratic and cubic functions (see also Figures \ref{SED CTA 102} to \ref{SED OJ 287}), extracted the peak frequencies and luminosities, and created the scatterplots of observed seed factors (see also Figure \ref{Scatter2}). The scatter distributions still vary significantly under two different function fits. Figure \ref{Scatter2} also shows that the observed seed factors of the same LSPs are variable in different flare states. Some scatters are positioned in red areas, while others are not, indicating multiple locations of $\gamma$-ray emission regions in the same blazar. Some previous studies also support our results. \cite{Pati18} collected multi-wavelength light curves for 3C 279 over 6 years and divided them into three flaring periods. They analyzed the time delays and $\gamma$-ray spectral index, then found that the dominant radiation mechanism and $\gamma$-ray emission regions varied in different periods. Similar to the above LSPs, \cite{Deng23} located the $\gamma$-ray emission region of OT 081 at the edge of BLR during the 2016 multiwavelength flare and at about 1-10 pc away from the black hole during the 2009-2012 orphan X-ray flare. Given to the variability of blazars, broadtime analysis has become a typical approach of {\it{Fermi}}-LAT to reduce the impact of some short-lived flares \citep{Arsioli18b}.

On September 22 2017, the IceCube Observatory detected a $\sim$290 TeV neutrino from the direction of TXS 0506+056 \citep{IceCube18, Padovani18}. The neutrino was produced in the photopion process, in which the $\triangle^{+}(1232)$ resonance contributes the main cross section. Using Delta-approximation, we could derive the external soft photon energy in the AGN frame, 
\begin{equation}
	E_{\rm{0, soft}} \simeq \frac{50 {\rm{eV}}}{1+z} \left(\frac{\delta}{\Gamma}\right) \left( \frac{290{\rm{TeV}}}{E^{\rm{obs}}_{\nu}}\right).
\end{equation}
It shows that the dissipation region was in the BLR. This neutrino event was closely followed by two very high energy $\gamma$-ray flares of TXS 0506+056 \citep{Ansoldi18,Sahakyan18}. These flares were denoted as Flare 2 and Flare 3 in Figure \ref{SED TXS 0506+056}. The seed factors are presented in Figure \ref{Scatter2}, corresponding to serial number 2 and 3. Result of quadratic function depects that the dissipation region was in the DT, contradicting the above calculation. But that of cubic function approximately supports the BLR. This manifests the difference between two functions and possible superiority of cubic function.

\subsection{Model Comparison and Parameter Analysis}

\begin{figure}
	\centering
	\hspace{-0.25in}
    \includegraphics[width=\textwidth]{ 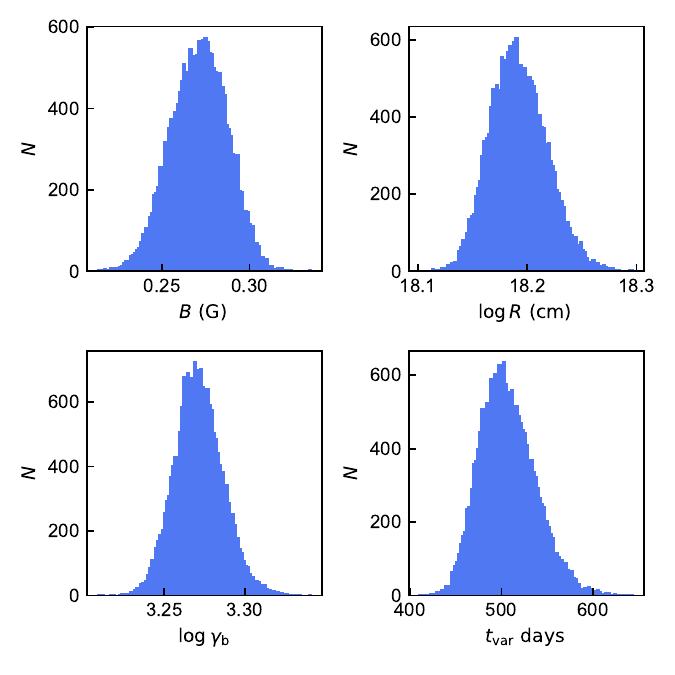}
    \caption{Parameter distribution of OD 166 when Doppler factor is known. Possible value counts of each parameter is 19200 (=32$\times$(17000-2000)/25) in total. The number of bins is 100. We present only one sample here, others are available in machine-readable form.}
    \label{distribution 1}
\end{figure}

We found that different authors adopted various functions to fit the SEDs. \cite{Abdo10SEDfitting} employed the cubic function, while \cite{Chen14,Yang22,Yang23} used a simpler quadratic function. \cite{Xue16_Curvature} fitted the Synchrotron hump with both functions and found the derived Synchrotron luminosity with cubic function is lower. In this work, we tried both of the two to further illuminate the difference. The AICc of each hump derived from two functions are presented on the figures in Appendix. For the low-energy humps, 33 (45 in total, one is null because $n=k+1$) AICc of cubic function are smaller. As for high-energy humps, 35 (45 in total) AICc of cubic function are smaller. This demonstrates a dominance of cubic function over quadratic function both in two kinds of humps. We found that the humps with perfect symmetry could be fitted well (i.e., smaller AICc) with quadratic function, such as low-energy humps in PKS 2123-463, OJ 287 Flare 3, OJ 287 Flare 4, etc., and high-energy humps in PKS 0208-512 Flare 1, PKS 0402-362 Flare 2, PKS 0420+022, etc. The asymmetrical SEDs, which take the most part of our samples, could be explained well with cubic function. On the other hand, we found some quadratic function curves deviate badly from the GeV data points, such as PKS 0208-512 Flare 2, TXS 0506+056 Flare 1, and NARO 512. It is worth noting that GeV data is rarer than other wavebands' in practical, given that the $\gamma$-ray telescopes are relatively scarce. Thus the goodness of fit of data points in GeV band has superiority. For 3C 279 Flare 4 and CTA 102 Flare 3, the quadratic function curves in high-energy hump even continuedly increase, which breaks the physical reality. Therefore, we recommend the cubic function to fit the SEDs.

However, cubic function is not suitable to all the SEDs, such as high-energy humps in TXS 0536+145 and CTA 102 Flare 3 (see Appendix). The hard $\gamma$-ray spectra and lack of higher-energy data make it difficult to form a full peak. We also noticed that the conventional one-zone leptonic or hadronic model could hardly explain the SEDs of bright TeV blazars, which results in extreme values \citep{Abdo11, Cerruti15, Li22}. In this case, some models of inhomogeneous jet were constructed, such as spine-layer model \citep{Ghisellini05, Tavecchio14}, two-zone leptonic model \citep{Shukla15} and two-zone leptohadronic model \citep{Aguilar22}.

Besides collecting the quasi-simultaneous data and fitting the SEDs, we collected the corresponding Doppler factor from \cite{Liodakis18} or observed variability timescales from the original work, and obtained the distribution of some related parameters using formulas in Section \ref{parameter analysis}. For example, Figure \ref{distribution 1} presents the derived parameter distribution of OD 166 when the corresponding Doppler factor from \cite{Liodakis18} is known. There are 19200 possible values for each parameter from the MCMC fitting. The 16-50-84 percentile rule is employed to characterize parameter uncertainty. The detailed results of parameter analysis are listed in Table \ref{table2}. The average value for variability timescale, Doppler factor, magnetic field strength, radius of emission region, and break Lorentz factor of electrons are 36.58 days, 23.40, 0.674 G, $3.540\times10^{16}$ cm, and 845.28, respectively. Table \ref{table2} shows that the physical parameters of a specific blazar changed in different flaring epochs. \cite{Feng22} reproduced the SEDs of various flares belonging to 3C 454.3 under a one-zone leptonic scenario and found similar variations in physical parameters. Interestingly, Table \ref{table2} demonstrates that parameters under two different function fits are analogous, which is unlike the scatterplots. This could be attributed to the effect of symmetry in two humps. It causes more influences in peak frequencies which are mainly used to calculate the observed seed factors, but less in the integral of two humps which are used to calculate the other parameters. We also constrained the Doppler factors and $r$ for the collected LSPs whose observed timescales are known. Table \ref{table2} depicts that all the former derived Doppler factors are consistent with their corresponding lower limit. We noticed that some lower limits of $r$ equal to $0$. Because the results of derivation in Section \ref{Constraint of the Internal gamma gamma absorption} are negative numbers, but $r$ must be non-negative numbers in real world. This indicates that the constraint of internal $\gamma\gamma$-absorption on $r$ is relatively weak.

\renewcommand{\arraystretch}{1.25}
\begin{table*}
    \centering
    \caption{Results of parameter analysis.}
    \scalebox{0.9}{
    \begin{tabular}{ccccccccccc}
        \hline\hline   
         Fermi Name & Source Name  & Time & $t_{\rm{var}}$ & $\delta$ & $B$ & $\log R$ & $\log \gamma_{\rm{b}}$ & $\delta_{\rm{low}}$ & $r_{\rm{low}}$ & Ref \\
         &  & & (days) &  & (G) & (cm) & &   & (pc) & \\
         (1) & (2) & (3) & (4) & (5) & (6) & (7) & (8) & (9) & (10) & (11) \\
         \hline
        4FGL J0242.3+1102$^{*}$ & OD 166 & 2013.04.18-2013.07.27 & $504.12 ^{+32.78}_{-27.82}$ & $4.37$ & $0.272^{+0.015}_{-0.017}$& $18.191^{+0.027}_{-0.025}$ & $3.271^{+0.015}_{-0.014}$&  &  & 1    \\
        &&& $627.67^{+15.68}_{-15.72}$ & & $0.270^{+0.007}_{-0.007} $ & $18.286^{+0.011}_{-0.011}$ & $3.006^{+0.014}_{-0.014}$&  &  & \\
        4FGL J1354.8-1041$^{*}$ & PKS 1352-104 & 2008.08.11-2008.09.11 & $7.45 ^{+0.20 }_{-0.20 }$& $6.85$ & $0.229^{+0.006}_{-0.006}$ & $16.997^{+0.012}_{-0.012}$ &  $3.657^{+0.027}_{-0.027}$ &  &  & 2   \\
        &&& $6.84^{+0.23}_{-0.22}$ &  & $0.283^{+0.009}_{-0.009}$ & $16.959^{+0.014}_{-0.014}$ & $3.280^{+0.044}_{-0.043}$ & & & \\
        4FGL J1549.5+0236$^{*}$ & PKS 1546+027&2010.01.18-2010.03.18 & $3.80 ^{+0.09}_{-0.09}$ & $10.75$ & $0.221^{+0.005}_{-0.006}$ & $16.874^{+0.011}_{-0.011}$ & $3.021^{+0.010}_{-0.010}$&  &  & 3  \\
        &&& $3.39^{+0.09}_{-0.08}$ &  & $0.247^{+0.006}_{-0.006}$ & $16.825^{+0.011}_{-0.010}$ & $2.991^{+0.012}_{-0.011}$ & & & \\
        4FGL J1635.2+3808$^{*}$ & 4C +38.41&2010.02.07-2010.04.07& $0.58^{+0.02}_{-0.02}$ & $26.96$ & $1.274^{+0.057}_{-0.050}$ & $16.160^{+0.017}_{-0.019}$ &$ 2.376^{+0.010}_{-0.011}$ &  &  &  3    \\
        &&& $0.65^{+0.05}_{-0.04}$&  & $1.222^{+0.077}_{-0.080}$ & $16.208^{+0.029}_{-0.027}$ & $2.412^{+0.016}_{-0.014}$& & & \\
        4FGL J1640.4+3945$^{*}$ & NRAO 512 &2010.07.07-2010.09.07& $0.31^{+0.02}_{-0.02}$ & $31.29$ & $0.960^{+0.070}_{-0.059}$ & $15.973^{+0.029}_{-0.032}$ & $2.518^{+0.017}_{-0.018}$ &  &  &  3     \\
        &&& $0.32^{+0.02}_{-0.01}$ &  & $0.974^{+0.048}_{-0.044}$ & $15.980^{0.020}_{-0.021}$ & $2.515^{+0.012}_{-0.012}$ & &  &  \\
        4FGL J0438.4-1254$^{*}$ & PKS 0436-129 &2010.06.01-2010.08.01& $0.05^{+0.00}_{-0.00}$ &  $46.93$ & $0.549^{+0.017}_{-0.018}$ & $15.423^{+0.014}_{-0.014}$ & $2.579^{+0.017}_{-0.016}$ &  &  &  3    \\
        &&& $0.04^{+0.00}_{0.00}$ &  & $0.623^{+0.030}_{-0.032}$ & $15.335^{+0.023}_{-0.021}$ & $2.494^{+0.020}_{-0.020}$ &  &  & \\
        4FGL J2110.2-1021c$^{*}$ & PKS 2107-105 & unknown & $337.97_{-13.39}^{+14.27}$ & $4.03$ & $0.143^{+0.006}_{-0.006}$ & $18.004_{-0.018}^{+0.018}$ & $3.482^{+0.024}_{-0.022}$ & & & 4 \\
        &&& $359.10_{-13.47}^{+13.19}$ & & $0.142_{-0.006}^{+0.006}$ & $18.030_{-0.016}^{0.016}$ & $3.478_{-0.026}^{+0.025}$ & & & \\
        4FGL J1345.5+4453$^{*}$ & B3 1343+451 &2013.07.27& $0.10^{+0.00}_{-0.00}$ & $45.61$ & $0.647^{+0.014}_{-0.013}$ & $15.542^{+0.009}_{-0.009}$ &  $2.624^{+0.010}_{-0.011}$ &  &  &  1     \\
        &&& $0.10^{+0.00}_{0.00}$ &  & $0.628^{+0.014}_{-0.014}$ & $15.517^{+0.009}_{-0.009}$ & $2.538^{+0.017}_{-0.018}$ &  &  &\\

        4FGL J0210.7-5101 & PKS 0208-512 & 2016.10.20-2016.11.10 & $8.45$ & $10.67^{+0.19}_{-0.22}$ & $0.414^{+0.026}_{-0.021}$ & $17.066^{+0.008}_{-0.009}$ & $2.993^{+0.008}_{-0.013}$ &  $3.06$ & $0$  & 5   \\
        &&&& $10.81^{+0.35}_{-0.25}$ & $0.398^{+0.030}_{-0.036}$ & $17.072^{+0.014}_{-0.010}$ & $2.990^{+0.021}_{-0.016}$ & $3.16$ & $0$ &     \\
        & & 2019.12.14-2019.12.25 & $5.21$ & $12.74^{+0.07}_{-0.06}$ & $0.197^{+0.003}_{-0.003}$ & $16.934^{+0.002}_{-0.002}$& $3.160^{+0.002}_{-0.009}$ & $4.30$ & $0.064$  & 5   \\
        &&&& $12.26^{+0.07}_{-0.07}$ & $0.220^{+0.004}_{-0.004}$ & $16.917^{+0.003}_{-0.002}$ & $3.130^{+0.011}_{-0.010}$& $4.75$ & $0.069$ &    \\

        4FGL J0403.9-3605 & PKS 0402-362 & 2010.01.20-2010.03.01 & $2.20$ & $17.92^{+0.08}_{-0.08}$ & $0.480^{+0.008}_{-0.008}$ & $16.625^{+0.002}_{-0.002}$ & $2.771^{+0.008}_{-0.007}$ & $7.23$ & $0.043$  & 6    \\
        &&&& $17.79^{+0.08}_{-0.08}$ & $0.435^{+0.008}_{-0.007}$ & $16.622^{+0.002}_{-0.002}$& $2.735^{+0.012}_{-0.012}$& $7.23$ & $0.043$ &   \\ 
        & & 2010.03.01-2010.03.21 & $2.20$ & $18.58^{+0.11}_{-0.11}$ & $0.455^{+0.010}_{-0.010}$ & $16.640^{+0.003}_{-0.002}$& $2.948^{+0.012}_{-0.012}$& $7.34$ & $0.059$ & 6   \\
        &&&& $18.63^{+0.11}_{-0.11}$ & $0.460^{+0.010}_{-0.010}$ & $16.641^{+0.002}_{-0.003}$ & $2.804^{+0.021}_{-0.020}$ & $7.59$ & $0.06$  &   \\  
        & & 2011.09.20-2011.10.04 & $2.75$ & $23.22^{+0.11}_{-0.11}$ & $0.371^{+0.008}_{-0.008}$ & $16.834^{+0.002}_{-0.002}$ & $2.765^{+0.011}_{-0.010}$& $5.70$ & $0.044$  & 6  \\
        &&&& $23.17^{+0.10}_{-0.10}$ & $0.384^{+0.007}_{-0.006}$ & $16.642^{+0.003}_{-0.003}$ & $2.804^{+0.021}_{-0.020}$ & $5.88$ & $0.045$ &  \\ 
        & & 2014.07.31-2014.08.17 & $3.57$ & $19.64^{+0.08}_{-0.08}$ & $0.562^{+0.009}_{-0.009}$ & $16.875^{+0.002}_{-0.002}$ & $2.756^{+0.011}_{-0.011}$& $5.01$ & $0.023$ & 6    \\
        &&&& $19.61^{+0.09}_{-0.09}$ & $0.558^{+0.009}_{-0.010}$ & $16.873^{+0.002}_{-0.002}$ & $2.688^{+0.010}_{-0.011}$ & $5.01$ & $0.023$  &   \\

        4FGL J0530.9+1332 & PKS 0528+134 & 2009.09.08 & $0.94$ & $27.06^{+0.27}_{-0.28}$ & $0.648^{+0.025}_{-0.024}$& $16.335^{+0.004}_{-0.005}$& $2.976^{+0.031}_{-0.029}$& $6.48$ & $0.039$  & 7  \\
        &&&& $28.21^{+0.65}_{-0.62}$ & $1.290^{+0.153}_{-0.136}$& $16.353^{+0.010}_{-0.010}$& $2.829^{+0.029}_{-0.029}$& $6.73$ & $0.041$ &  \\
        4FGL J0539.6+1432 &  TXS 0536+145 & 2012.03.04-2012.04.04 & $1.00$ & $30.33^{+0.46}_{-0.45}$& $0.907^{+0.054}_{-0.054}$&$16.328^{+0.007}_{-0.006}$&$2.928^{+0.035}_{-0.033}$& $7.71$ & $0.061$ & 8   \\
        &&&& $29.90^{+0.62}_{-0.59}$&$0.721^{+0.049}_{-0.045}$ & $16.322^{+0.009}_{-0.009}$ & $3.051^{+0.036}_{-0.036}$& $9.45$ & $0.071$ &  \\
        4FGL J2253.9+1609 & 3C 454.3 & 2011.01.27-2011.02.08 & $0.78$ & $24.39^{+0.11}_{-0.11}$& $0.758^{+0.015}_{-0.015}$& $16.425^{+0.002}_{-0.002}$& $2.715^{+0.007}_{-0.007}$&  $7.47$ & $0$ & 9  \\
        &&&& $25.27^{+0.22}_{-0.21}$& $0.933^{+0.038}_{-0.036}$ & $16.440^{+0.004}_{-0.004}$& $2.739^{+0.008}_{-0.007}$& $7.85$ & $0$  &\\
        & & 2010.11.11-2010.12.06 & $0.78$ & $26.98^{+0.14}_{-0.14}$&$0.387^{+0.009}_{-0.009}$&$16.468^{+0.002}_{-0.002}$& $3.007^{+0.018}_{-0.016} $& $8.32$ & $0.066$  & 9  \\
        &&&& $27.02^{+0.13}_{-0.13}$& $0.386^{+0.009}_{-0.009}$& $16.469^{+0.002}_{-0.002}$& $3.037^{+0.018}_{-0.018}$& $7.77$ & $0.063$ &  \\
        & & 2011.05.19-2012.09.30 & $0.56$ & $19.20^{+0.14}_{-0.14}$& $0.698^{+0.017}_{-0.017}$& $16.172^{+0.003}_{-0.003}$& $2.809^{+0.008}_{-0.008}$& $5.21$ & $0$  & 10   \\
        &&&& $20.15^{+0.14}_{-0.14}$& $0.711^{+0.016}_{-0.016}$ & $16.193^{+0.003}_{-0.003}$& $2.583^{+0.015}_{-0.015}$& $5.38$ & $0$   &  \\
        
        4FGL J2329.3-4955 & PKS 2326-502 & 2010.07.31-2010.09.29 & $0.53$ & $17.05^{+0.23}_{-0.24}$& $0.173^{+0.012}_{-0.012}$& $16.190^{+0.006}_{-0.006}$& $2.995^{+0.025}_{-0.024}$& $2.86$ & $0$ & 11   \\
        &&&& $20.03^{+1.16}_{-1.18}$& $0.456^{+0.148}_{-0.119}$& $16.260^{+0.024}_{-0.026}$& $2.945^{+0.050}_{-0.058}$& $3.87$ & $0$  &  \\
        & & 2012.06.25-2012.07.05 & $1.50$ & $18.39^{+0.41}_{-0.37}$ & $0.220^{+0.024}_{-0.020}$& $16.674^{+0.010}_{-0.009}$& $2.934^{+0.026}_{-0.027}$& $1.88$ & $0$  & 11    \\
        &&&& $20.10^{+0.38}_{-0.33}$ & $0.437^{+0.024}_{-0.025}$& $16.713^{+0.008}_{-0.007}$& $3.014^{+0.017}_{-0.016}$& $1.98$ & $0$  &  \\

        4FGL J2126.3-4605 & PKS 2123-463 & 2011.12.10-2011.12.19 & $1.97$ & $19.57^{+0.26}_{-0.26}$ & $0.280^{+0.013}_{-0.012}$ & $16.572^{+0.006}_{-0.006}$ & $3.456^{+0.057}_{-0.051}$& $4.20$ & $0$ & 12  \\
        &&&& $19.82^{+0.30}_{-0.28}$ & $0.269^{+0.014}_{-0.013}$ & $16.578^{+0.007}_{-0.006}$ & $3.467^{+0.062}_{-0.062}$ & $4.35$ & $0$  &  \\
        4FGL J0108.6+0134 & 4C +01.02 & 2015.11.23-2015.12.15 & $0.66$ & $31.85^{+0.11}_{-0.11}$ & $0.271^{+0.003}_{-0.003}$ & $16.245^{+0.002}_{-0.002}$ & $2.695^{+0.003}_{-0.003}$ & $6.84$ & $0.051$ & 13 \\
        &&&& $33.00^{+0.12}_{-0.12}$ & $0.343^{+0.004}_{-0.004}$ & $16.260^{+0.0062}_{-0.002}$ & $2.803^{+0.001}_{-0.003}$& $8.43$ & $0.061$  &  \\

         \hline
    \end{tabular}
    }
    \label{table2}
\end{table*}

\addtocounter{table}{-1}

\begin{table*}
    \centering
    \caption{-$continued$.}
    \scalebox{0.9}{
    \begin{tabular}{ccccccccccc}
        \hline\hline   
         Fermi Name & Source Name  & Time & $t_{\rm{var}}$ & $\delta$ & $B$ & $\log R$ & $\log \gamma_{\rm{b}}$ & $\delta_{\rm{low}}$ & $r_{\rm{low}}$ & Ref \\
         &  & & (days) &  & (G) & (cm) & &   & (pc) & \\
        (1) & (2) & (3) & (4) & (5) & (6) & (7) & (8) & (9) & (10) & (11) \\
         \hline
        4FGL J1256.1-0547 & 3C 279 & 2014.04.03-2014.04.07 & $0.09$ & $41.50^{+0.22}_{-0.23}$ & $1.354^{+0.030}_{-0.030}$ & $15.799^{+0.002}_{-0.002}$& $2.445^{+0.013}_{-0.014}$& $7.96$ & $0.005$ & 14  \\
        &&&& $42.00^{+0.24}_{-0.24}$ & $1.181^{+0.030}_{-0.029}$ & $15.804^{+0.002}_{-0.002}$ & $2.422^{+0.002}_{-0.012}$ & $7.96$ & $0.005$ &  \\
        & & 2012.04.03 & $0.08$ & $44.36^{+0.30}_{-0.29}$ & $0.562^{+0.012}_{-0.013}$ & $15.795^{+0.003}_{-0.002}$& $2.695^{+0.013}_{-0.013}$& $8.79$ & $0$ & 15 \\
        &&&& $43.58^{+0.27}_{-0.27}$ & $0.593^{+0.014}_{-0.014}$ & $15.787^{+0.003}_{-0.003}$ & $2.684^{+0.002}_{-0.011}$ & $9.63$ & $0$ & \\
        & & 2011.02.08-2011.04.12 & $3.49$ & $13.49^{+0.12}_{-0.13}$ & $1.096^{+0.044}_{-0.043}$ & $16.900^{+0.009}_{-0.009}$ & $2.526^{+0.028}_{-0.028} $ & $2.87$ & $0$  & 16\\
        &&&& $14.33^{+0.35}_{-0.34}$ & $1.514^{+0.181}_{-0.167}$ & $16.926^{+0.010}_{-0.011}$ & $2.530^{+0.010}_{-0.031}$ & $2.87$ & $0$  & \\
        & & 2011.06.01-2011.06.08 & $3.49$ & $15.01^{+0.46}_{-0.39}$ & $0.456^{+0.038}_{-0.039}$ & $16.945^{+0.012}_{-0.013}$ & $3.389^{+0.020}_{-0.021} $ & $3.30$ & $0$ & 16  \\
        &&&& $14.88^{+0.15}_{-0.16}$ & $0.931^{+0.034}_{-0.032}$ & $16.942^{+0.004}_{-0.005}$ & $2.523^{+0.004}_{-0.009}$ & $3.30$ & $0$  &  \\
        & & 2014.03.25-2014.04.02 & $11.61$ & $11.73^{+0.06}_{-0.06}$ & $0.358^{+0.007}_{-0.006}$ & $17.361^{+0.002}_{-0.002}$ & $3.046^{+0.009}_{-0.010 }$ & $2.76$ & $0$ & 16  \\
        &&&& $11.23^{+0.05}_{-0.05}$ & $0.402^{+0.007}_{-0.007}$ & $17.342^{+0.002}_{-0.002}$ & $3.011^{0.002}_{-0.008}$ & $3.07$ & $0$  &  \\
        & & 2015.06.16 & $3.17$ &  $18.02^{+0.13}_{-0.11}$ & $0.106^{+0.003}_{-0.003}$ & $16.973^{+0.003}_{-0.003}$ & $3.260^{+0.016}_{-0.017}$ & $4.31$ & $0$  & 16 \\
        &&&& $16.39^{+0.11}_{-0.10}$ & $0.131^{+0.003}_{-0.003}$ & $16.943^{+0.003}_{-0.003}$ & $3.244^{+0.003}_{-0.015}$ & $4.44$ & $0$  &   \\
        & & 2010.01.14-2010.06.28 & $2.89$ & $12.24^{+0.08}_{-0.08}$ & $0.745^{+0.022}_{-0.021}$ & $16.776^{+0.004}_{-0.004}$ & $2.552^{+0.011}_{-0.011} $ & $3.48$ & $0$  & 10 \\
        &&&& $12.75^{+0.09}_{-0.10}$ & $0.708^{+0.025}_{-0.022}$ & $16.794^{+0.003}_{-0.003}$ & $2.416^{+0.003}_{-0.007}$ & $3.48$ & $0$  &  \\
        
        4FGL J2232+1143 & CTA 102 & 2016.12.23 & $0.56$ & $27.80^{+0.12}_{-0.12}$ & $0.771^{+0.013}_{-0.013}$ & $16.298^{+0.002}_{-0.002}$ & $2.940^{+0.002}_{-0.009}$ & $6.97$ & $0$  & 17 \\
        &&&& $27.25^{+0.32}_{-0.25}$ & $0.702^{+0.025}_{-0.025}$ & $16.289^{+0.005}_{-0.005}$ & $2.962^{+0.013}_{-0.013}$ &  $6.97$ & $0$ & \\
        & & 2012.09.18-2012.10.03 & $3.93$ & $21.20^{+0.15}_{-0.15}$ & $0.225^{+0.005}_{-0.005}$ & $17.022^{+0.003}_{-0.003}$ & $3.247^{+0.003}_{-0.009}$ &  $6.06$ & $0.105$ & 18 \\
        &&&& $20.83^{+0.27}_{-0.24}$ & $0.224^{+0.008}_{-0.009}$ & $17.018^{+0.005}_{-0.006}$ & $3.247^{+0.011}_{-0.011}$ & $5.98$ & $0.104$ &  \\
        & & 2017.04.19 & $0.17$ & $68.81^{+1.79}_{-1.81}$ & $0.170^{+0.013}_{-0.014}$ & $16.172^{+0.012}_{-0.011}$ & $3.135^{+0.014}_{-0.014}$ & $11.33$ & $0$ & 19  \\
        &&&& $57.05^{+2.91}_{-2.53}$ & $0.243^{+0.036}_{-0.035}$ & $16.091^{+0.020}_{-0.021}$ & $3.117^{+0.029}_{-0.027}$ & $11.70$ & 0  & \\
        & & 2017.01.08 & $0.05$ & $69.69^{+0.25}_{-0.25}$ & $1.487^{+0.027}_{-0.028}$ & $15.601^{+0.002}_{-0.002}$ & $3.061^{+0.006}_{-0.006}$ & $16.60$ & $0.040$ & 20 \\
        &&&& $76.81^{+0.31}_{-0.30}$ & $1.964^{+0.030}_{-0.030}$ & $15.643^{+0.002}_{-0.002}$ & $3.120^{+0.006}_{-0.006}$ & $15.95$ & $0.039$ &  \\
        & & 2016.12.30 & $0.50$ & $38.33^{+0.18}_{-0.18}$ & $0.573^{+0.012}_{-0.012}$ & $16.387^{+0.002}_{-0.002}$ & $3.371^{+0.009}_{-0.009}$ & $9.96$ & $0.072$ & 21 \\
        &&&& $31.29^{+0.67}_{-0.66}$ & $0.217^{+0.024}_{-0.022}$ & $16.299^{+0.009}_{-0.009}$ & $3.073^{+0.065}_{-0.066}$ & $10.45$ & $0.075$  &   \\
        & & 2016.12.26-2016.12.31 & $0.21$ & $49.42^{+0.17}_{-0.17}$ & $1.421^{+0.021}_{-0.022}$ & $16.121^{+0.001}_{-0.002}$ & $3.073^{+0.001}_{-0.006}$ & $16.15$ & $0.064$  & 22 \\
        &&&& $52.23^{+0.38}_{-0.37}$ & $1.954^{+0.069}_{-0.065}$ & $16.150^{+0.003}_{-0.003}$ & $2.663^{+0.026}_{-0.027}$ & $13.39$ & $0.555$  & \\
        & & 2011.09.04-2011.10.18 & $2.49$ & $14.77^{+0.11}_{-0.11}$ & $0.421^{+0.012}_{-0.012}$ & $16.669^{+0.003}_{-0.003}$ & $3.179^{+0.011}_{-0.011}$ & $3.46$ & $0$  & 10 \\ 
        &&&& $16.08^{+0.17}_{-0.16}$ & $0.433^{+0.019}_{-0.019}$ & $16.706^{+0.004}_{-0.004}$ & $2.848^{+0.036}_{-0.038}$ & $3.63$ & $0$  &  \\

        4FGL J0509.4+0542 & TXS 0506+056 & 2018.10.06 & $0.14$ & $25.07^{+0.14}_{-0.13}$ & $1.646^{+0.033}_{-0.033}$ & $15.833^{+0.002}_{-0.002}$ & $3.079^{+0.008}_{-0.008}$ &  $9.13$ & $0$ & 23\\
        &&&& $23.86^{+0.11}_{-0.12}$ & $1.529^{+0.029}_{-0.029}$ & $15.816^{+0.002}_{-0.002}$ & $3.359^{+0.014}_{-0.014}$ & $7.34$ & $0$  & \\
        & & 2017.10.03-2017.10.04 &  $1.16$ & $13.30^{+0.10}_{-0.10}$ & $1.040^{+0.025}_{-0.024}$ & $16.493^{+0.003}_{-0.003}$ & $3.819^{+0.046}_{-0.042}$ & $10.76$ & $0.028$ &  24 \\
        &&&& $14.02^{+0.06}_{-0.06}$ & $0.962^{+0.017}_{-0.017}$ & $16.498^{+0.002}_{-0.002}$ & $3.551^{+0.034}_{-0.038}$ & $9.83$ & $0.023$  & \\
        & & 2017.10.31 & $1.16$ & $13.87^{+0.27}_{-0.31}$ & $0.661^{+0.038}_{-0.047}$ & $16.493^{+0.008}_{-0.010}$ & $3.648^{+0.016}_{-0.016}$ & $9.97$ & $0.029$ &  24 \\ 
        &&&& $13.30^{+0.31}_{-0.35}$ & $0.765^{+0.059}_{-0.057}$ & $16.475^{+0.010}_{-0.011}$ & $3.669^{+0.019}_{-0.020}$ & $9.97$ & $0.028$  &   \\

        4FGL J0854.8+2006 & OJ 287 & 2015.12.03 & $1.00$ & $14.92^{+0.07}_{-0.07}$ & $0.899^{+0.017}_{-0.017}$ & $16.471^{+0.002}_{-0.002}$ & $3.107^{+0.029}_{-0.031}$ & $3.32$ & $0$ & 25 \\
        &&&& $14.87^{+0.14}_{-0.09}$ & $0.910^{+0.021}_{-0.028}$ & $16.470^{+0.004}_{-0.003}$ & $3.115^{+0.040}_{-0.036}$ & $3.07$ & $0$  &  \\
        & & 2008.08.11-2008.11.11 & $10.70$ & $7.40^{+0.03}_{-0.03}$ & $0.763^{+0.012}_{-0.011}$ & $17.196^{+0.002}_{-0.002}$ & $3.210^{+0.008}_{-0.008}$ & $2.17$ & $0$ & 26 \\
        &&&& $7.78^{+0.03}_{-0.03}$ & $0.961^{+0.017}_{-0.017}$ & $17.218^{+0.002}_{-0.002}$ & $3.157^{+0.007}_{-0.007}$ & $1.83$ & $0$ & \\
        & & 2009.10.20-2009.10.27 &  $2.50$ & $10.16^{+0.12}_{-0.12}$ & $0.492^{+0.026}_{-0.024}$ & $16.702^{+0.005}_{-0.005}$ & $3.261^{+0.059}_{-0.053}$ & $4.25$ & $0$ & 27 \\
        &&&& $10.30^{+0.14}_{-0.14}$ & $0.501^{+0.029}_{-0.028}$ & $16.708^{+0.006}_{-0.005}$ & $3.267^{+0.046}_{-0.043}$ & $4.02$ & $0$  &  \\
        & & 2009.10.27-2009.11.17 &  $2.50$ & $11.57^{+0.17}_{-0.17}$ & $1.344^{+0.076}_{-0.070}$ & $16.759^{+0.006}_{-0.006}$ & $2.846^{+0.037}_{-0.034}$ & $3.04$ & $0$  &  27  \\ 
        &&&& $12.81^{+1.66}_{-1.12}$ & $1.925^{+1.594}_{-0.712}$ & $16.803^{+0.053}_{-0.040}$ & $2.375^{+0.534}_{-0.352}$ & $3.04$ & $0$  &  \\
        & & 2009.11.17-2009.12.19 &  $2.50$ & $11.11^{+0.19}_{-0.20}$ & $1.821^{+0.111}_{-0.101}$ & $16.741^{+0.007}_{-0.008}$ & $2.832^{+0.035}_{-0.031}$ & $3.04$ & $0$  &  27  \\ 
        &&&& $11.04^{+0.20}_{-0.19}$ & $1.812^{+0.113}_{-0.110}$ & $16.738^{+0.008}_{-0.008}$ & $2.874^{+0.040}_{-0.041}$ & $3.04$ & $0$ &  \\
        
        \hline
    \end{tabular}
    }
\end{table*}

\addtocounter{table}{-1}
\begin{table*}
	    \begin{tablenotes}
        \item Note: Column (1) and (2) give the Fermi name and source name, respectively. The sources marked with $^{*}$ are the DT dominated LSPs with observed Doppler factors, and the others possess observed variability timescales. Column (3) gives the observed time period. Column (4) gives the observed or derived variability timescales. Column (5) gives the observed or derived Doppler factors. Column (6), (7) and (8) are the  derived magnetic field strength, the derived radius of emission region and the derived break Lorentz factor of relativistic electrons, respectively. Column (9) and (10) give the lower limits of Doppler factors and distance between the black hole and emission region, which are derived from internal $\gamma\gamma$ absorption. Column (11) is the related reference. The table contains some rows with double sub-rows. The upper one is the parameter obtained by fitting SEDs with quadratic function, and the lower one is the parameter obtained by fitting SEDs with cubic function. 
        \item References: (1) \cite{Sahakyan20}; (2) \cite{Ghisellini10_SED}; (3) \cite{Tan20}; (4) \cite{Sahakyan20}; (5) \cite{Ammenadka22}; (6) \cite{Das23}; (7) \cite{Palma11}; (8) \cite{Orienti14}; (9) \cite{Das20}; (10) \cite{Roy21}; (11) \cite{Dutka17}; (12) \cite{DAmmando12}; (13) \cite{Malik22}; (14) \cite{Patel21}; (15) \cite{Hayashida15}; (16) \cite{Fraija19}; (17) \cite{Zacharias17}; (18) \cite{Pacciani14}; (19) \cite{Gasparyan18}; (20) \cite{Prince18}; (21) \cite{Zacharias19}; (22) \cite{Sahakyan20_CTA102}; (23) \cite{Acciari22}; (24) \cite{Sahakyan18}; (25) \cite{Oikonomou19}; (26) \cite{Chen10}; (27) \cite{Kushwaha13}.
    \end{tablenotes}
\end{table*}

\section{Conclusions}
\label{Conclusion}

In this work, we calculated the observed seed factors of 1138 LSPs and the characteristic seed factors of four external photon fields, then plotted the histogram distribution to locate the $\gamma$-ray emission region. SEDs related to historical flare states were collected to investigate the variable locations. These SEDs were fitted by both quadratic and cubic functions using the MCMC method. Furthermore, we derived some parameters of emission region and employed a constraint of internal $\gamma \gamma$-absorption to verify the derivation. Our main results are as follows:

\begin{itemize}
    \item[1.]We find that DT dominates the soft photon fields of LSPs and $\gamma$-ray emission regions of LSPs are mainly located at 1-10 pc. Histogram shows that the corresponding distribution of BL Lacs peaks at the area of BLR, but this area could also be covered by the actual value of DT. CMB and starlight make little contribution to the $\gamma$-ray emission of LSPs.

    \item[2.]The locations of $\gamma$-ray emission region of LSPs are variable in different flare epochs. Most $\gamma$-ray emission regions are within the DT, but the soft photon fields could also transition to BLR and CMB. 

    \item[3.]The cubic function is better than the quadratic function to fit the SEDs of blazars. We find that some high-energy humps of blazars cannot be fitted well by quadratic function due to the symmetry of the SEDs.
    
\end{itemize}

\begin{acknowledgement}
We thank the anonymous referee for valuable comments and constructive suggestions. This work is supported by the National Natural Science Foundation of China (NSFC) under Grant No. 12203043, 12203024 and 12263009.
\end{acknowledgement}

\section{Data Availability}

The data derived in this paper are available on Zenodo with a DOI 10.5281/zenodo.11119117 \footnote{https://zenodo.org/records/11119118}.
\printendnotes

\printbibliography

\appendix

\section{SED Fitting Results}

\begin{figure*}
    \caption{SED fitting results of the blazars dominated by dusty torus with Doppler factors.}
    \begin{subfigure}{.25\textwidth}
        \includegraphics[width = \linewidth]{ 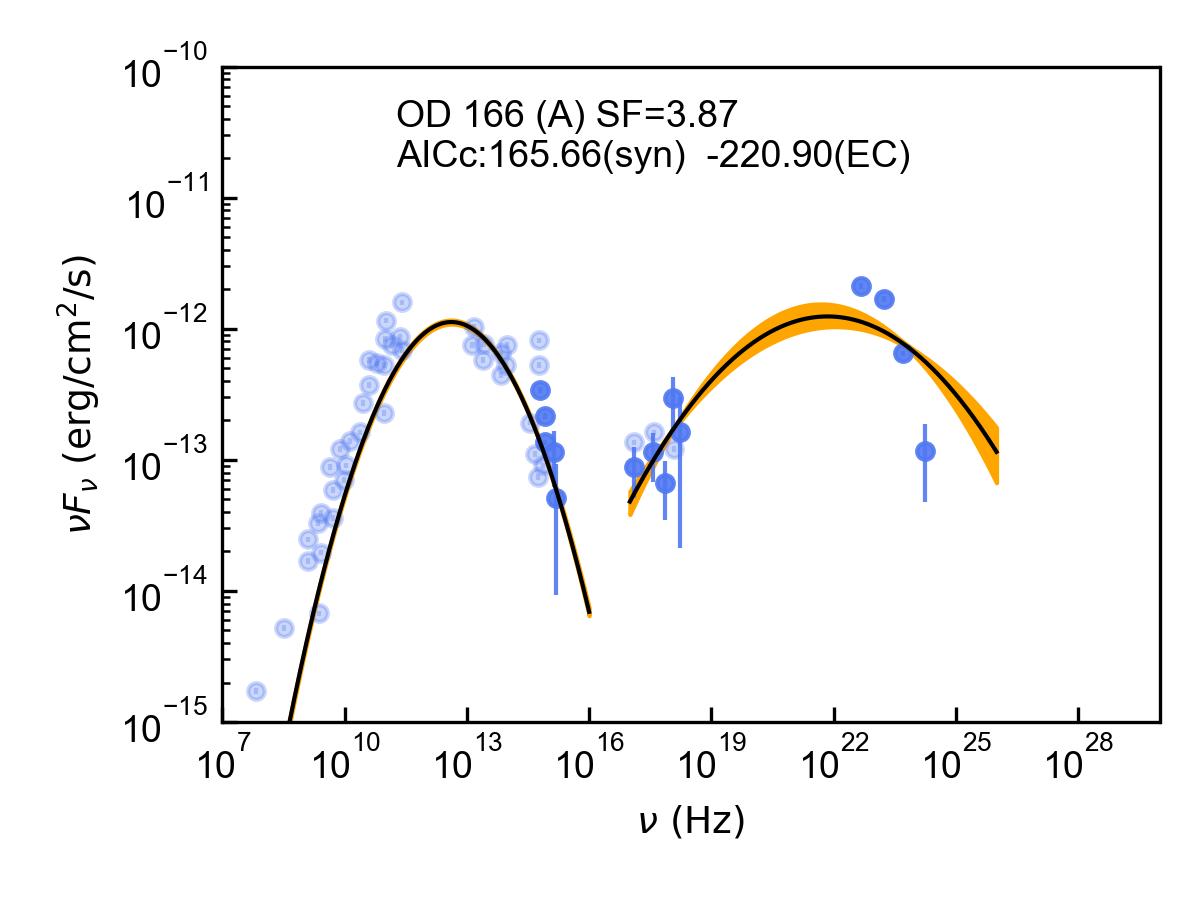}
    \end{subfigure}%
    \begin{subfigure}{.25\textwidth}
        \includegraphics[width = \linewidth]{ 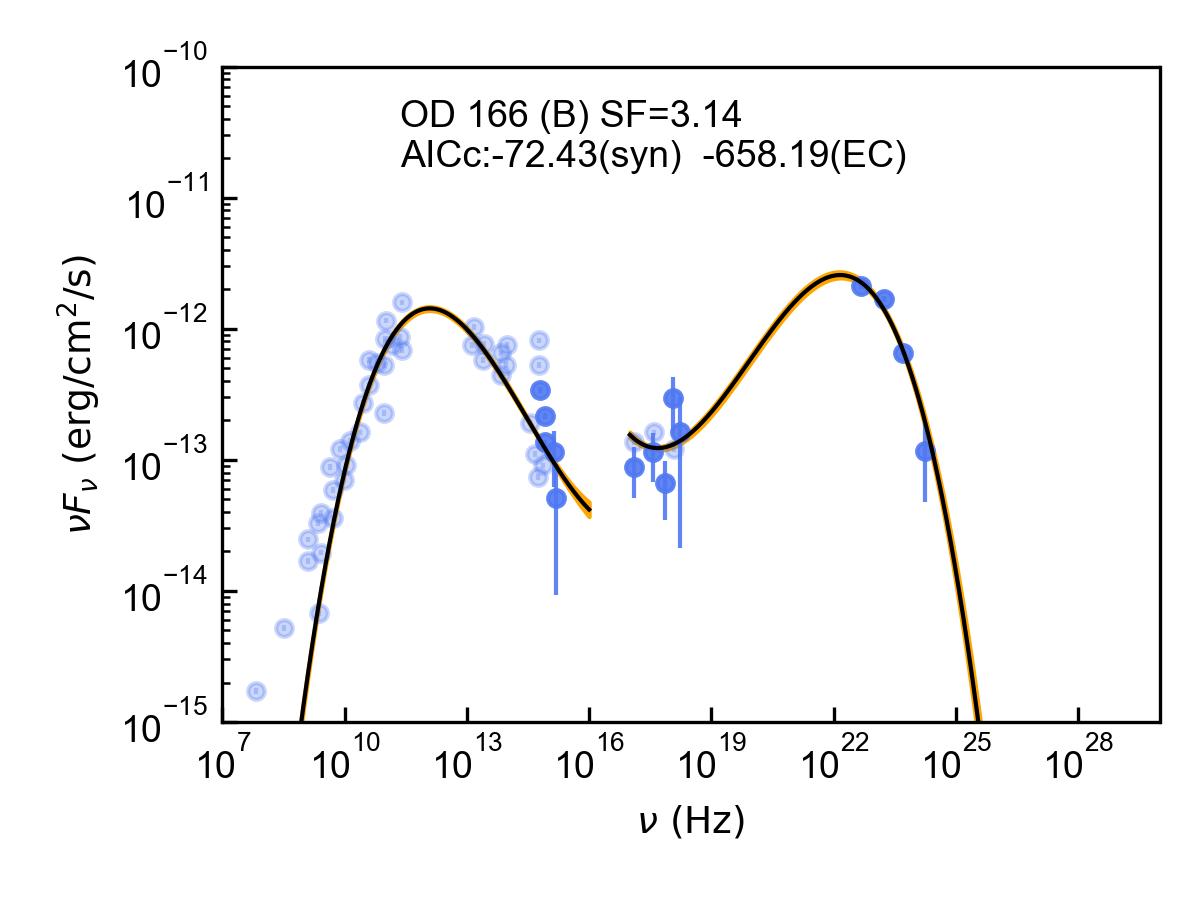}
    \end{subfigure}%
    \begin{subfigure}{.25\textwidth}
        \includegraphics[width = \linewidth]{ 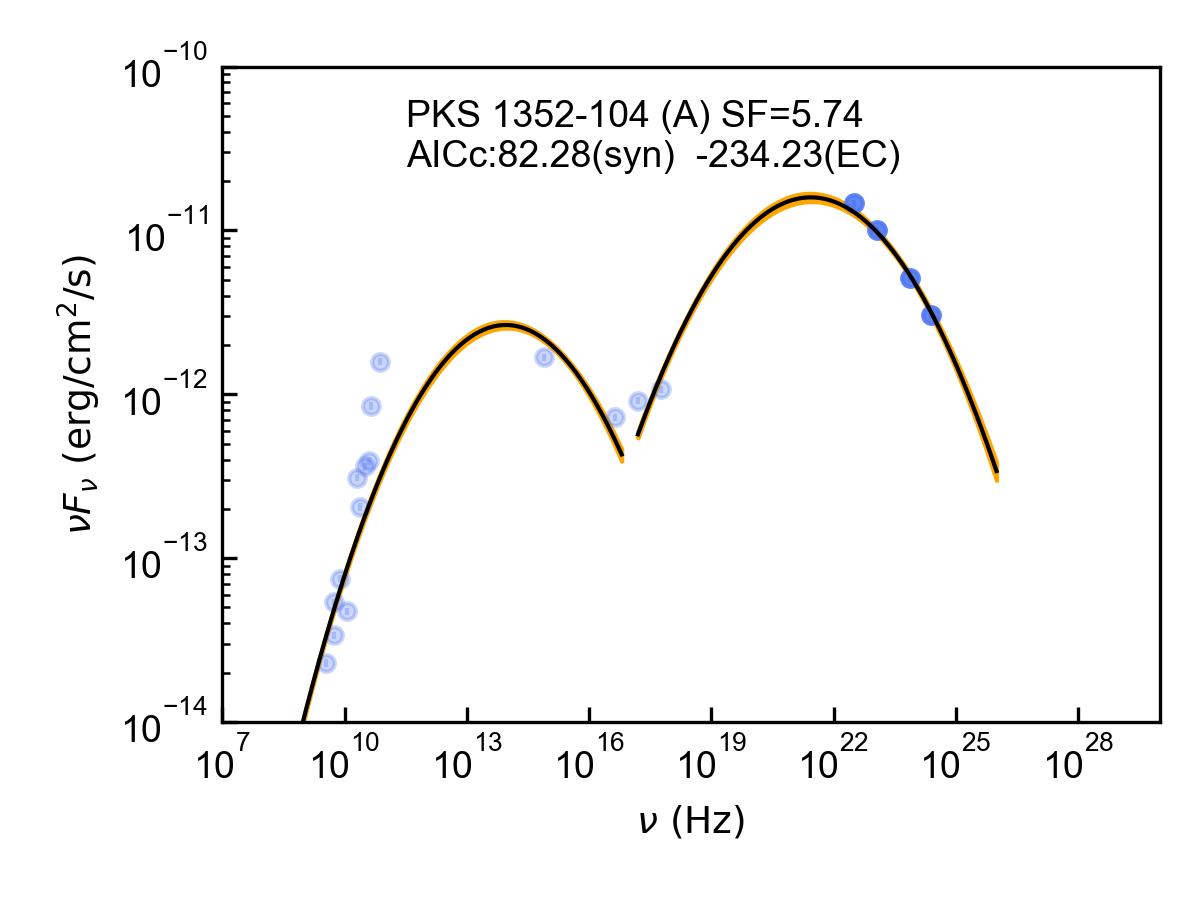}
    \end{subfigure}%
    \begin{subfigure}{.25\textwidth}
        \includegraphics[width = \linewidth]{ 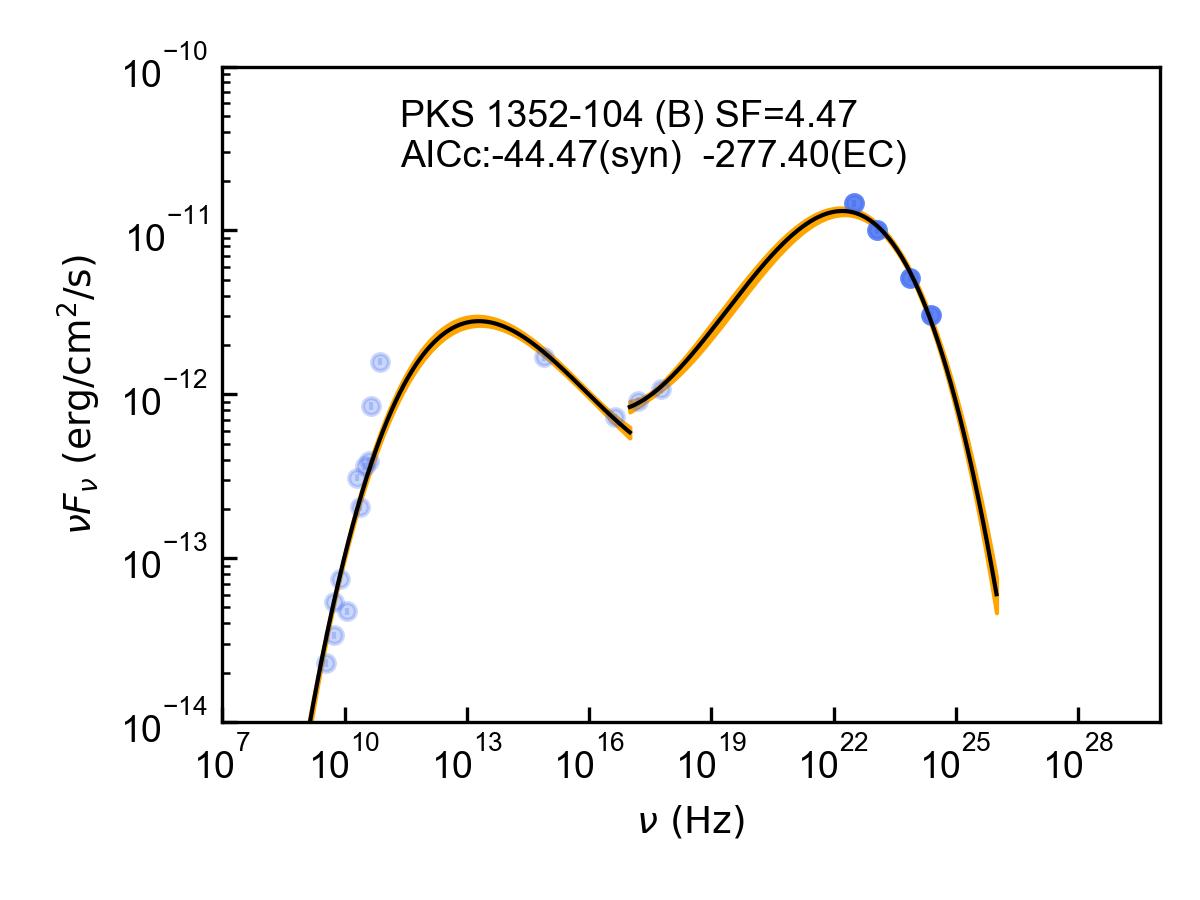}
    \end{subfigure}

    \begin{subfigure}{.25\textwidth}
        \includegraphics[width = \linewidth]{ 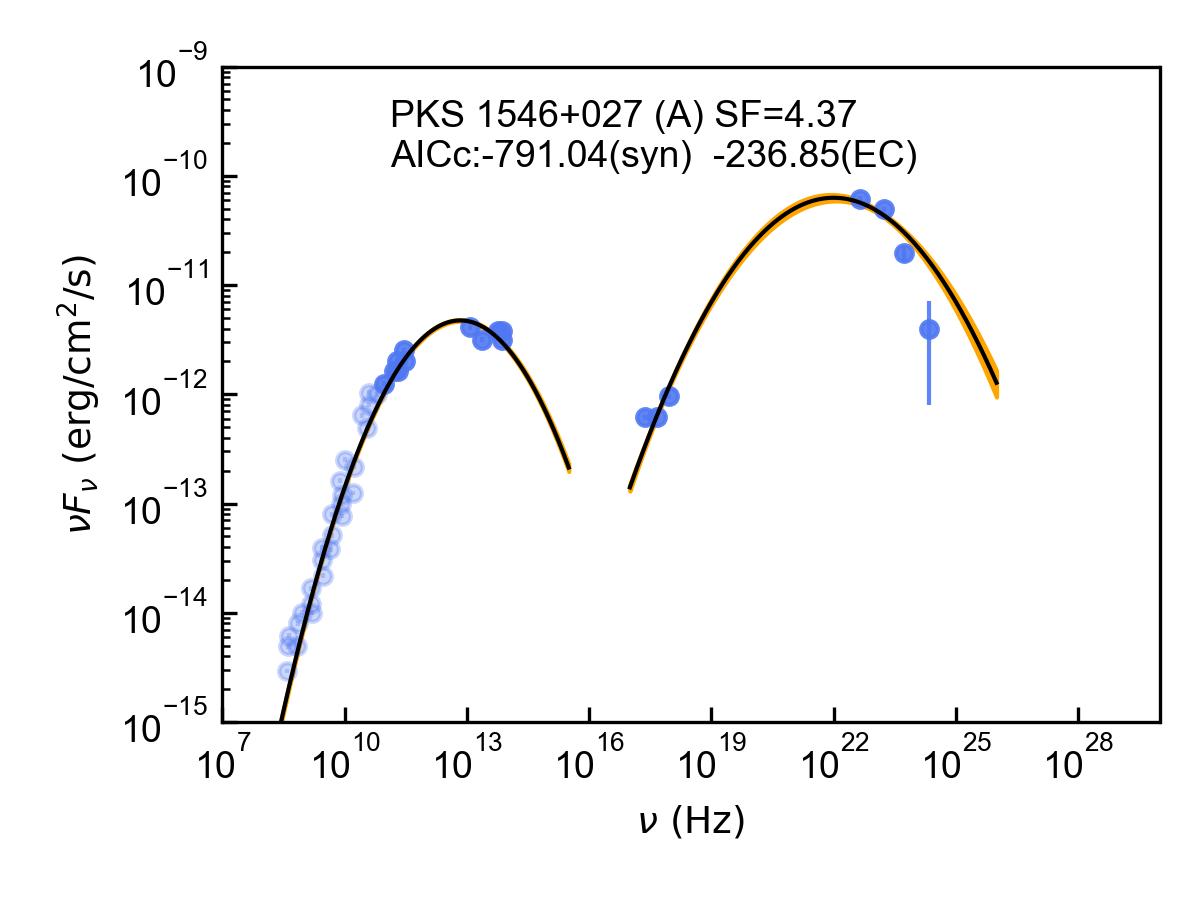}
    \end{subfigure}%
    \begin{subfigure}{.25\textwidth}
        \includegraphics[width = \linewidth]{ 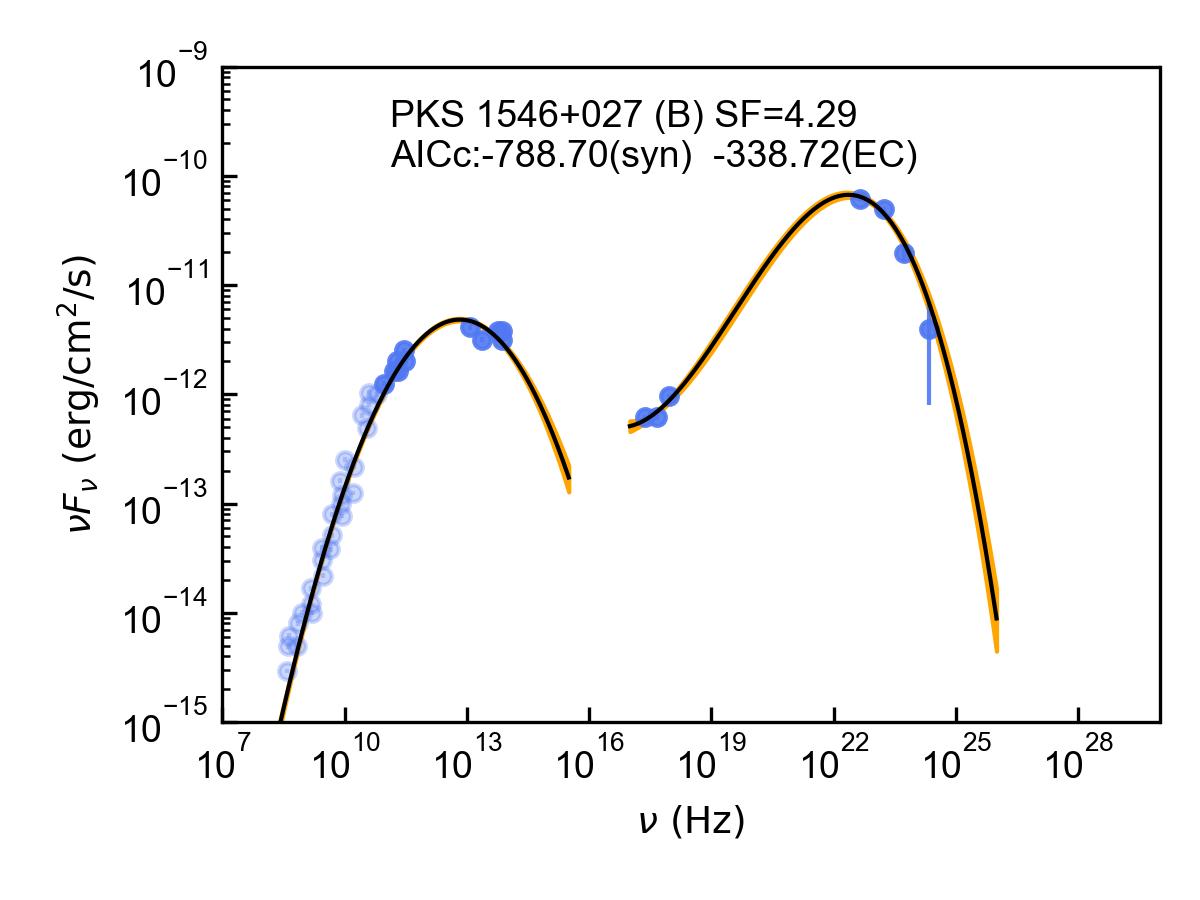}
    \end{subfigure}%
    \begin{subfigure}{.25\textwidth}
        \includegraphics[width = \linewidth]{ 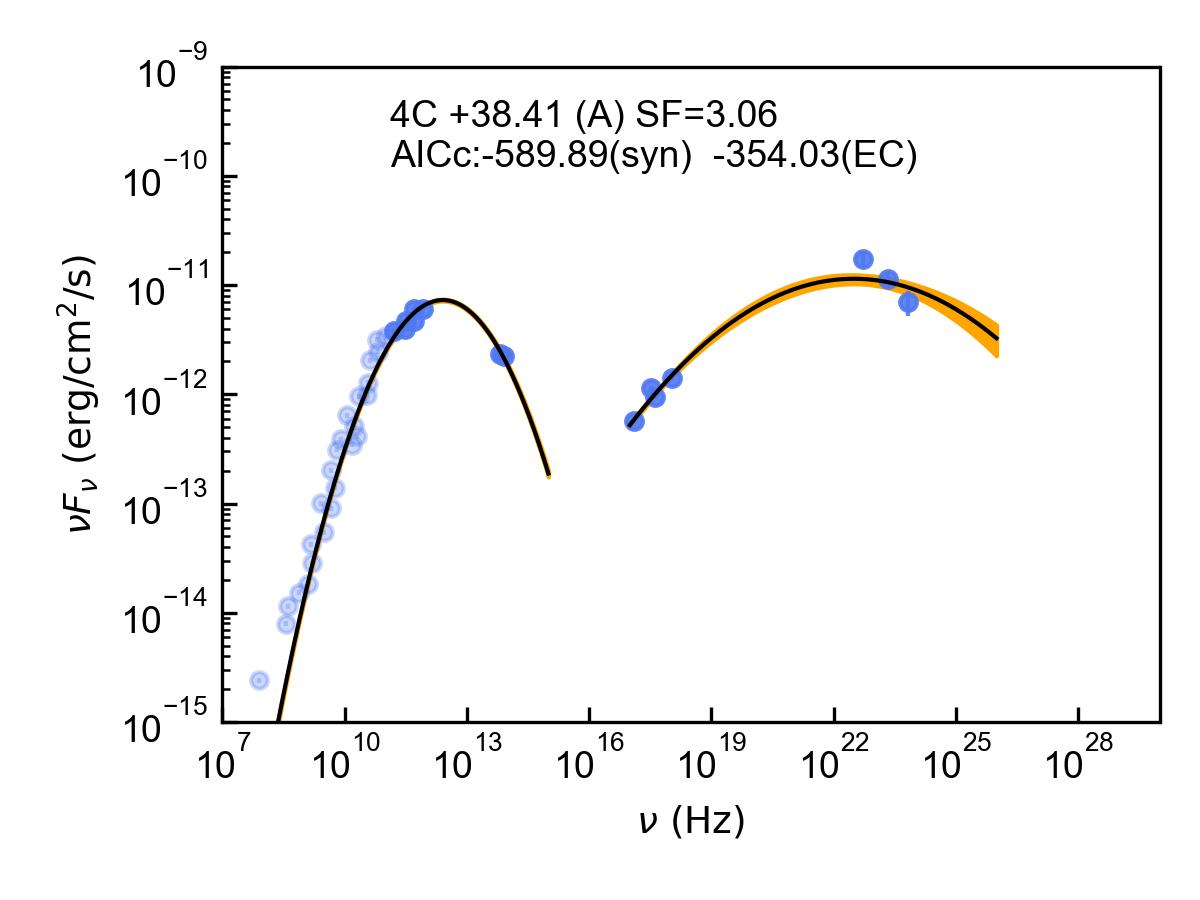}
    \end{subfigure}%
    \begin{subfigure}{.25\textwidth}
        \includegraphics[width = \linewidth]{ 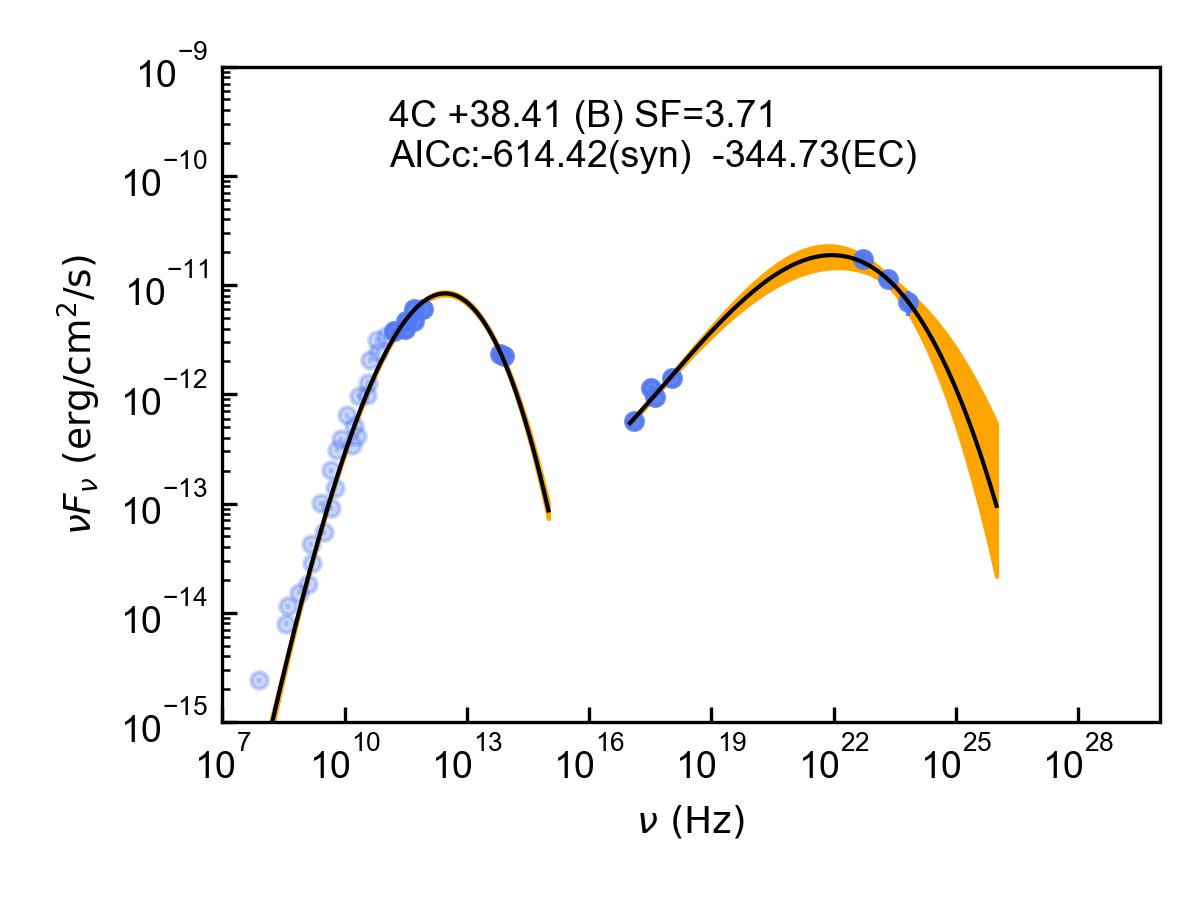}
    \end{subfigure}

    \begin{subfigure}{.25\textwidth}
        \includegraphics[width = \linewidth]{ 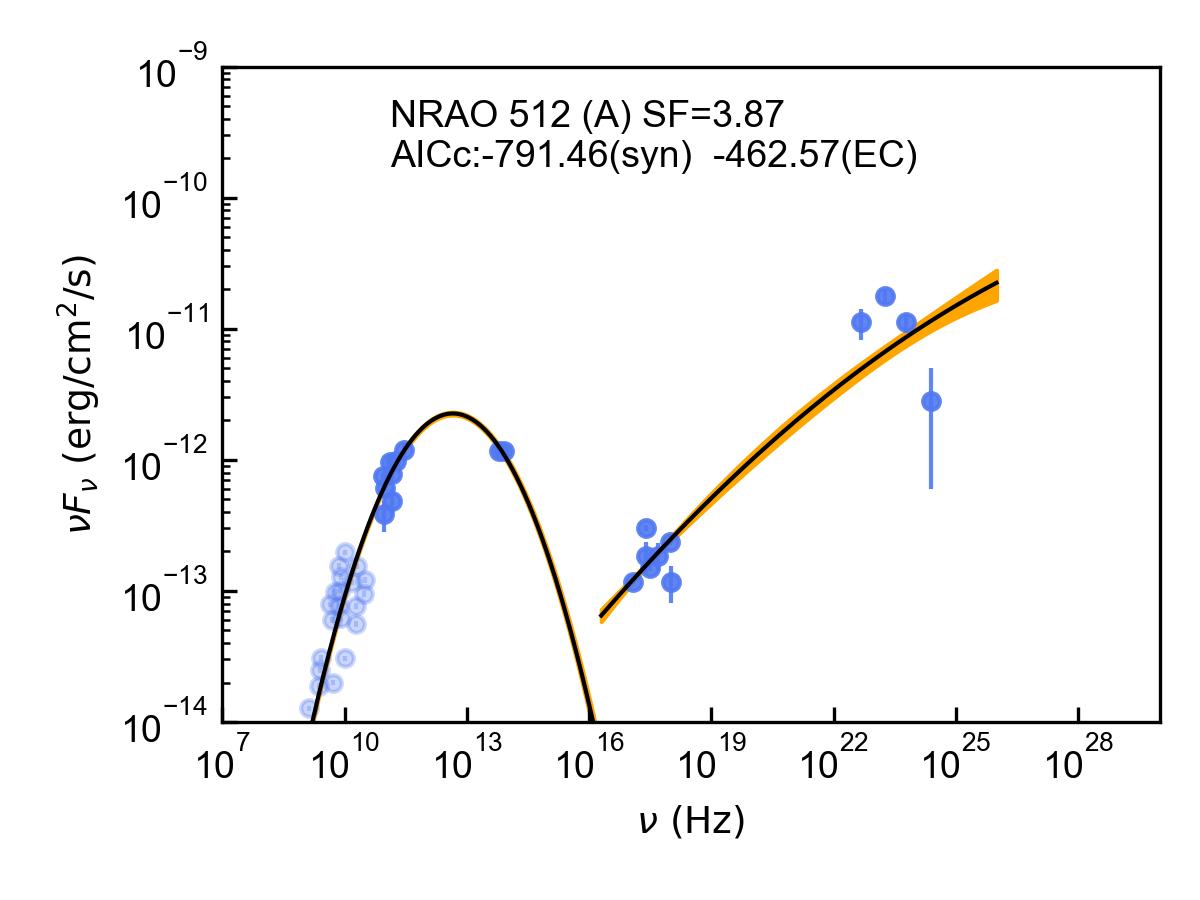}
    \end{subfigure}%
    \begin{subfigure}{.25\textwidth}
        \includegraphics[width = \linewidth]{ 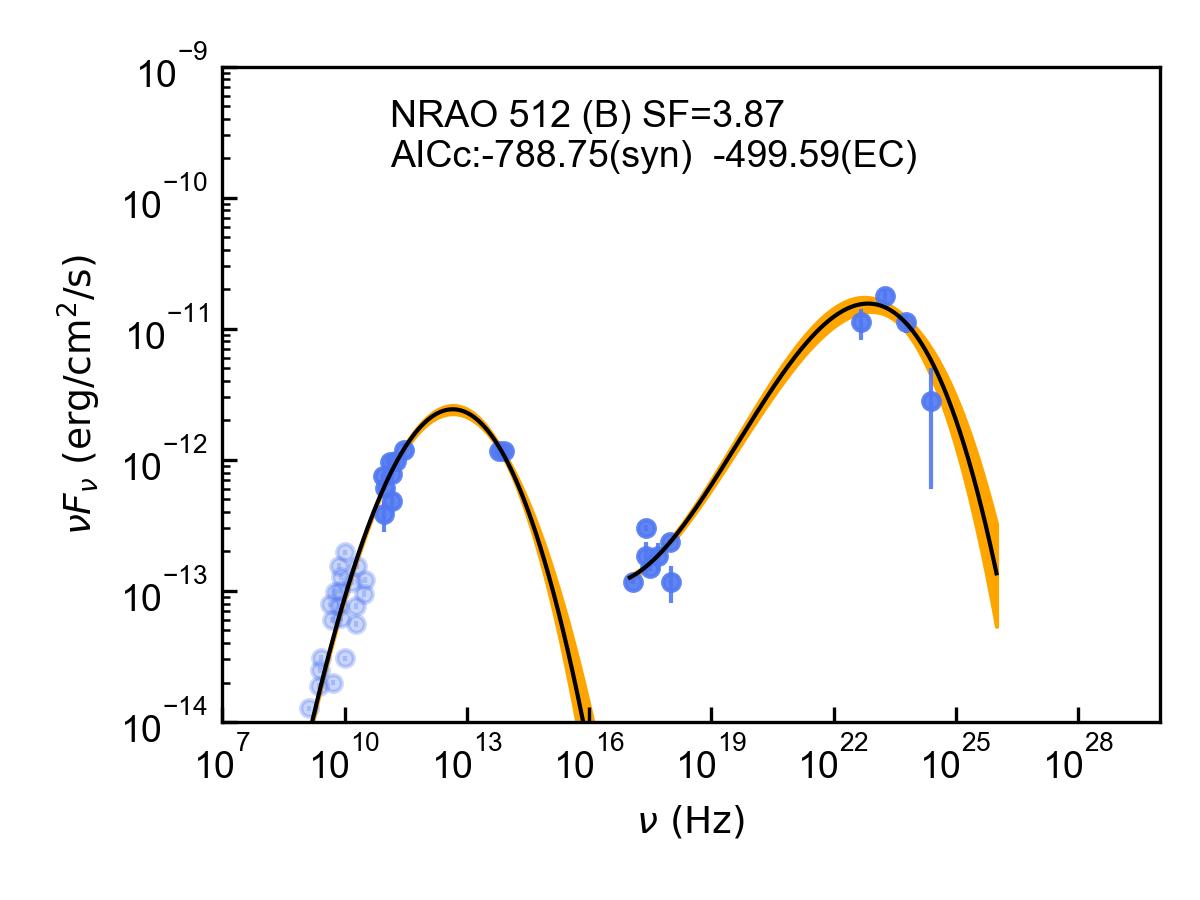}
    \end{subfigure}%
    \begin{subfigure}{.25\textwidth}
        \includegraphics[width = \linewidth]{ 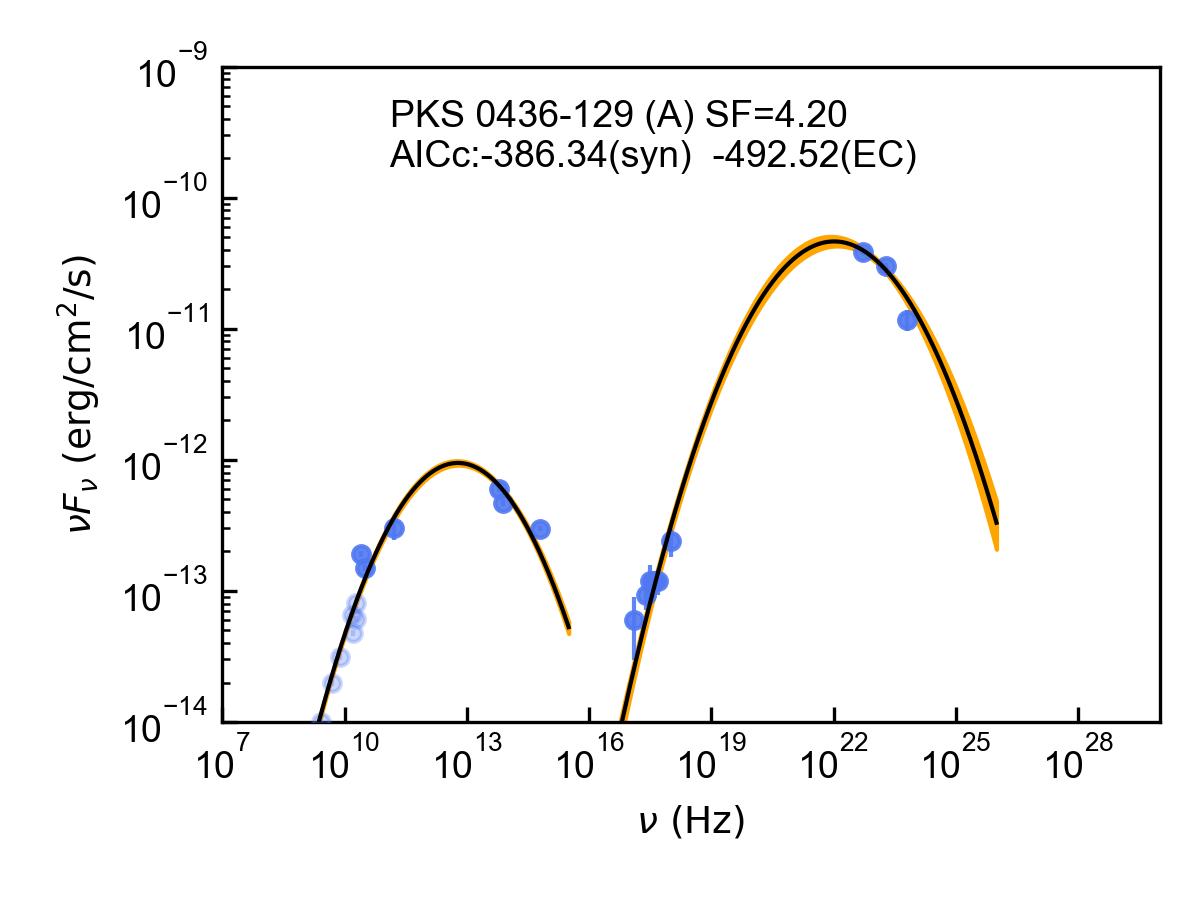}
    \end{subfigure}%
    \begin{subfigure}{.25\textwidth}
        \includegraphics[width = \linewidth]{ 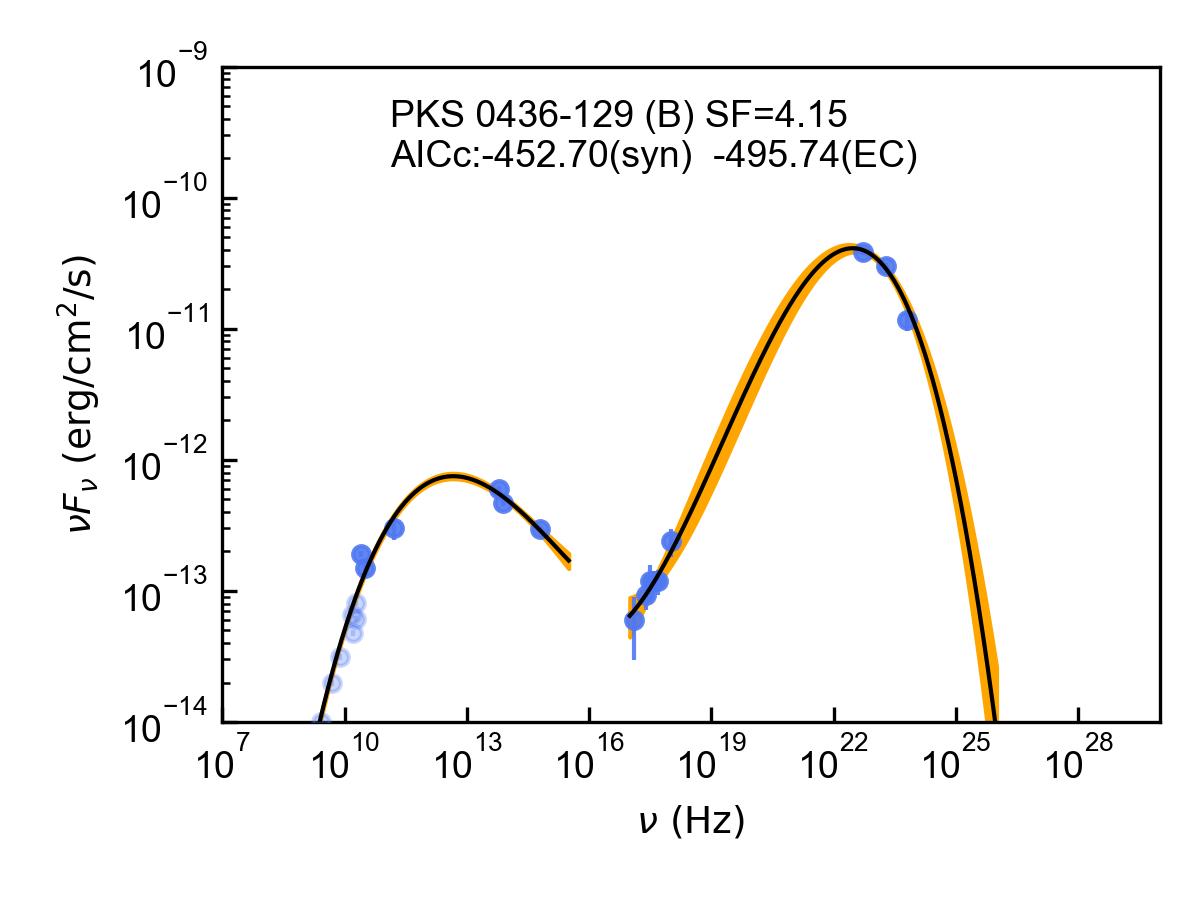}
    \end{subfigure}

    \begin{subfigure}{.25\textwidth}
        \includegraphics[width = \linewidth]{ 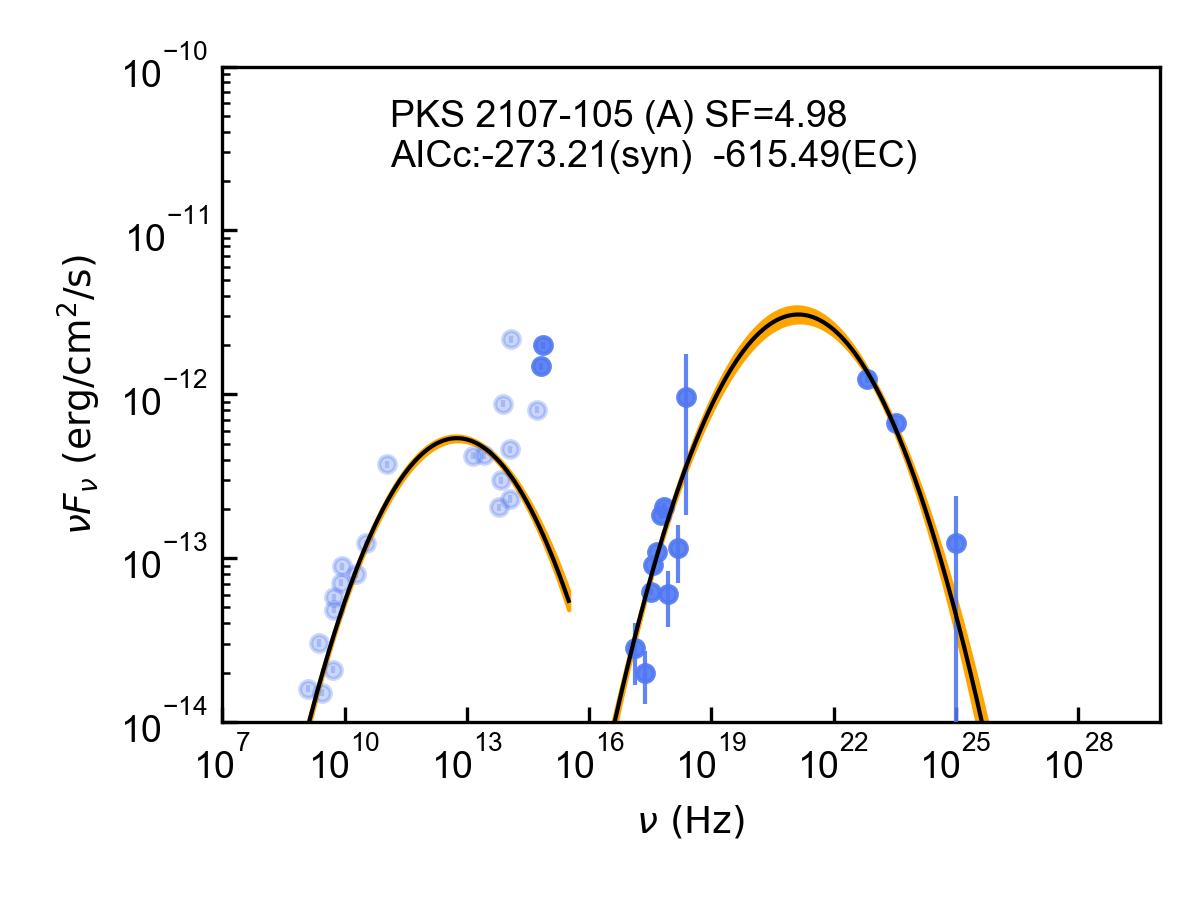}
    \end{subfigure}%
    \begin{subfigure}{.25\textwidth}
        \includegraphics[width = \linewidth]{ 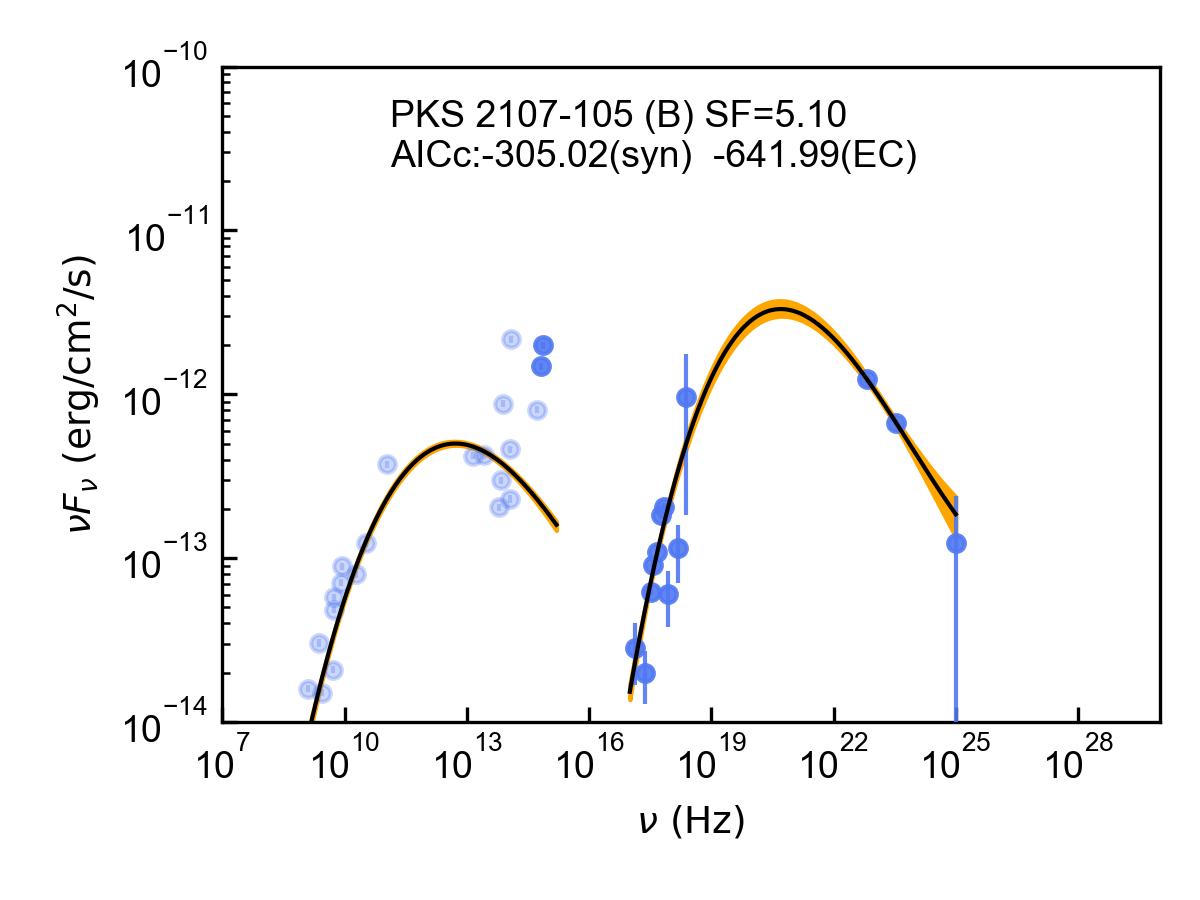}
    \end{subfigure}%
    \begin{subfigure}{.25\textwidth}
        \includegraphics[width = \linewidth]{ 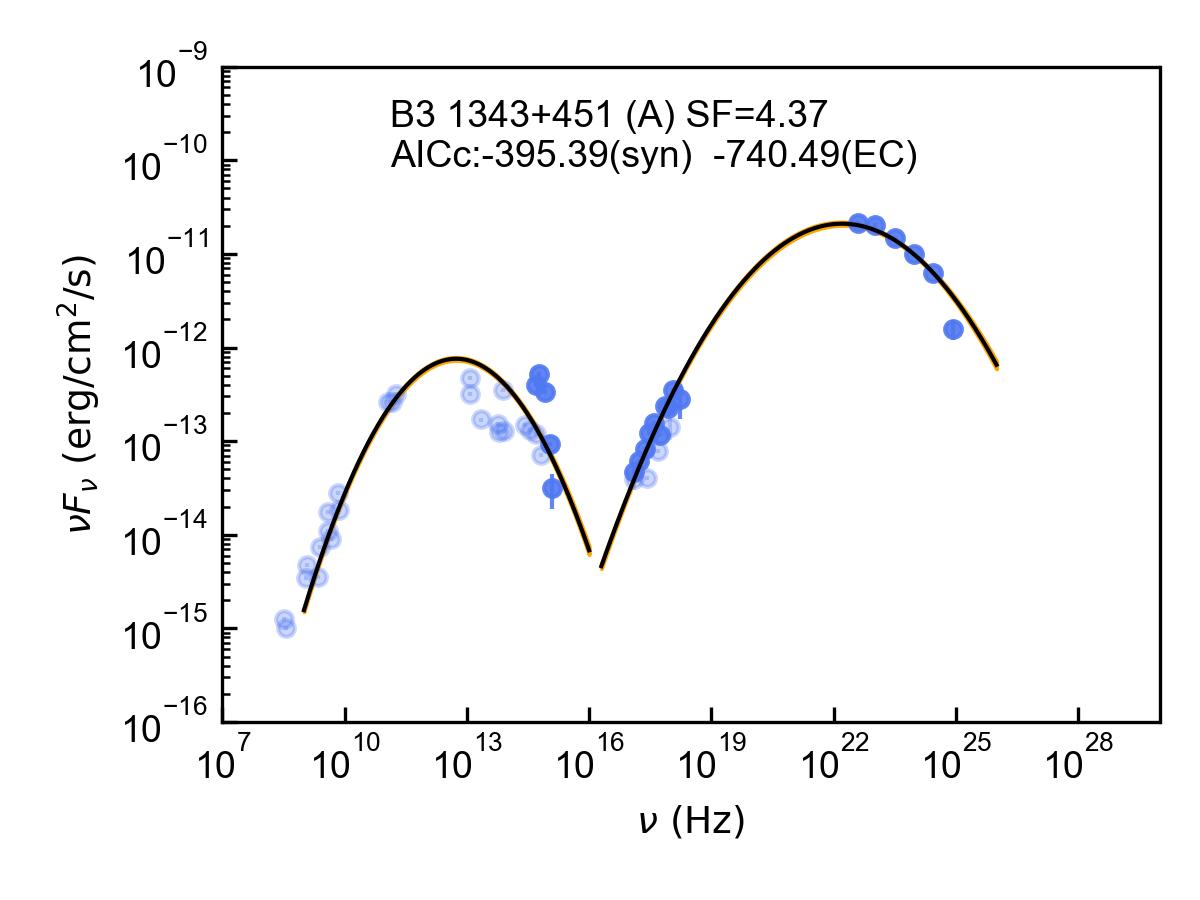}
    \end{subfigure}%
    \begin{subfigure}{.25\textwidth}
        \includegraphics[width = \linewidth]{ 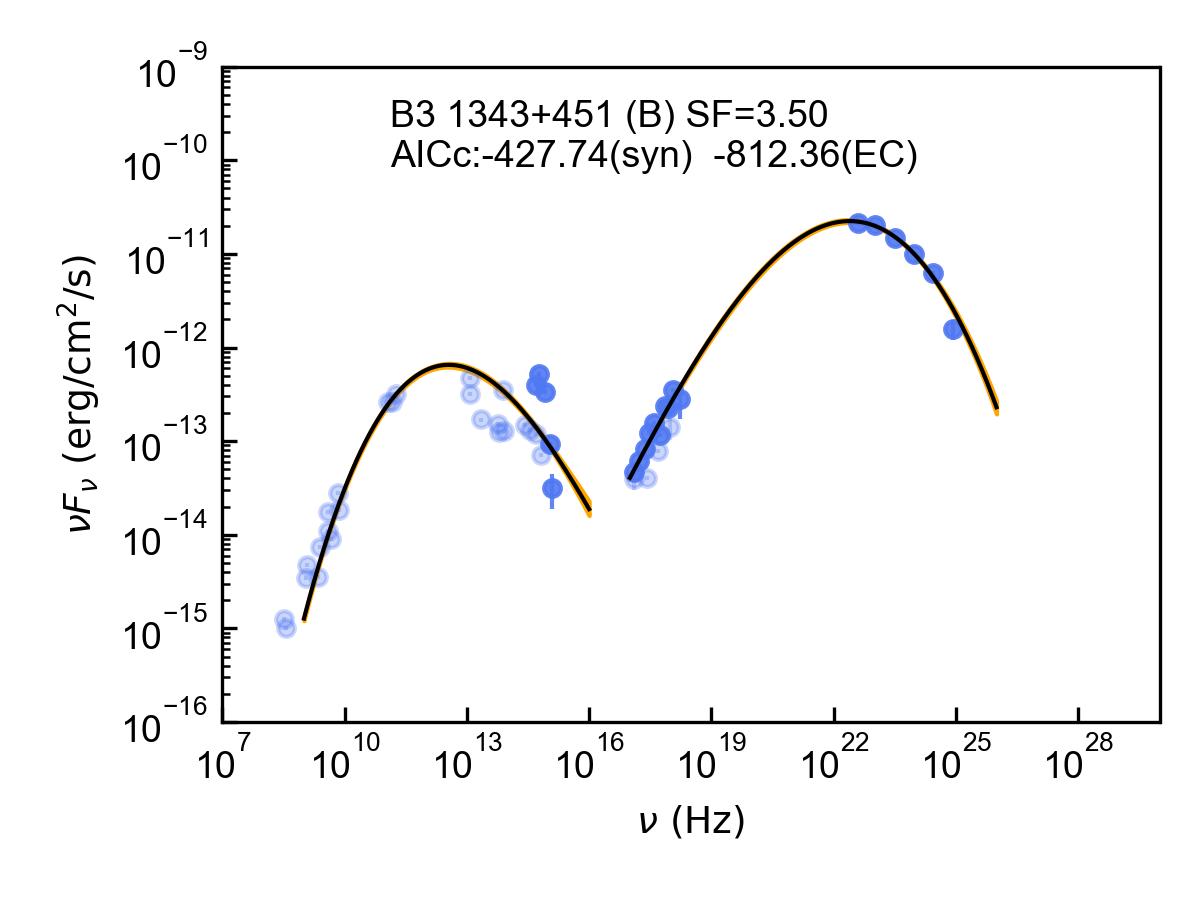}
    \end{subfigure}

    \label{SED DT Doppler}
\end{figure*}

\begin{figure*}
    \caption{SED fitting results of CTA 102.}
    \begin{subfigure}{.25\textwidth}
        \includegraphics[width = \linewidth]{ 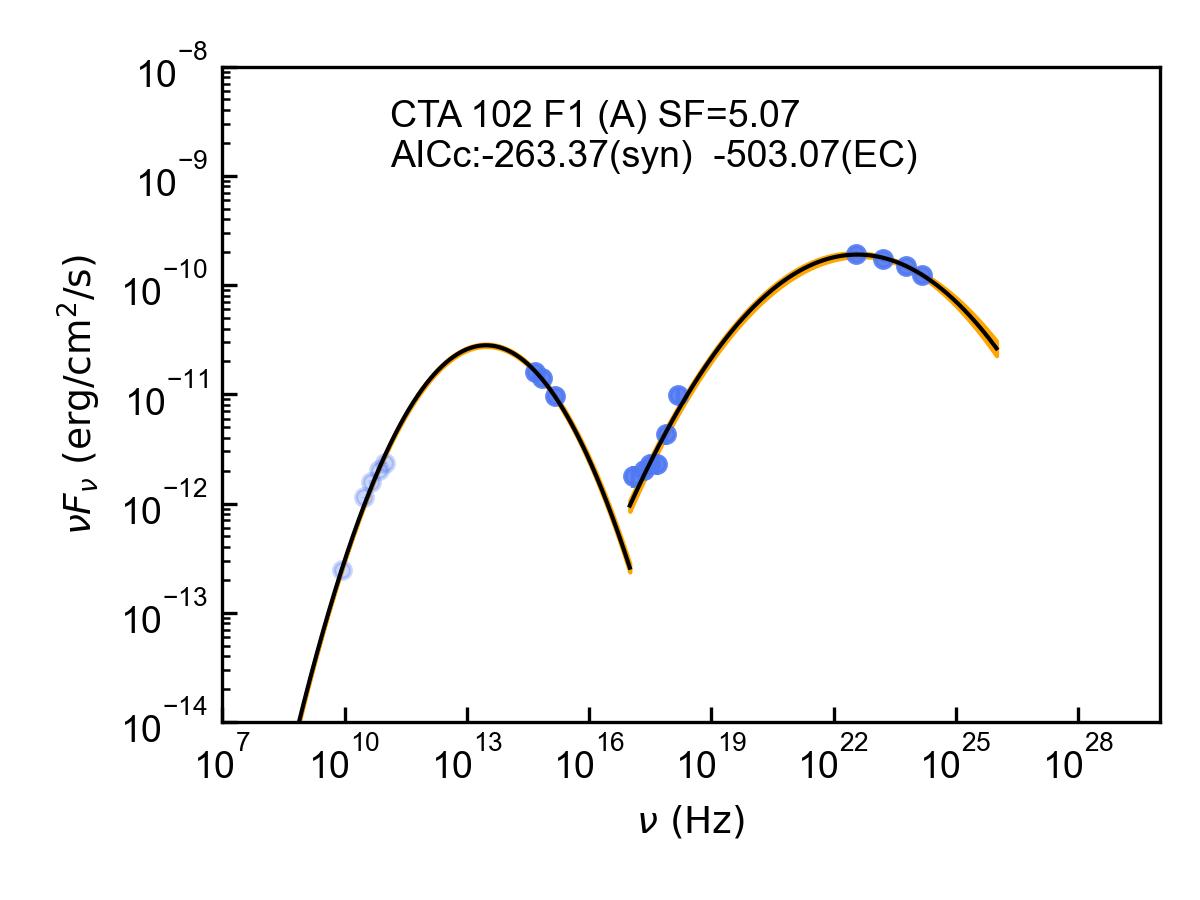}
    \end{subfigure}%
    \begin{subfigure}{.25\textwidth}
        \includegraphics[width = \linewidth]{ 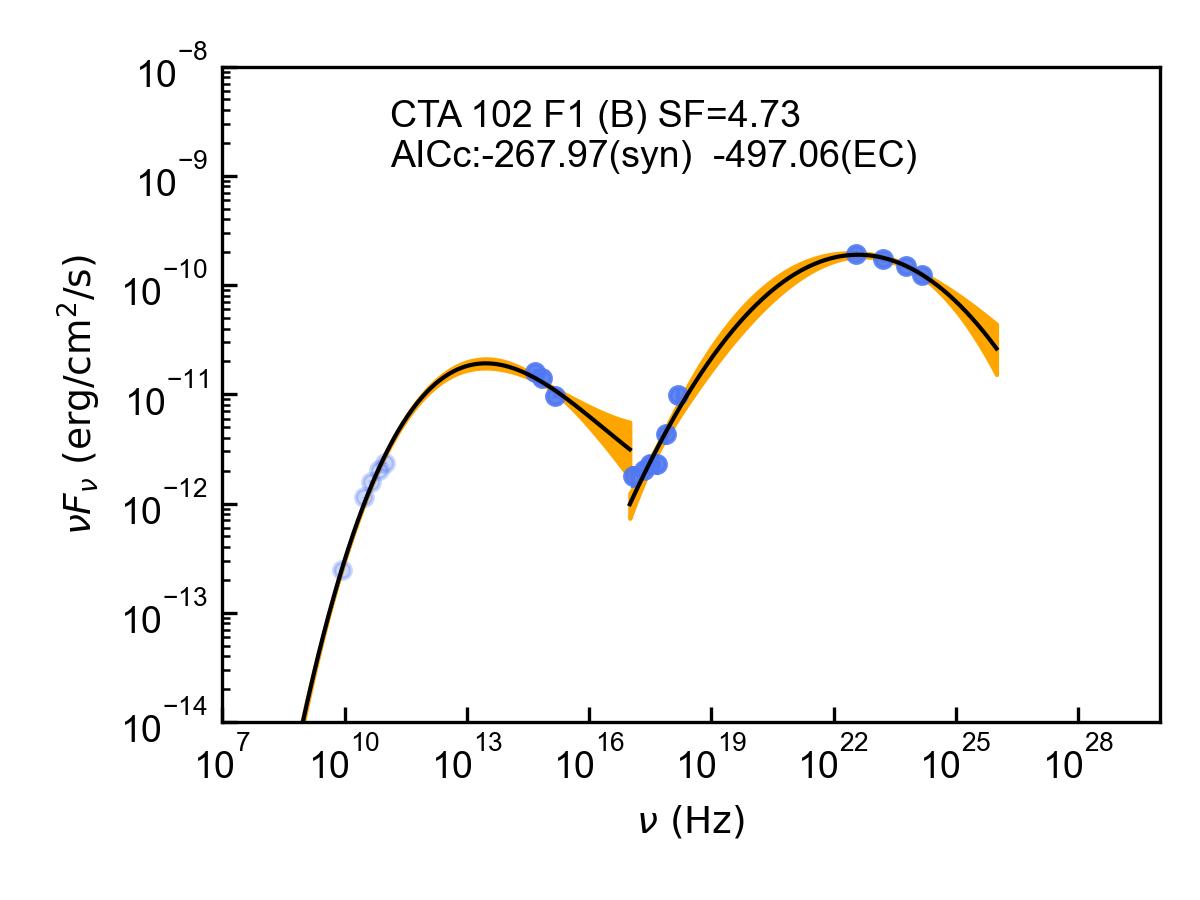}
    \end{subfigure}%
    \begin{subfigure}{.25\textwidth}
        \includegraphics[width = \linewidth]{ 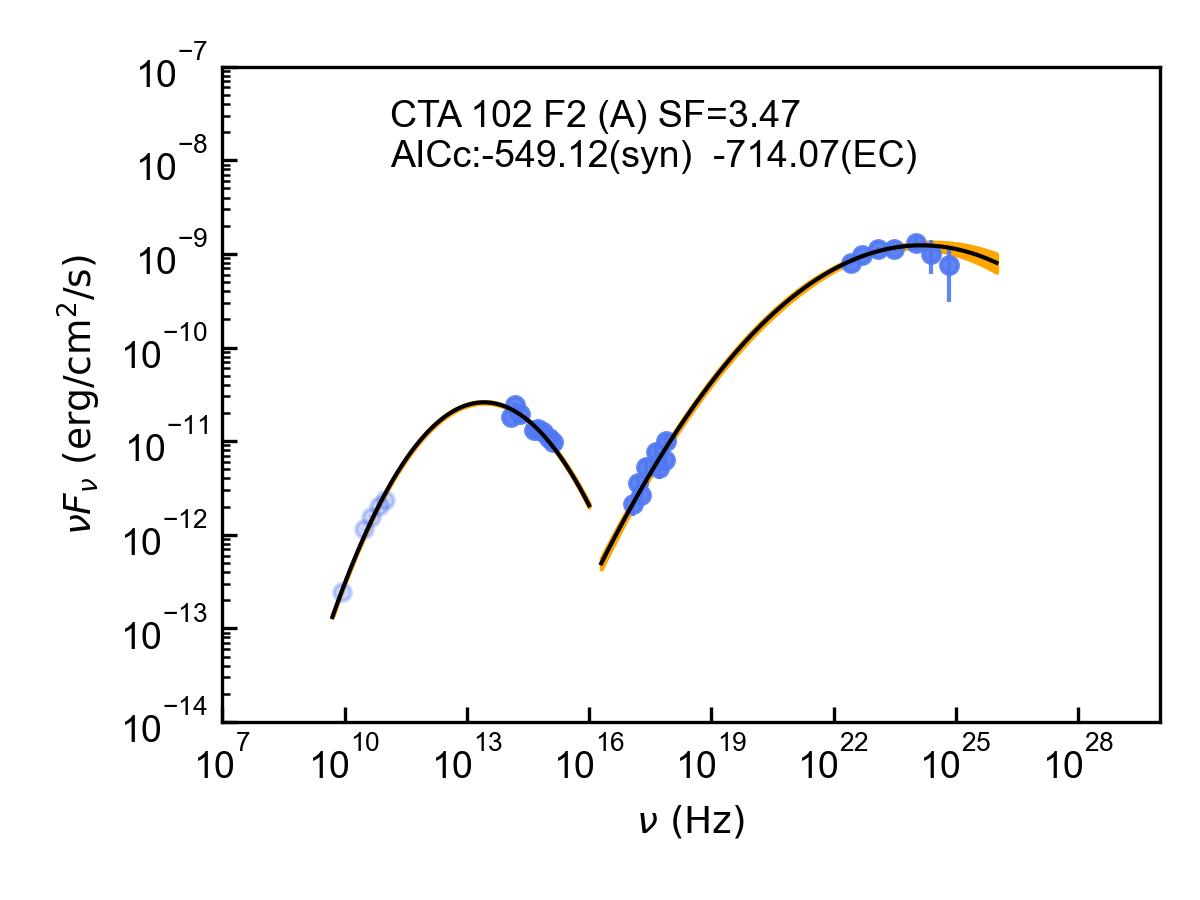}
    \end{subfigure}%
    \begin{subfigure}{.25\textwidth}
        \includegraphics[width = \linewidth]{ 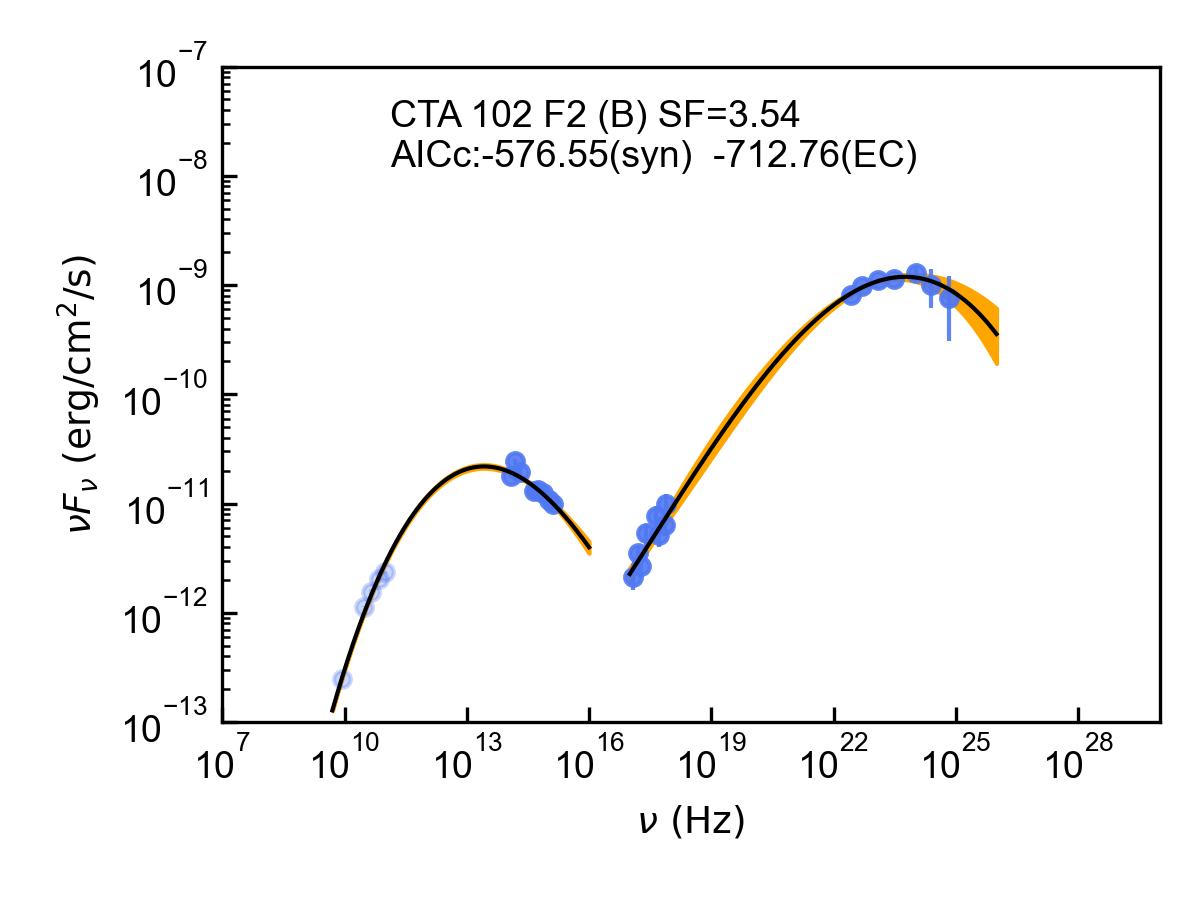}
    \end{subfigure}

    \begin{subfigure}{.25\textwidth}
        \includegraphics[width = \linewidth]{ 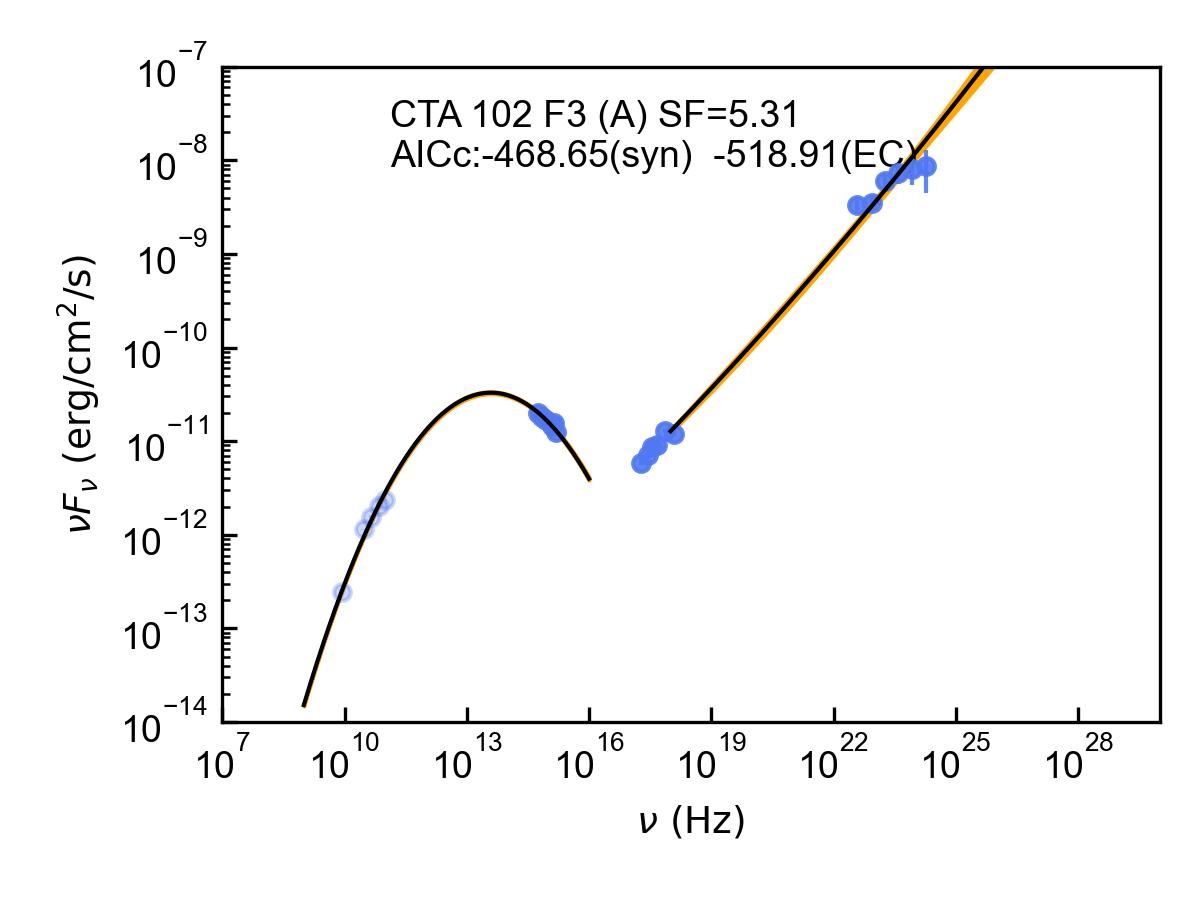}
    \end{subfigure}%
    \begin{subfigure}{.25\textwidth}
        \includegraphics[width = \linewidth]{ 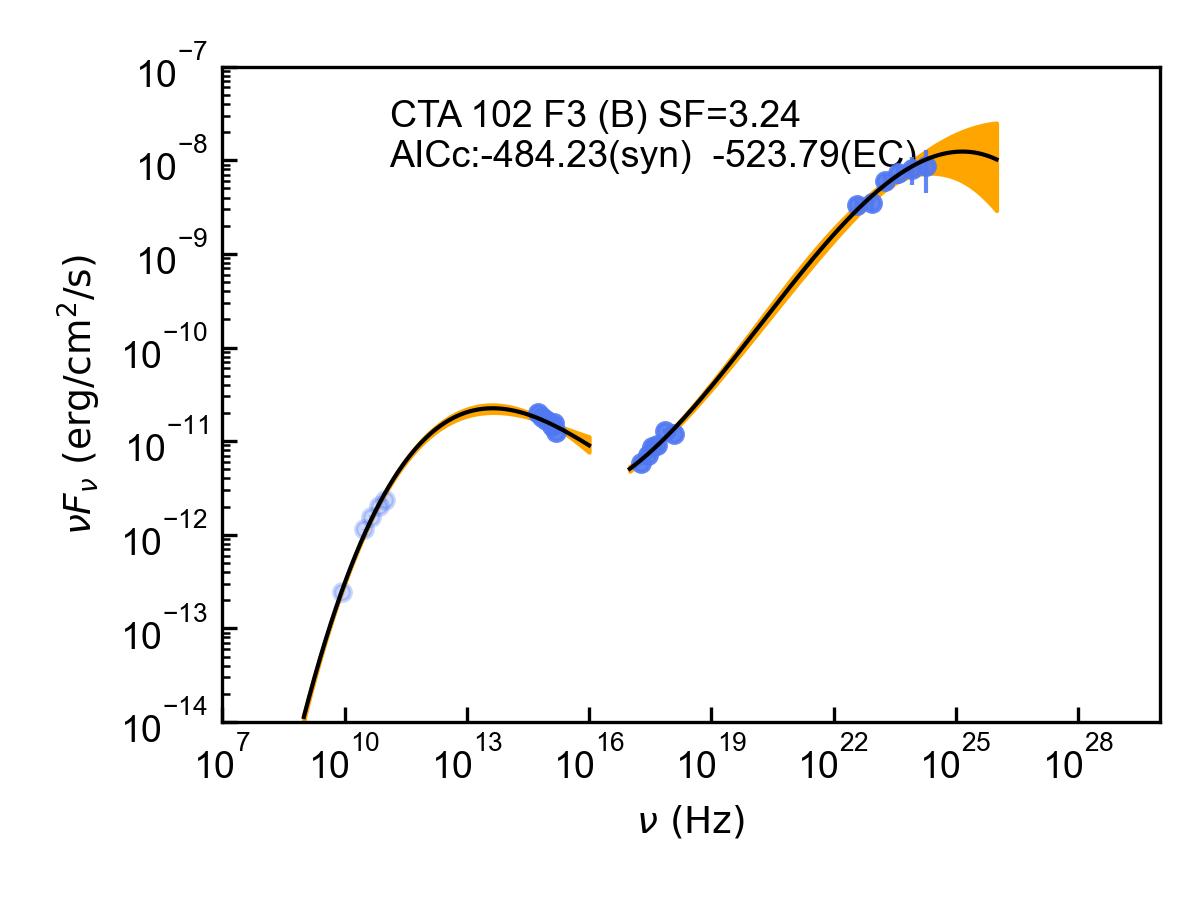}
    \end{subfigure}%
    \begin{subfigure}{.25\textwidth}
        \includegraphics[width = \linewidth]{ 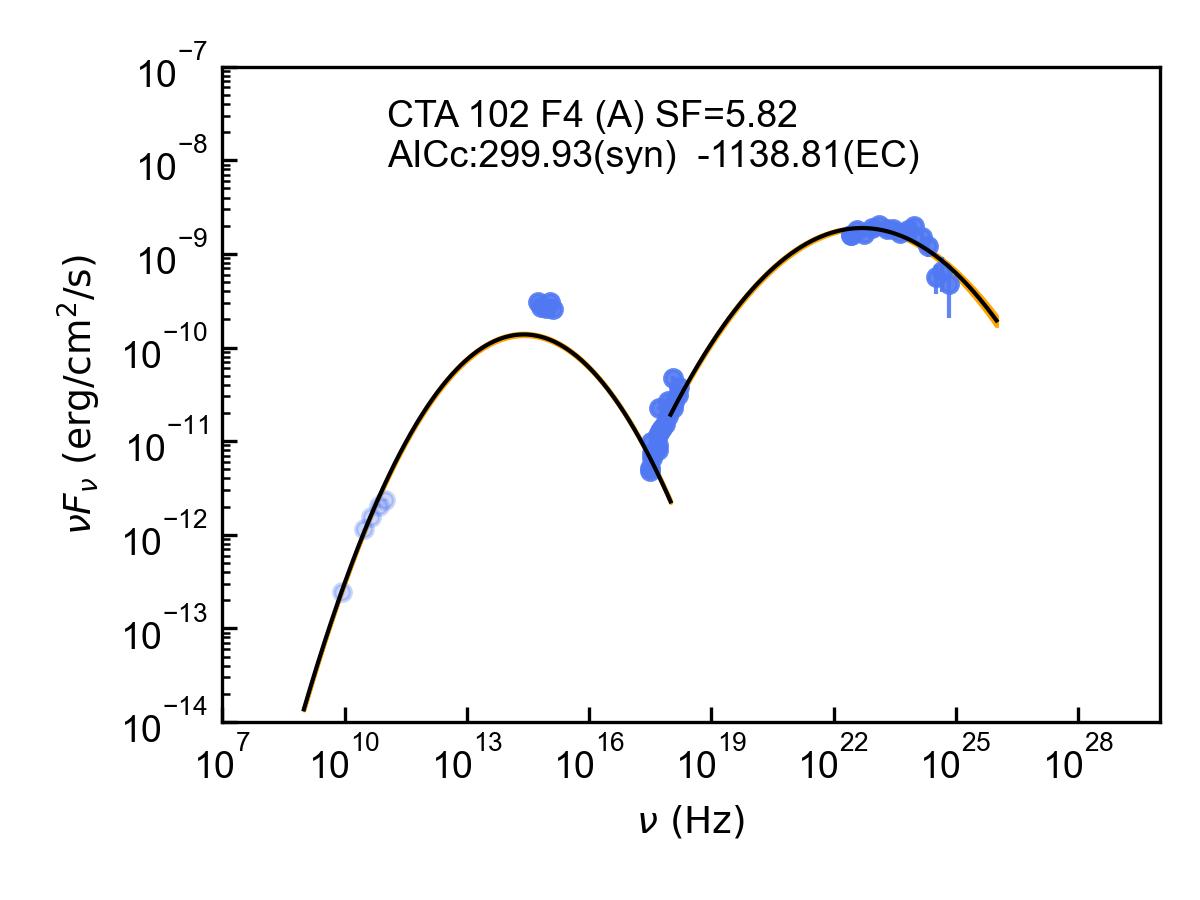}
    \end{subfigure}%
    \begin{subfigure}{.25\textwidth}
        \includegraphics[width = \linewidth]{ 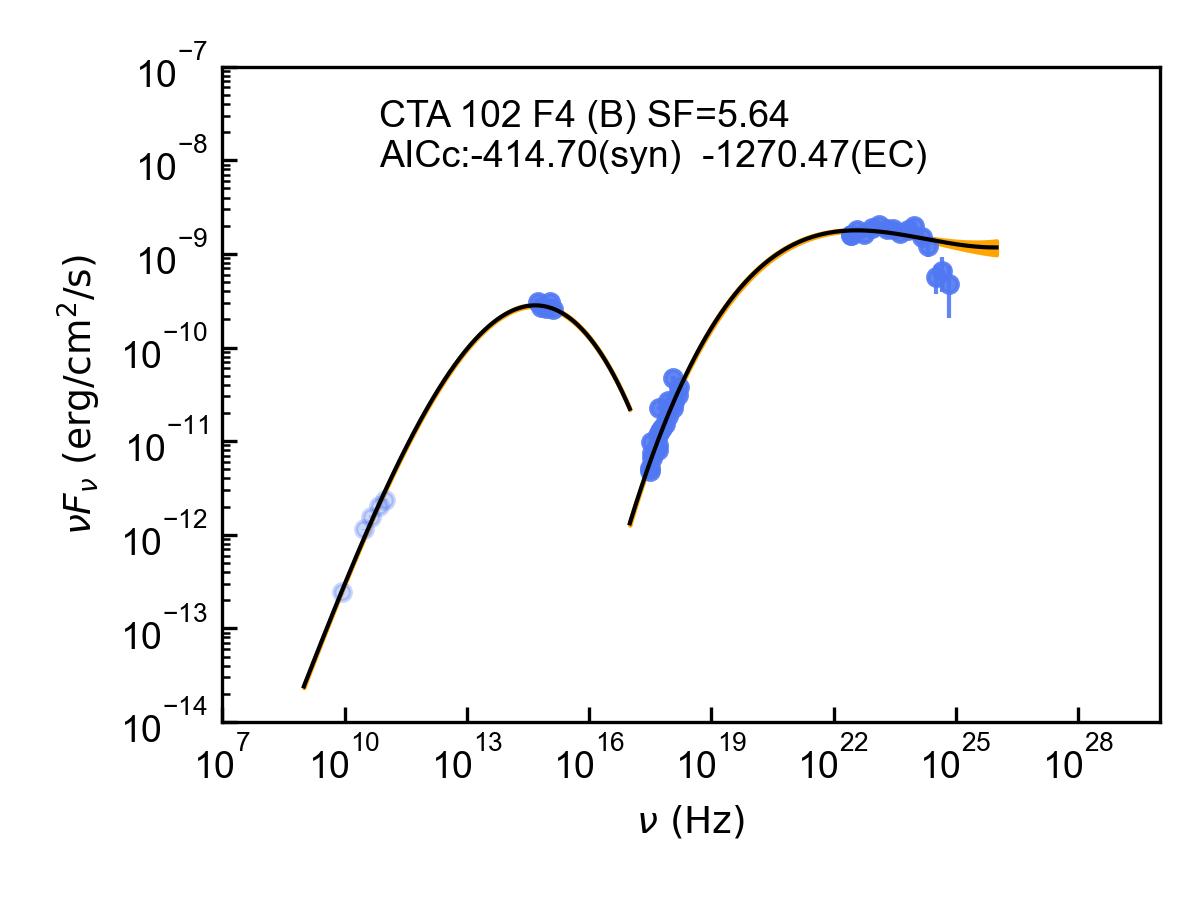}
    \end{subfigure}

    \begin{subfigure}{.25\textwidth}
        \includegraphics[width = \linewidth]{ 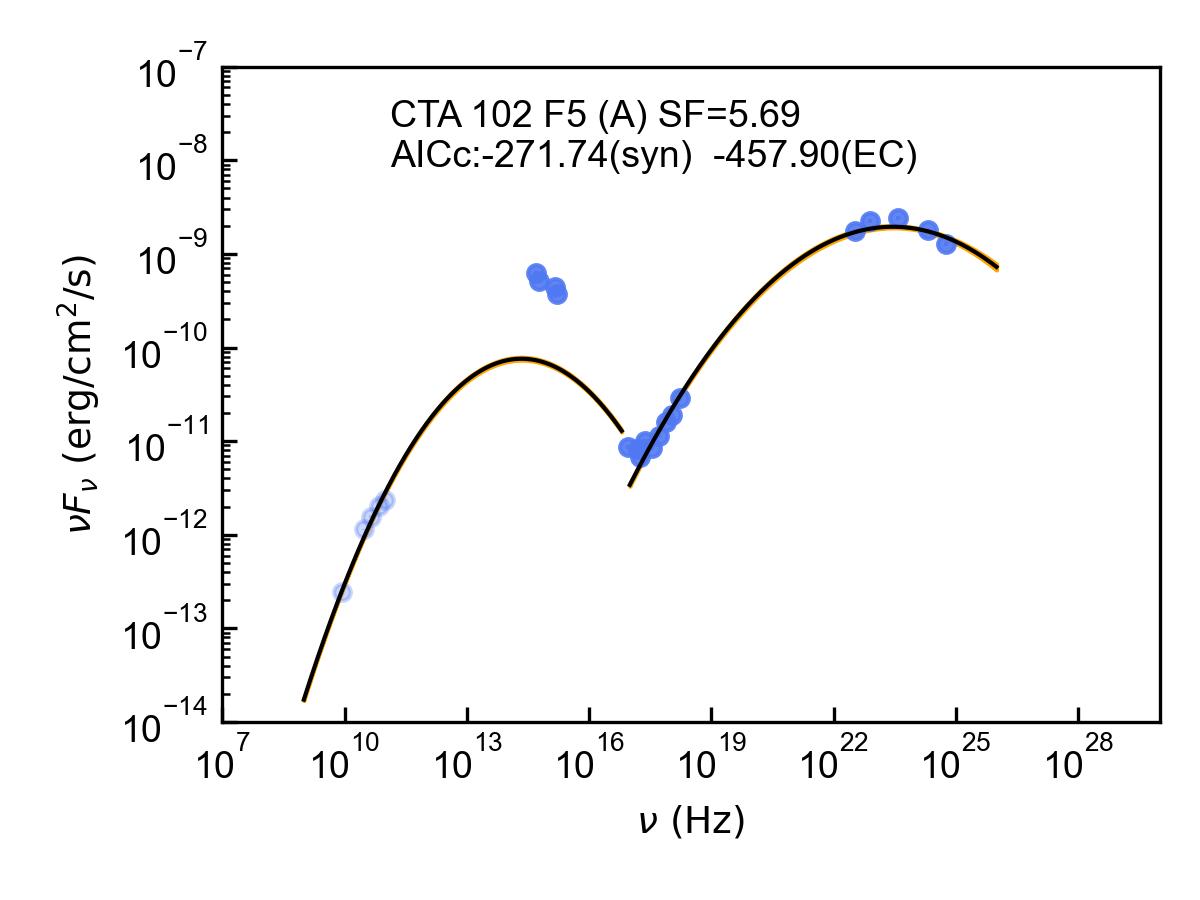}
    \end{subfigure}%
    \begin{subfigure}{.25\textwidth}
        \includegraphics[width = \linewidth]{ 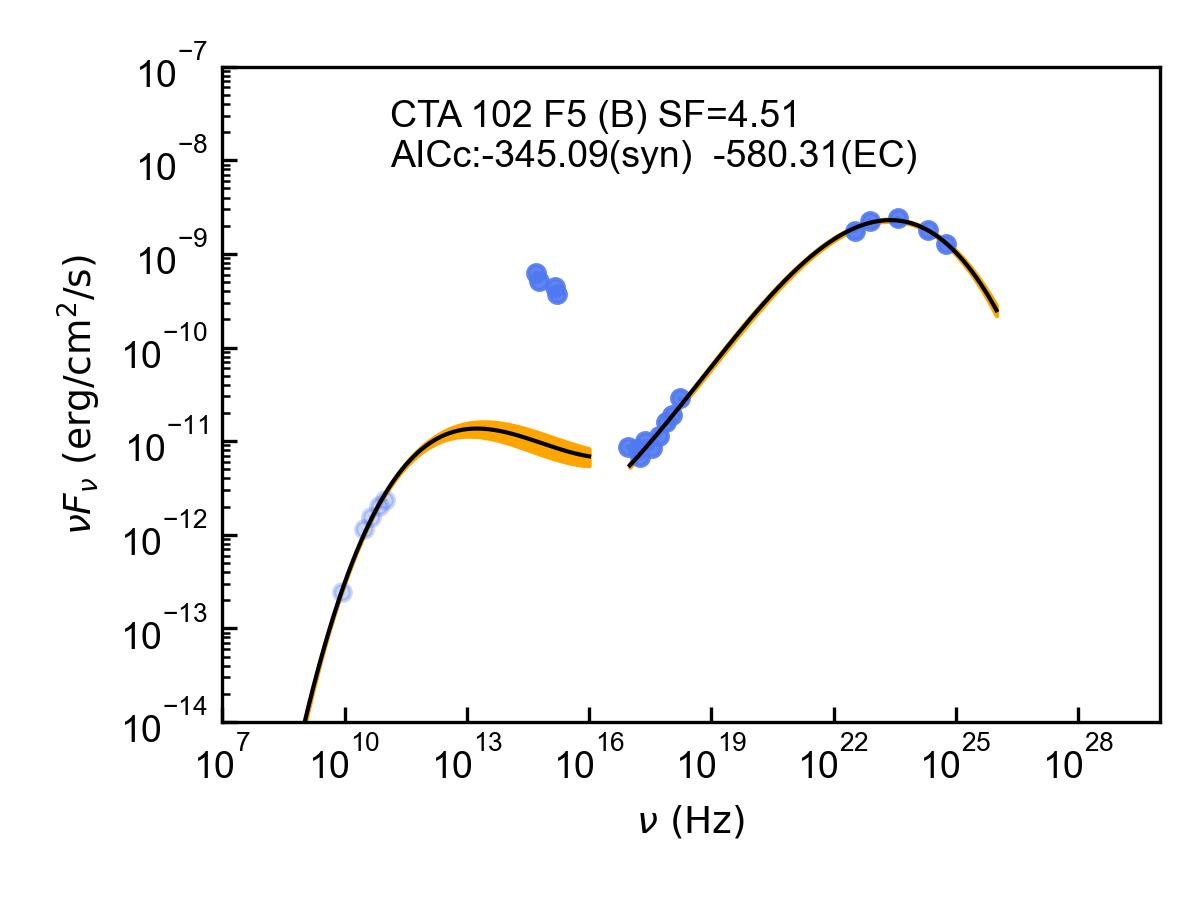}
    \end{subfigure}%
    \begin{subfigure}{.25\textwidth}
        \includegraphics[width = \linewidth]{ 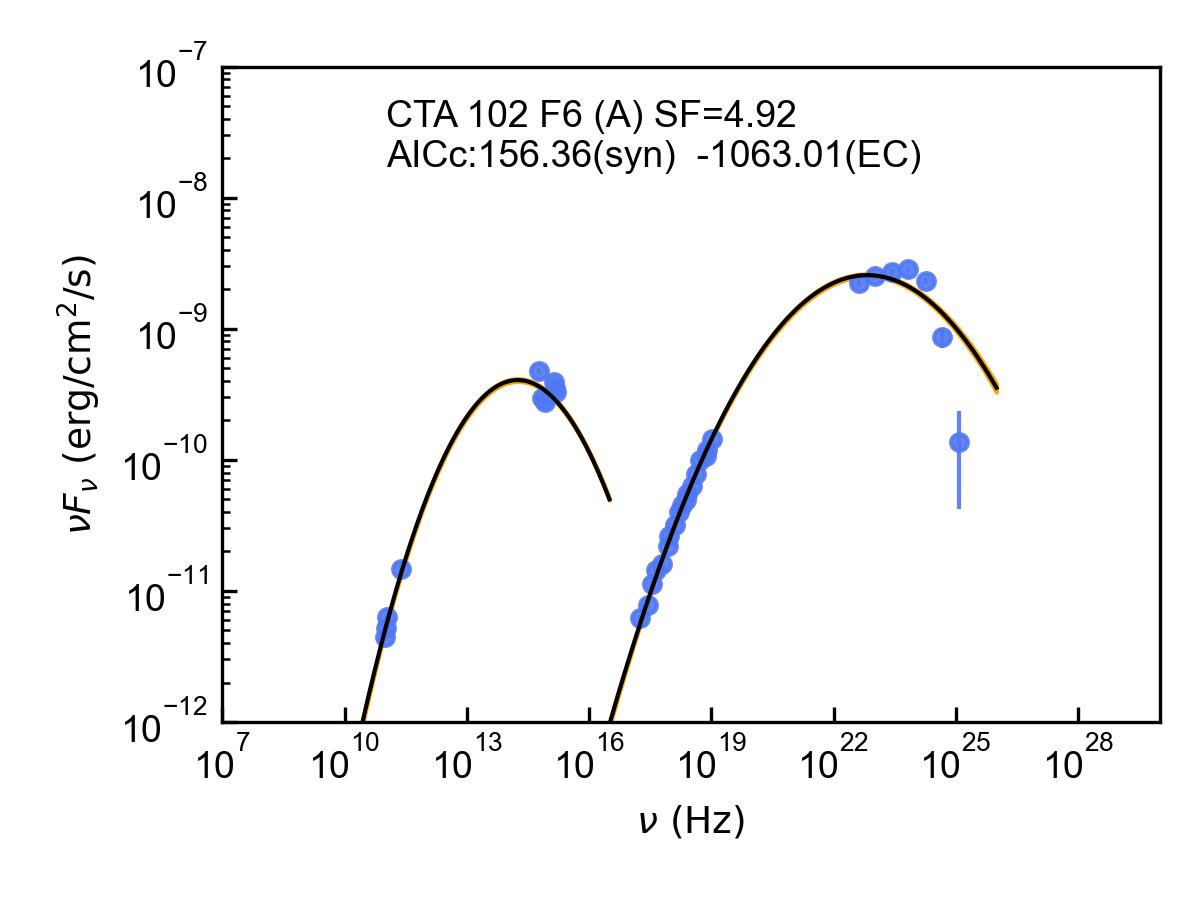}
    \end{subfigure}%
    \begin{subfigure}{.25\textwidth}
        \includegraphics[width = \linewidth]{ 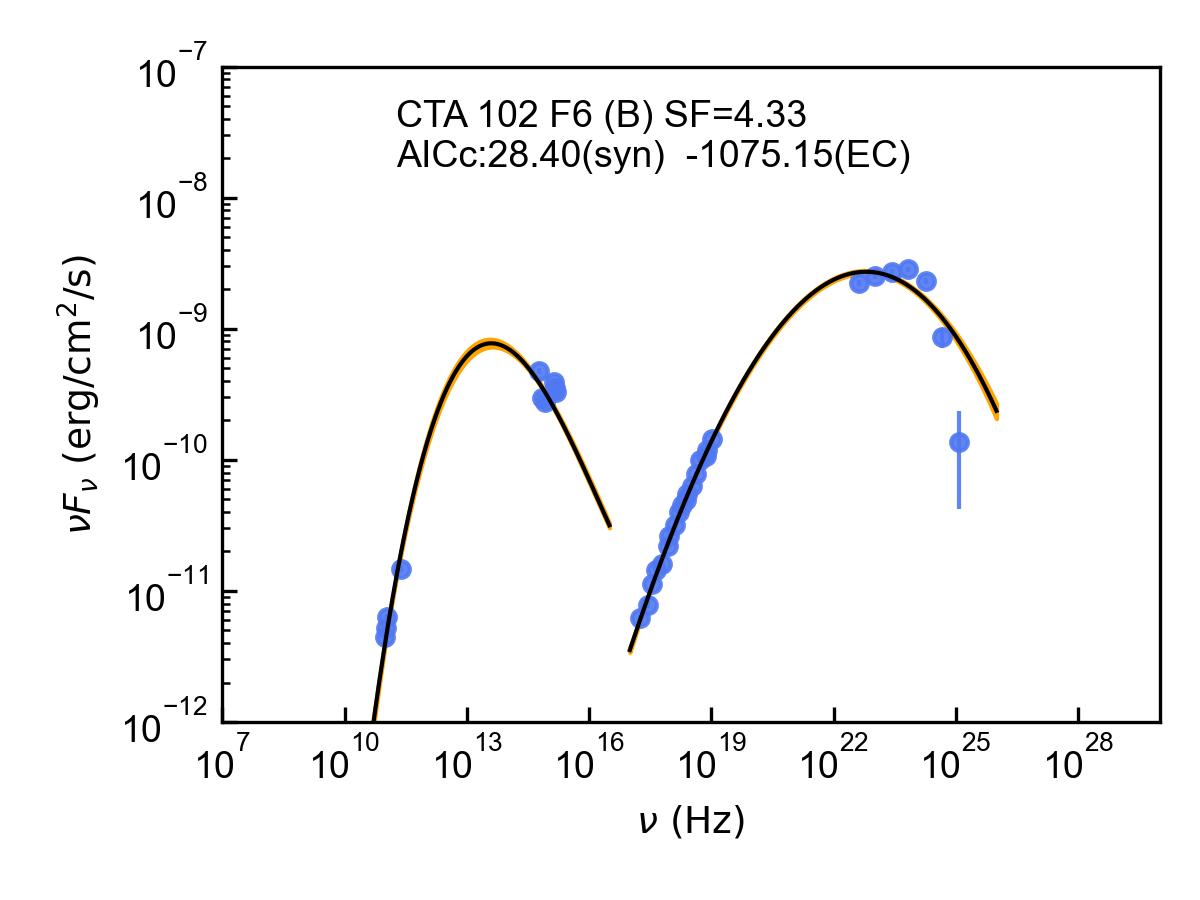}
    \end{subfigure}

    \raggedright
    \begin{subfigure}{.25\textwidth}
        \includegraphics[width = \linewidth]{ 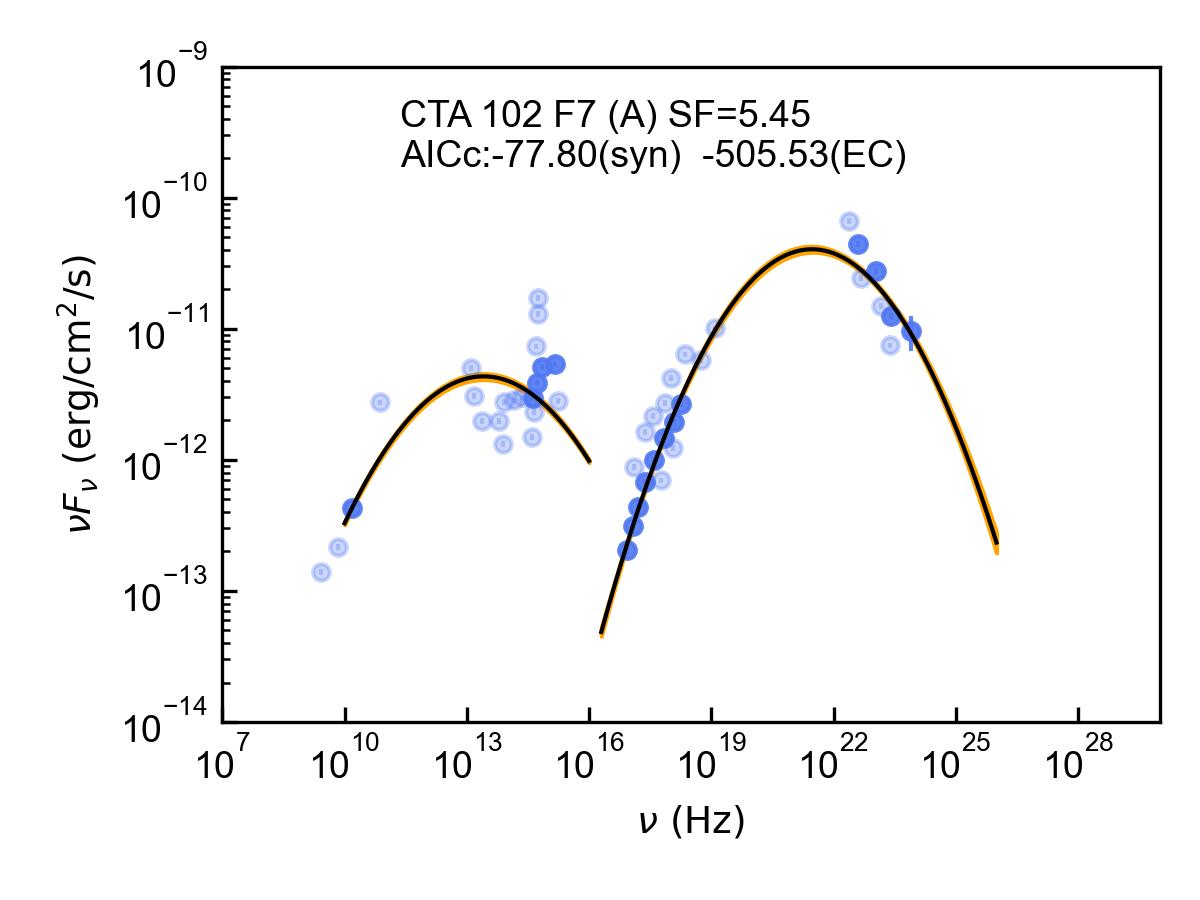}
    \end{subfigure}%
    \begin{subfigure}{.25\textwidth}
        \includegraphics[width = \linewidth]{ 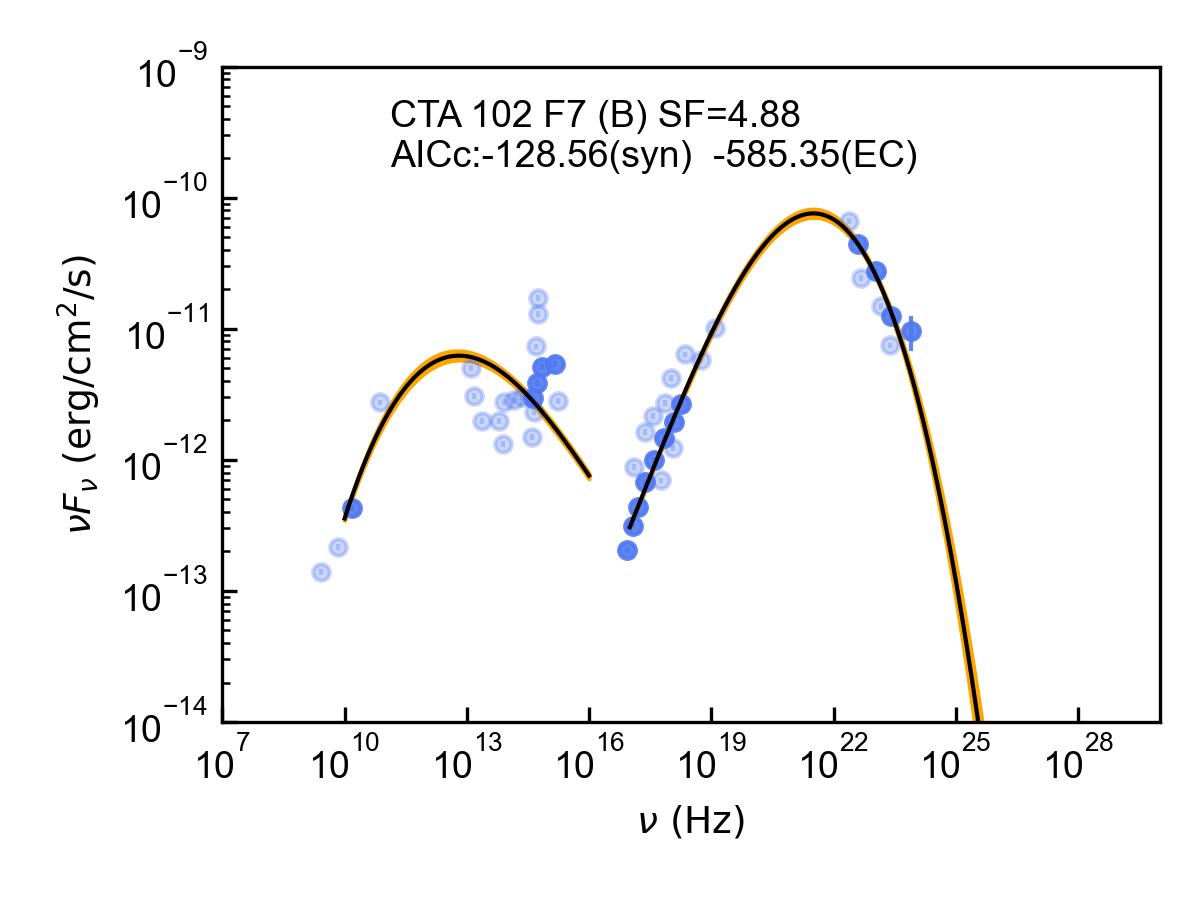}
    \end{subfigure}

    \label{SED CTA 102}
\end{figure*}

\begin{figure*}
    \caption{SED fitting results of 3C 279.}
    \begin{subfigure}{.25\textwidth}
        \includegraphics[width = \linewidth]{ 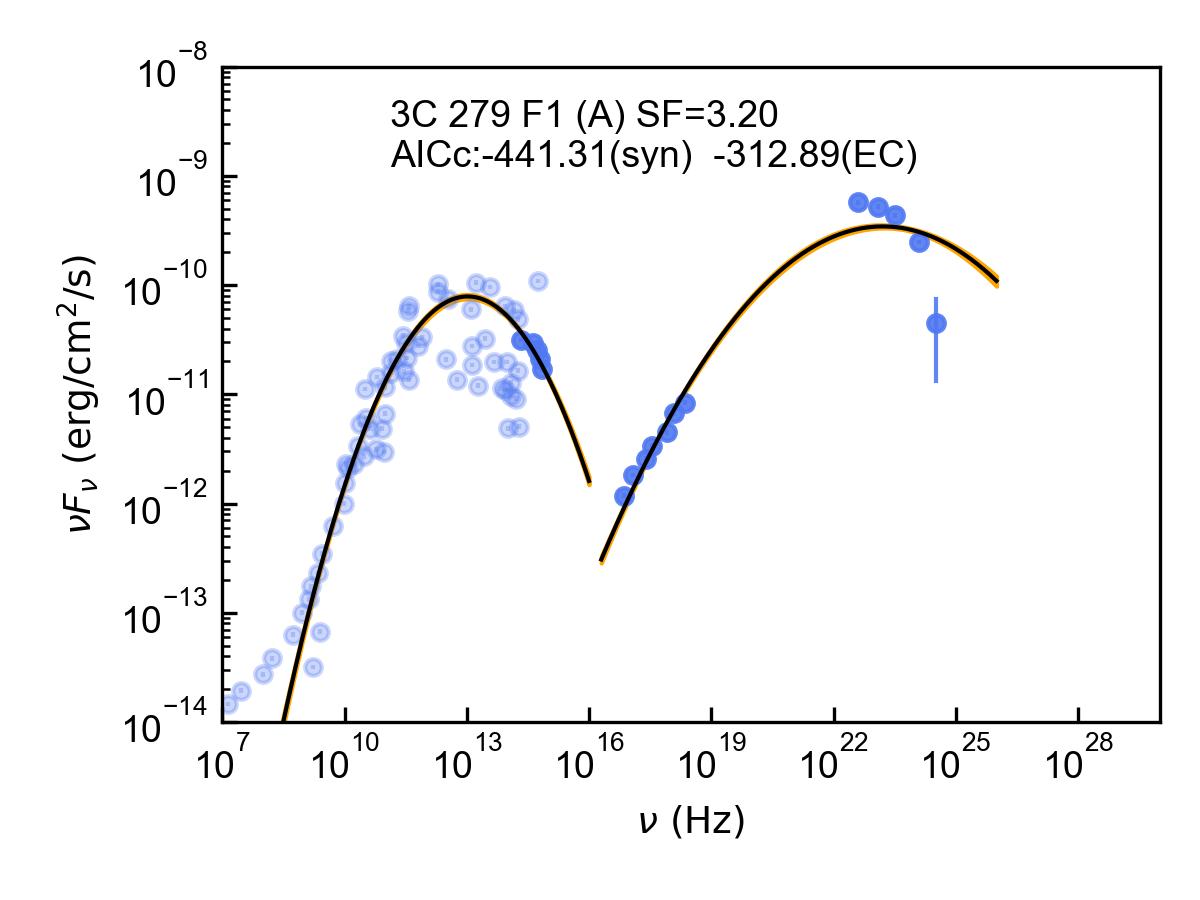}
    \end{subfigure}%
    \begin{subfigure}{.25\textwidth}
        \includegraphics[width = \linewidth]{ 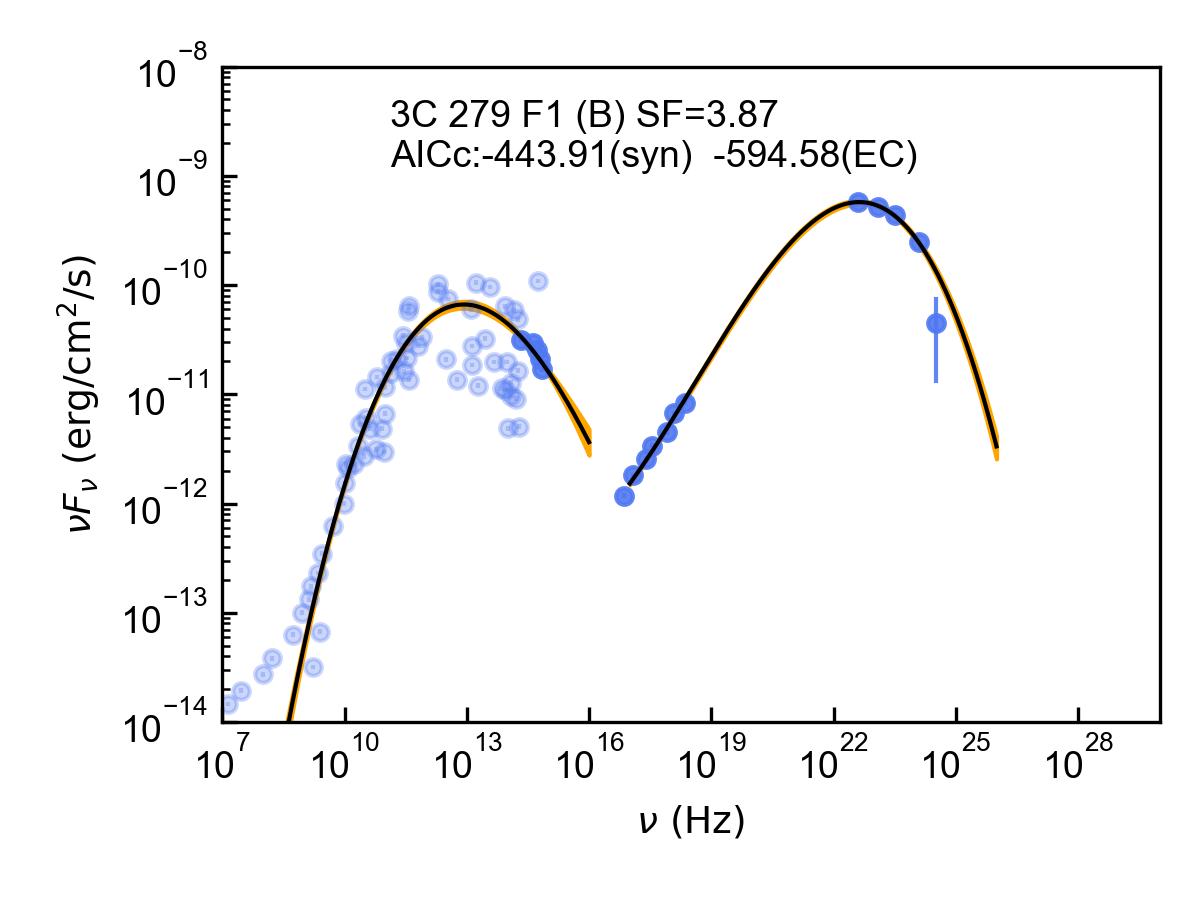}
    \end{subfigure}%
    \begin{subfigure}{.25\textwidth}
        \includegraphics[width = \linewidth]{ 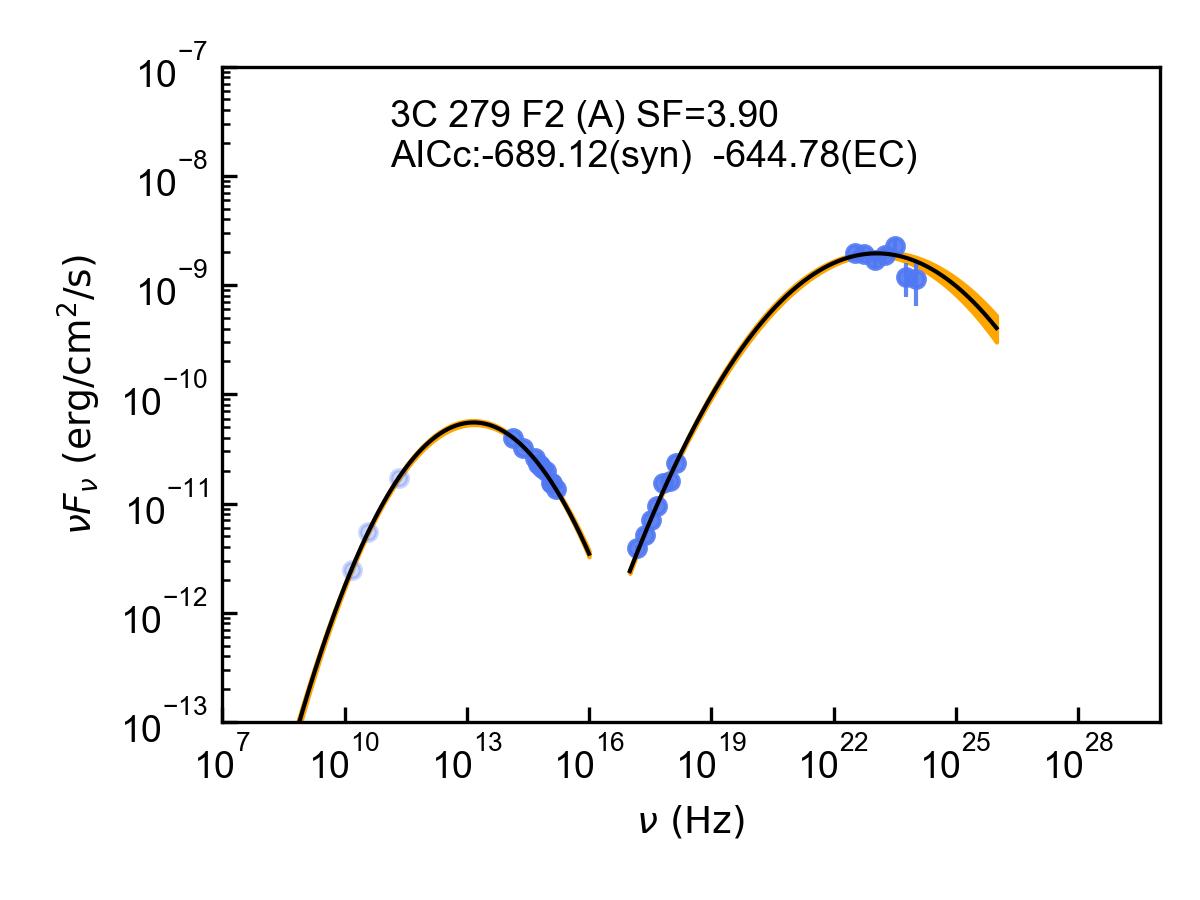}
    \end{subfigure}%
    \begin{subfigure}{.25\textwidth}
        \includegraphics[width = \linewidth]{ 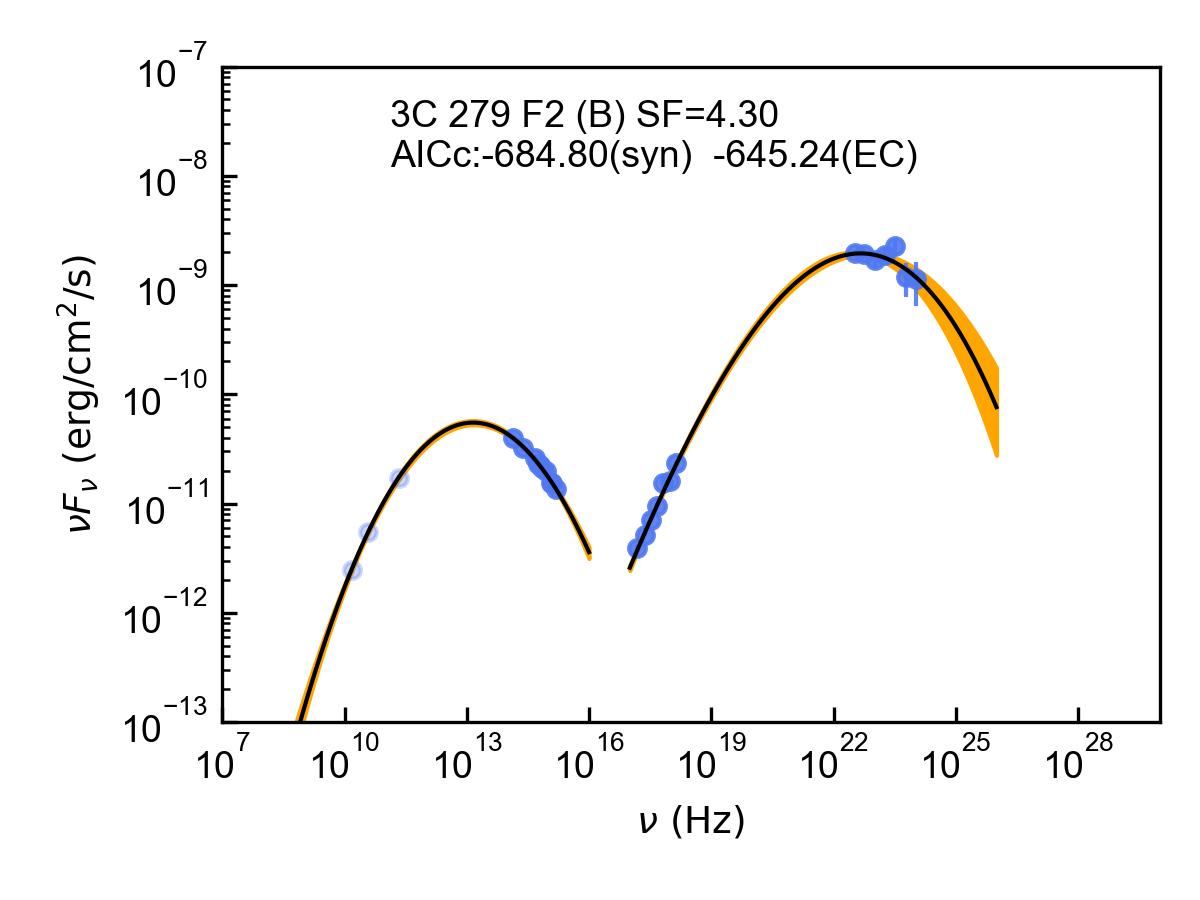}
    \end{subfigure}

    \begin{subfigure}{.25\textwidth}
        \includegraphics[width = \linewidth]{ 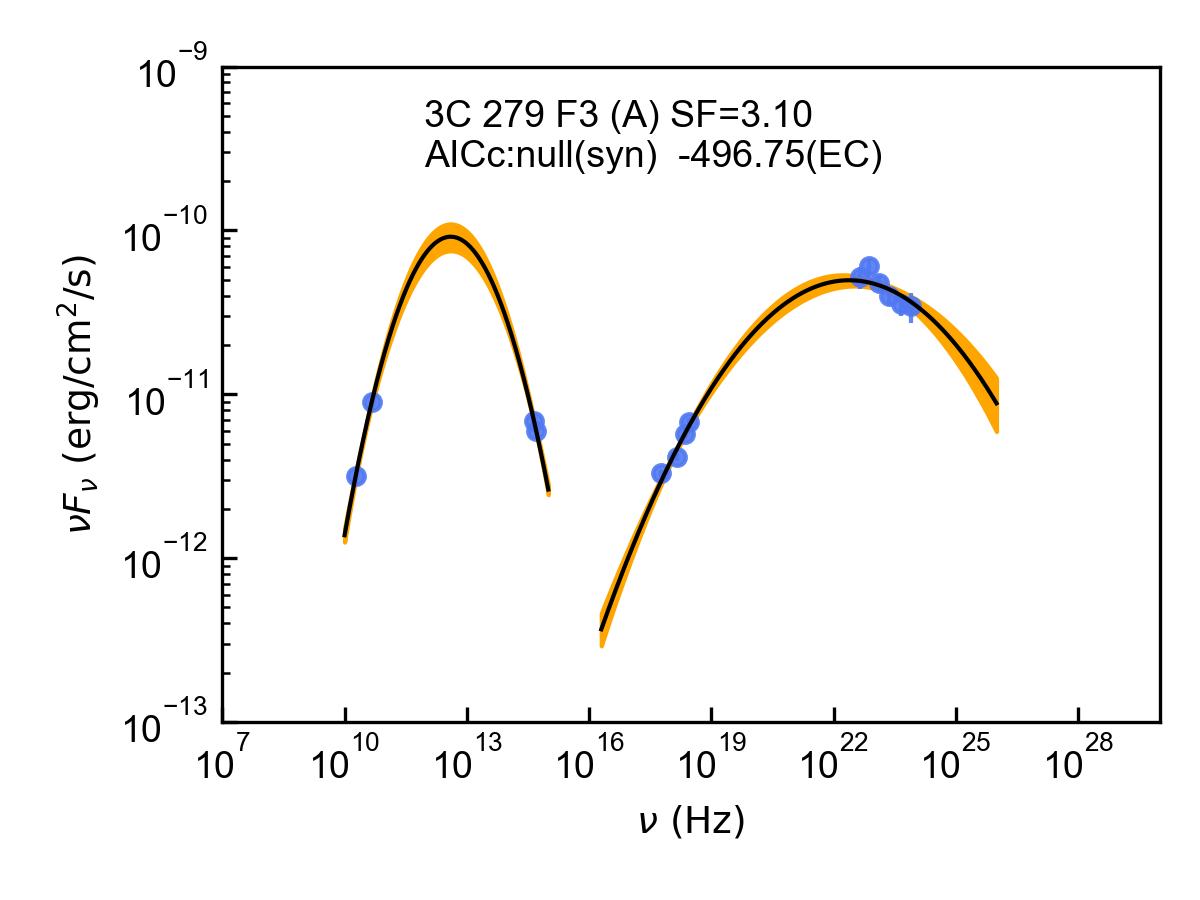}
    \end{subfigure}%
    \begin{subfigure}{.25\textwidth}
        \includegraphics[width = \linewidth]{ 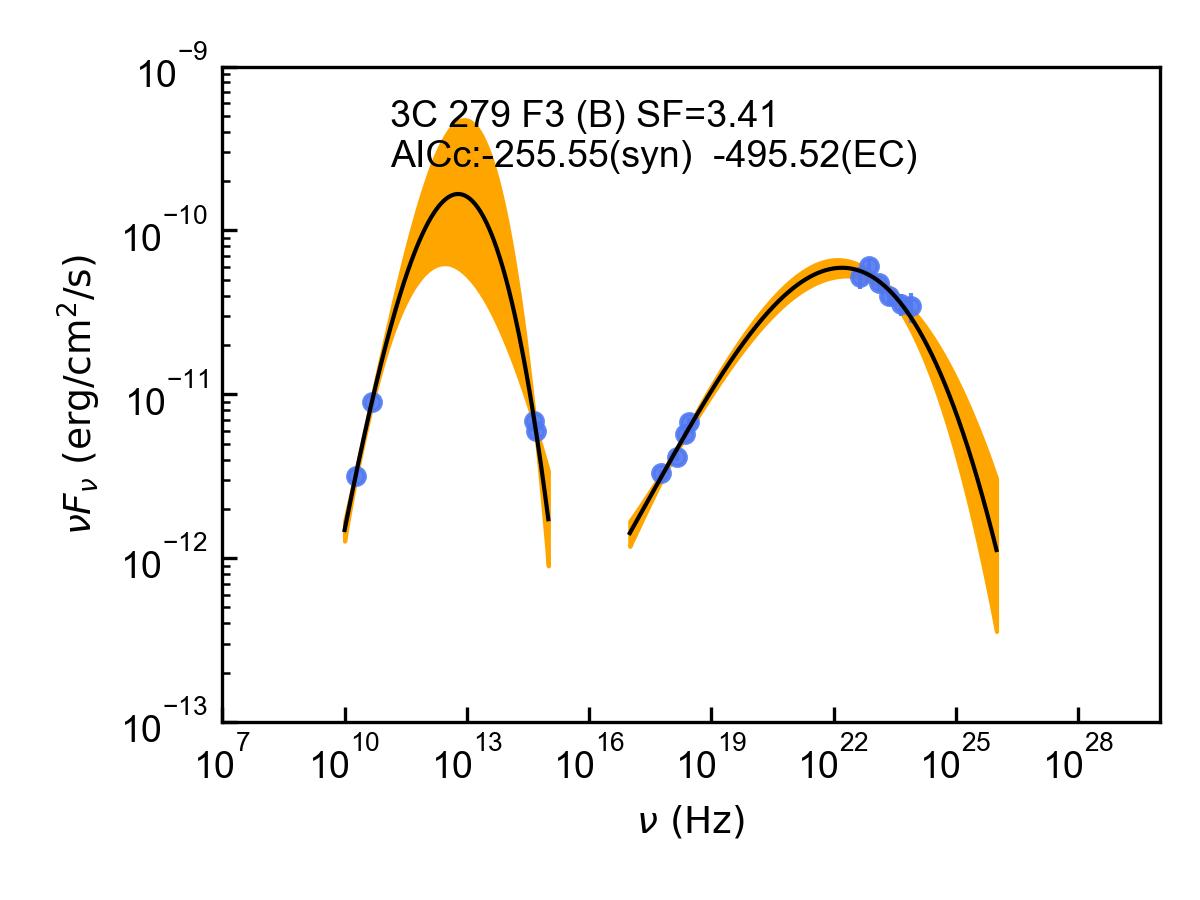}
    \end{subfigure}%
    \begin{subfigure}{.25\textwidth}
        \includegraphics[width = \linewidth]{ 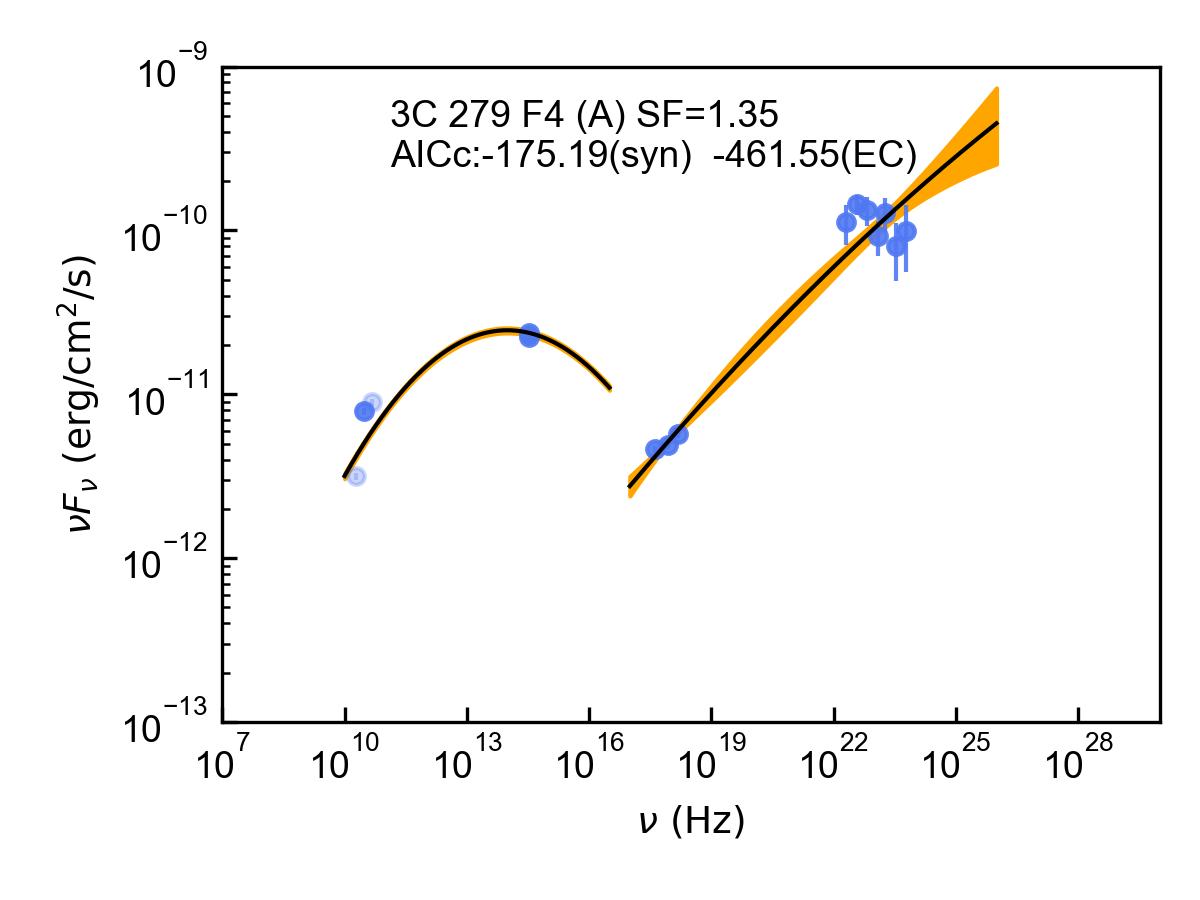}
    \end{subfigure}%
    \begin{subfigure}{.25\textwidth}
        \includegraphics[width = \linewidth]{ 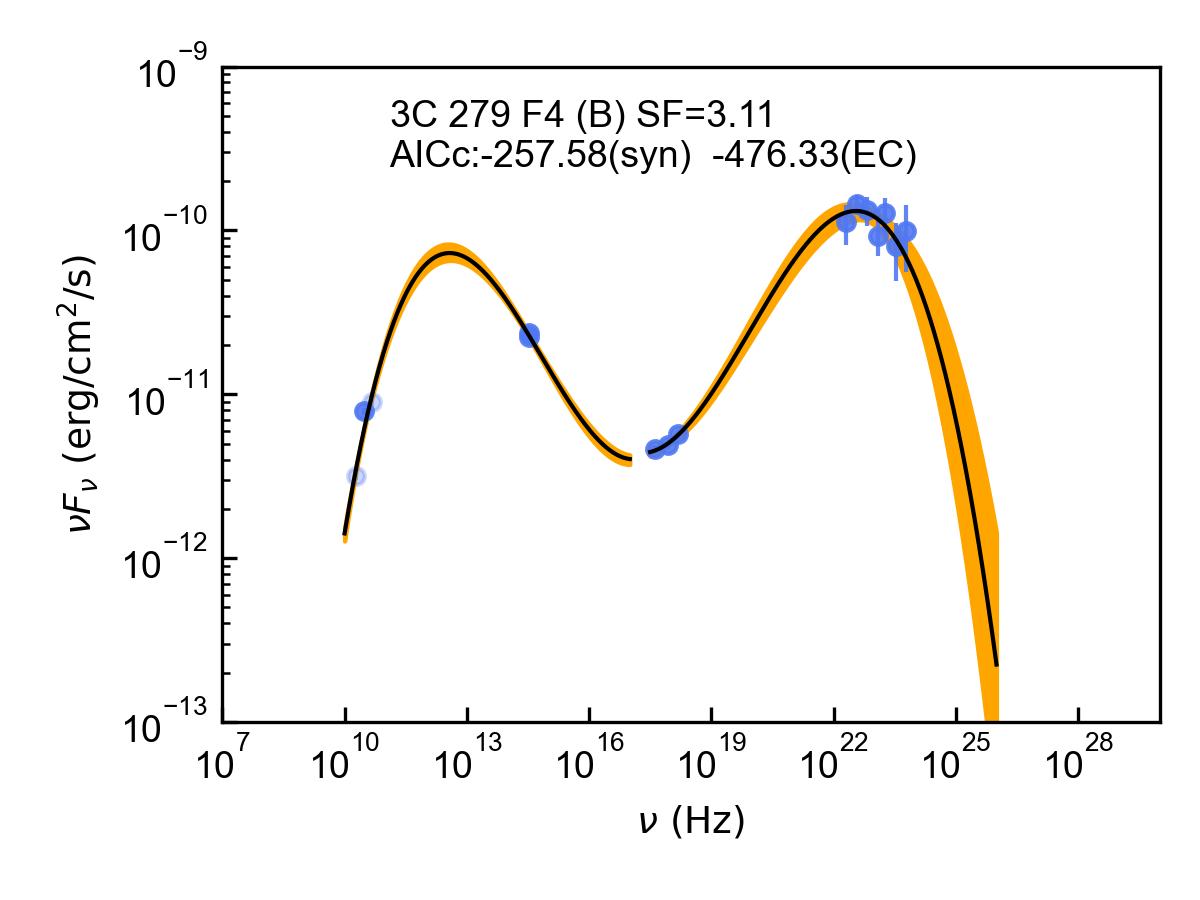}
    \end{subfigure}

    \begin{subfigure}{.25\textwidth}
        \includegraphics[width = \linewidth]{ 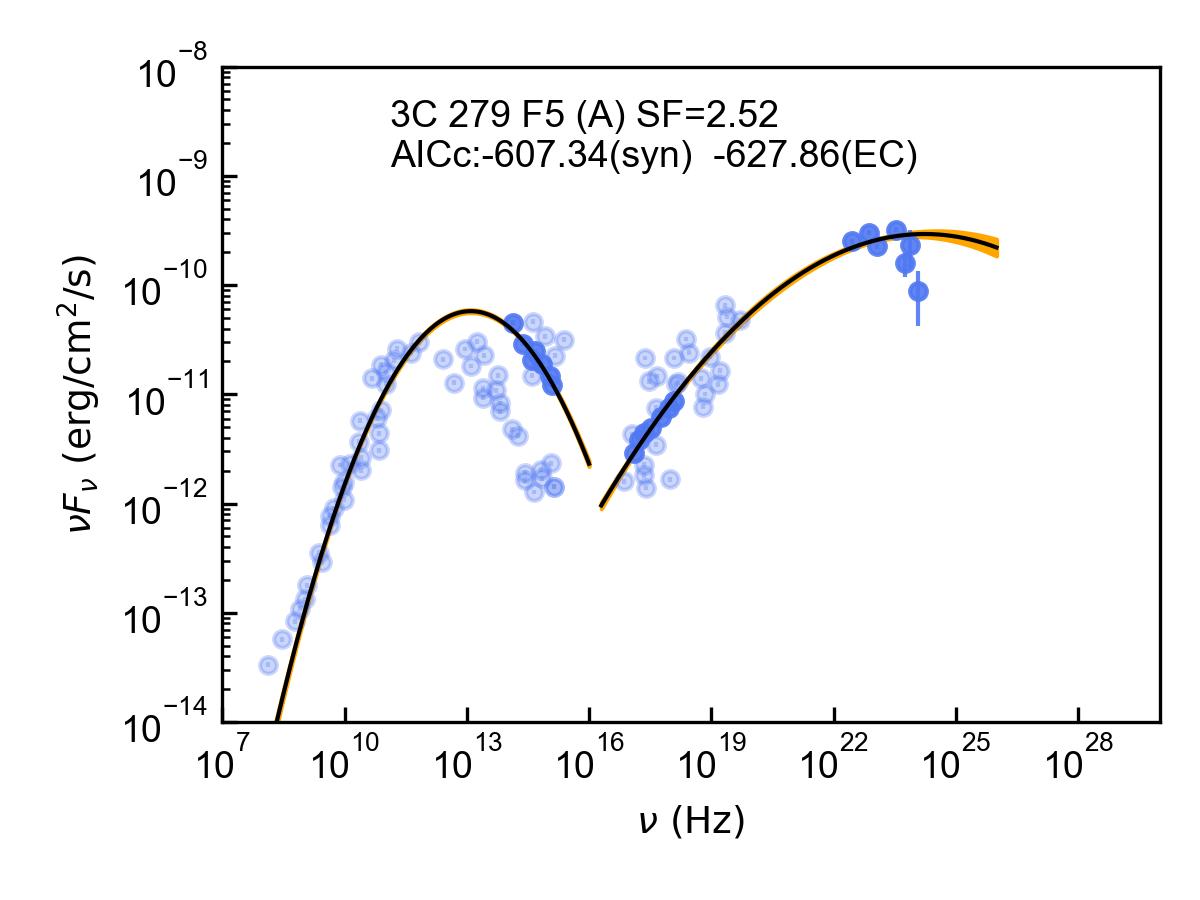}
    \end{subfigure}%
    \begin{subfigure}{.25\textwidth}
        \includegraphics[width = \linewidth]{ 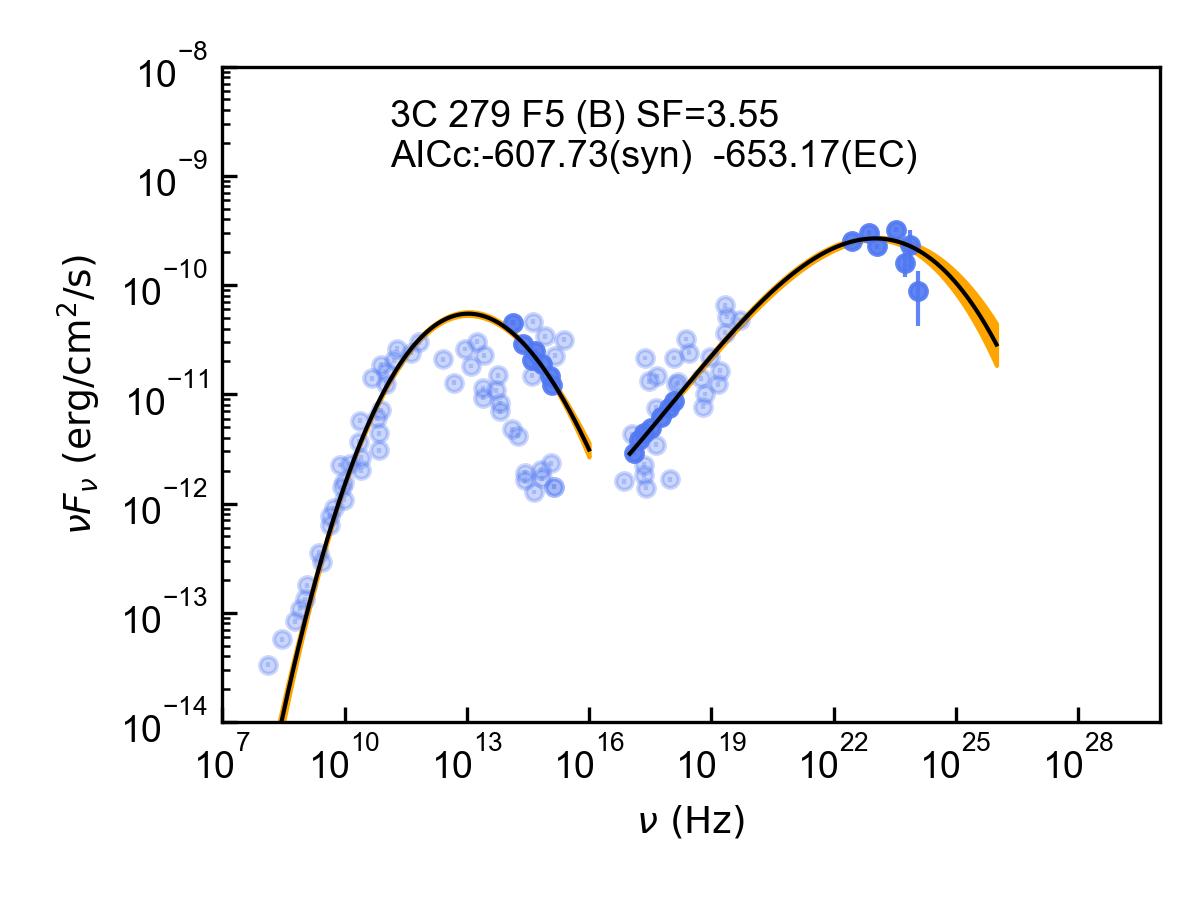}
    \end{subfigure}%
    \begin{subfigure}{.25\textwidth}
        \includegraphics[width = \linewidth]{ 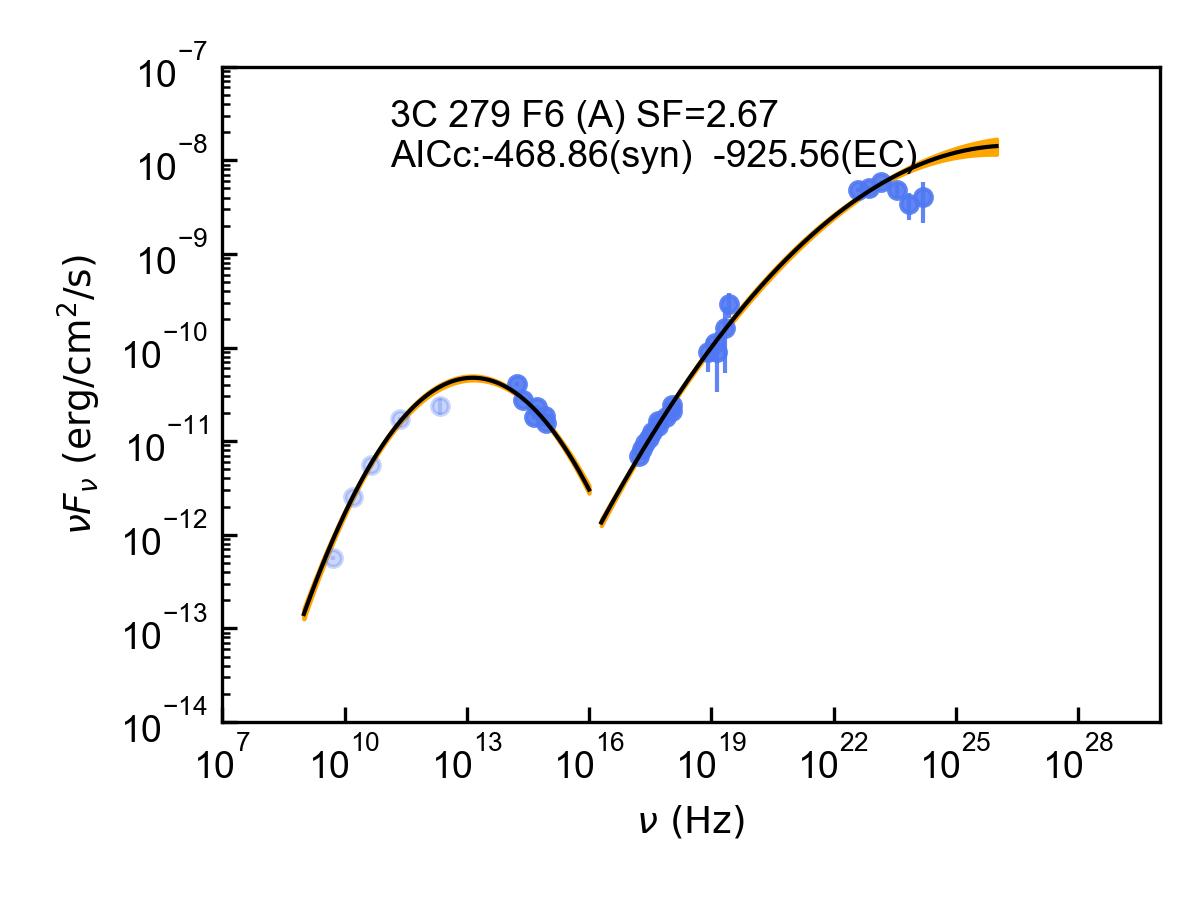}
    \end{subfigure}%
    \begin{subfigure}{.25\textwidth}
        \includegraphics[width = \linewidth]{ 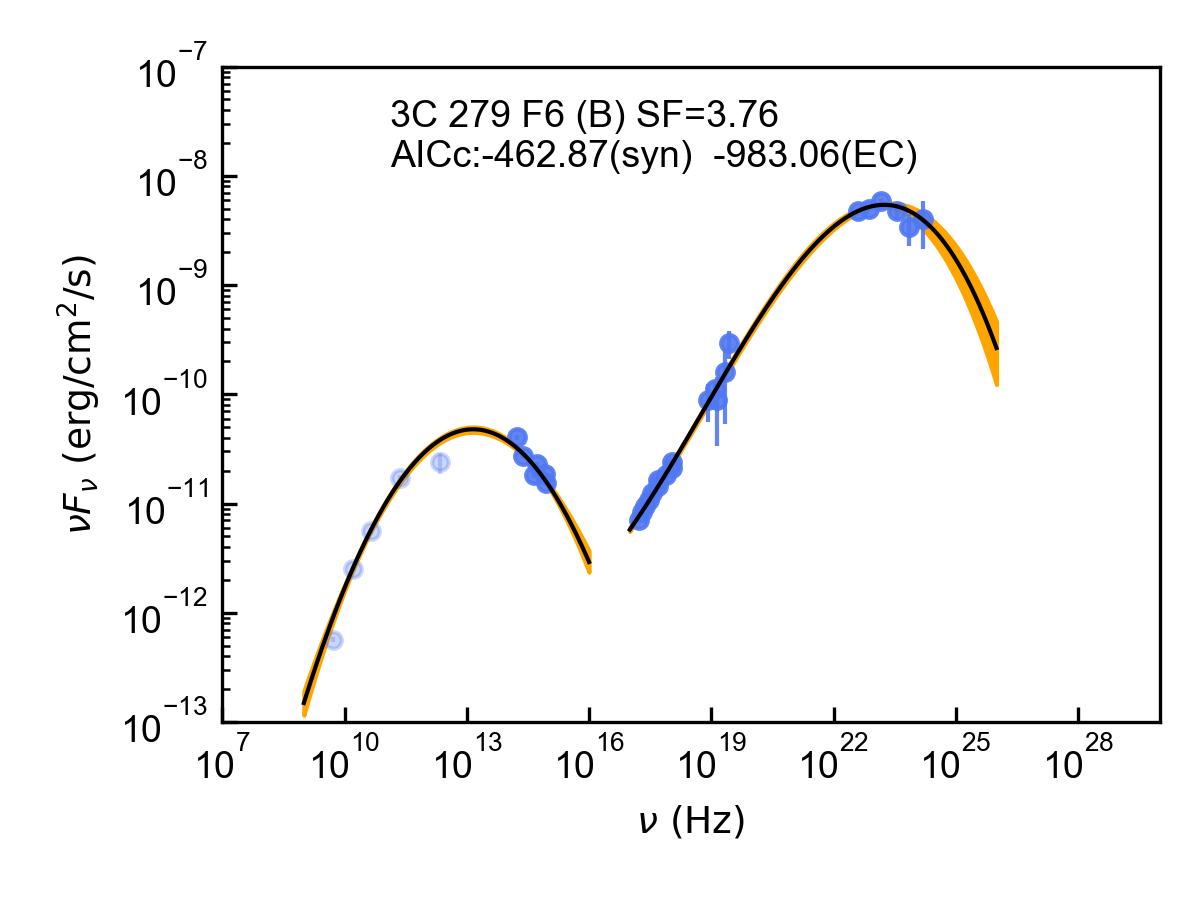}
    \end{subfigure}

    \raggedright
    \begin{subfigure}{.25\textwidth}
        \includegraphics[width = \linewidth]{ 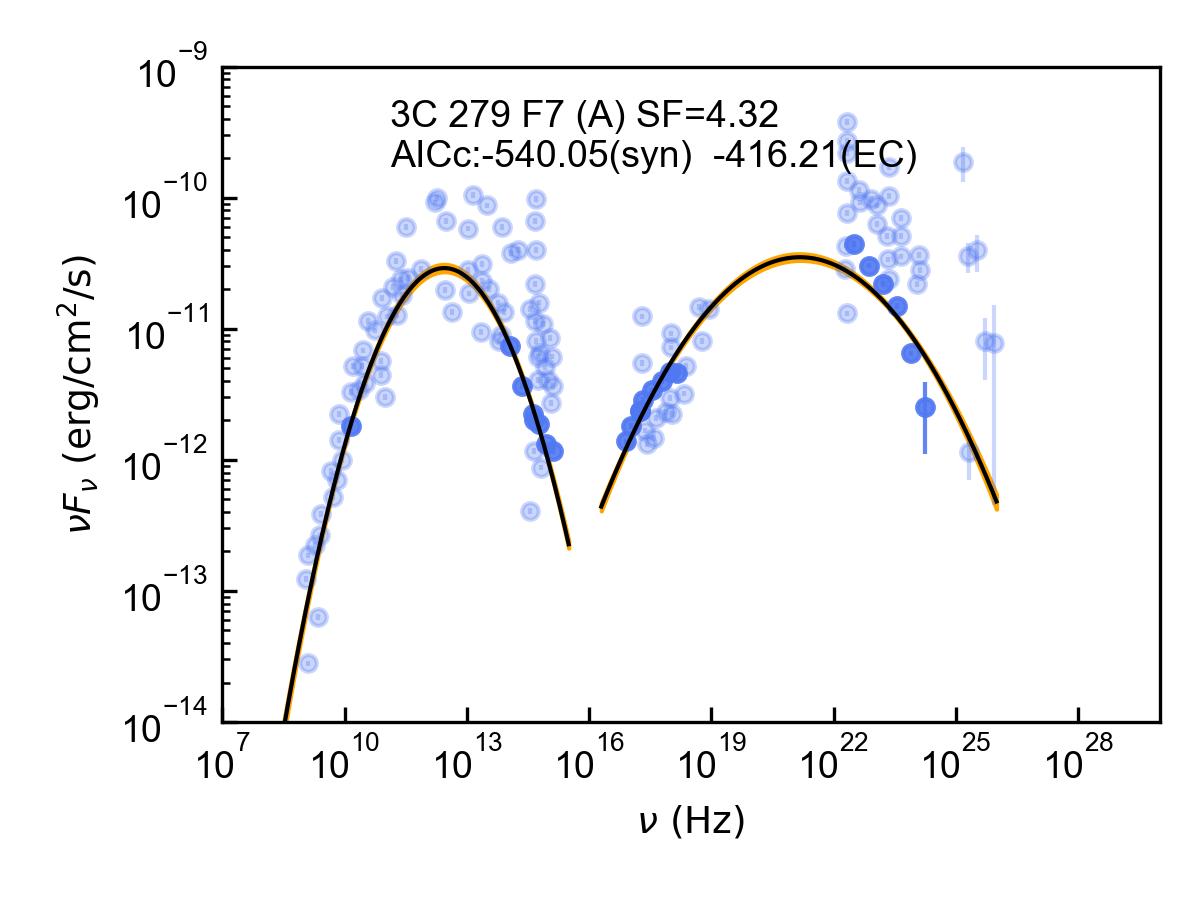}
    \end{subfigure}%
    \begin{subfigure}{.25\textwidth}
        \includegraphics[width = \linewidth]{ 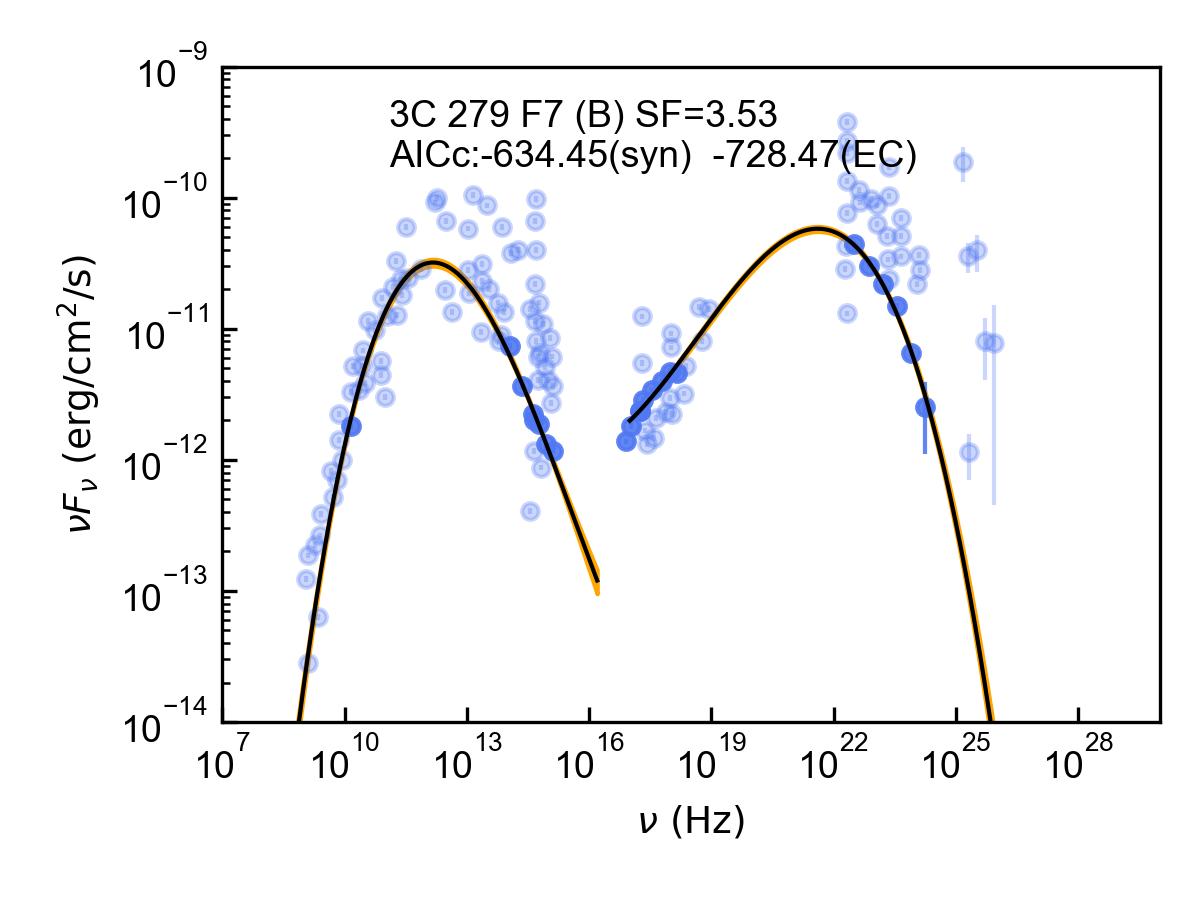}
    \end{subfigure}

    \label{SED 3C 279}
\end{figure*}

\begin{figure*}
    \caption{SED fitting results of TXS 0506+056.}
    \begin{subfigure}{.25\textwidth}
        \includegraphics[width = \linewidth]{ 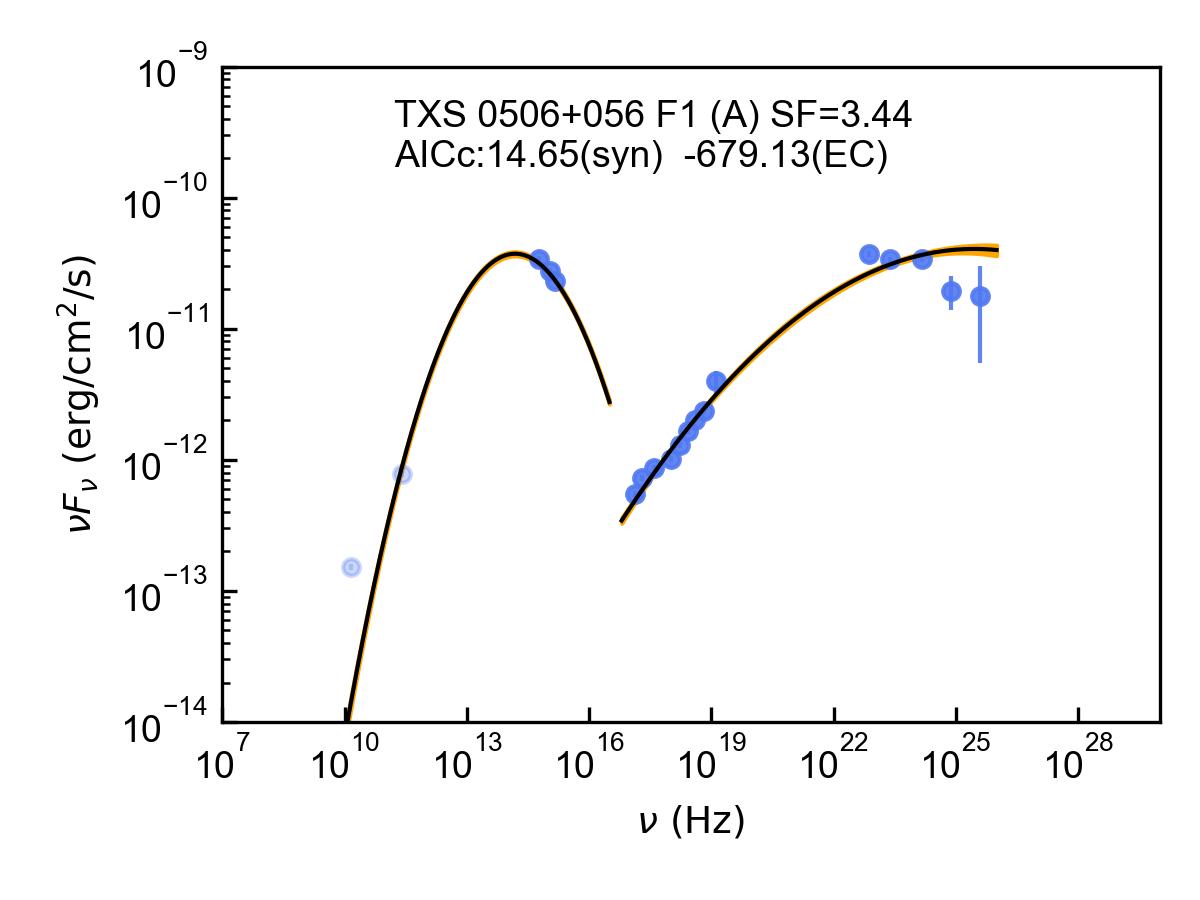}
    \end{subfigure}%
    \begin{subfigure}{.25\textwidth}
        \includegraphics[width = \linewidth]{ 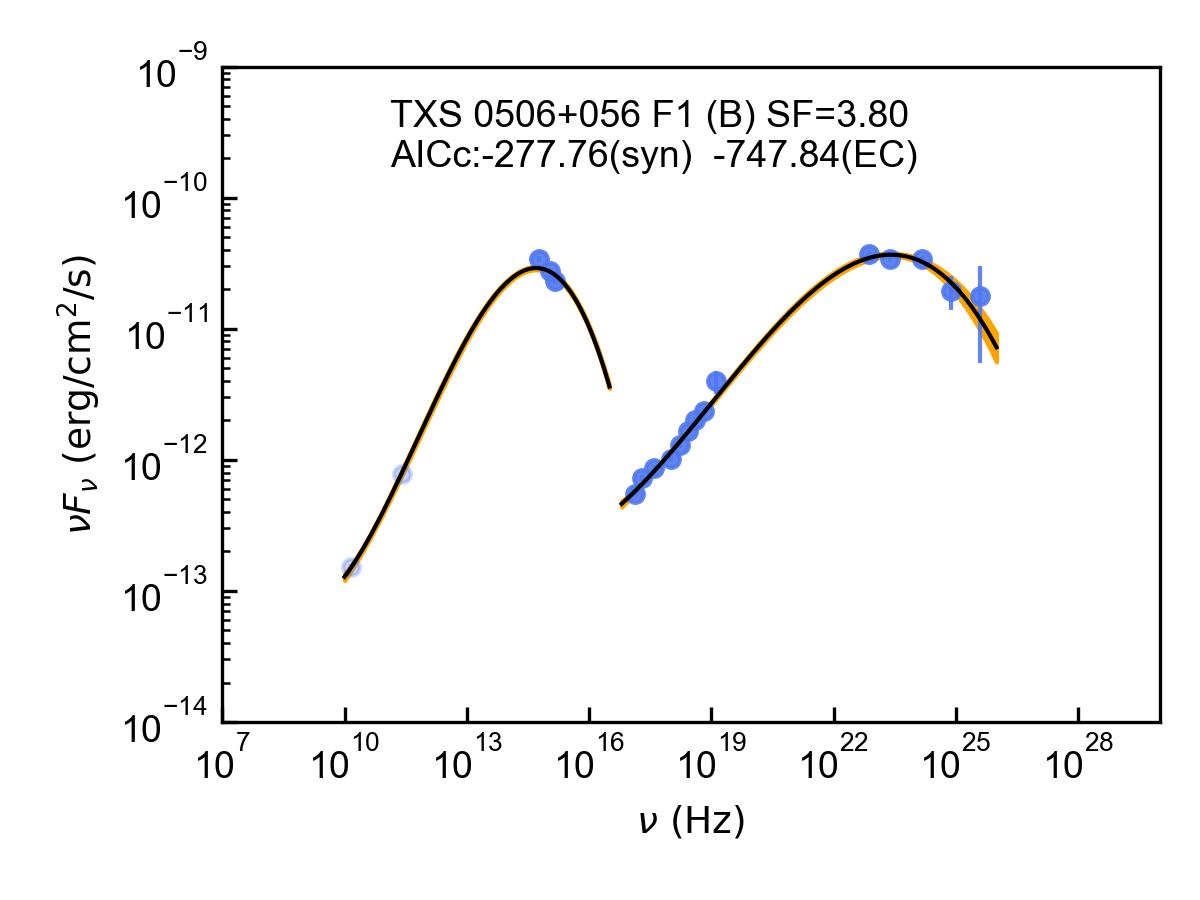}
    \end{subfigure}%
    \begin{subfigure}{.25\textwidth}
        \includegraphics[width = \linewidth]{ 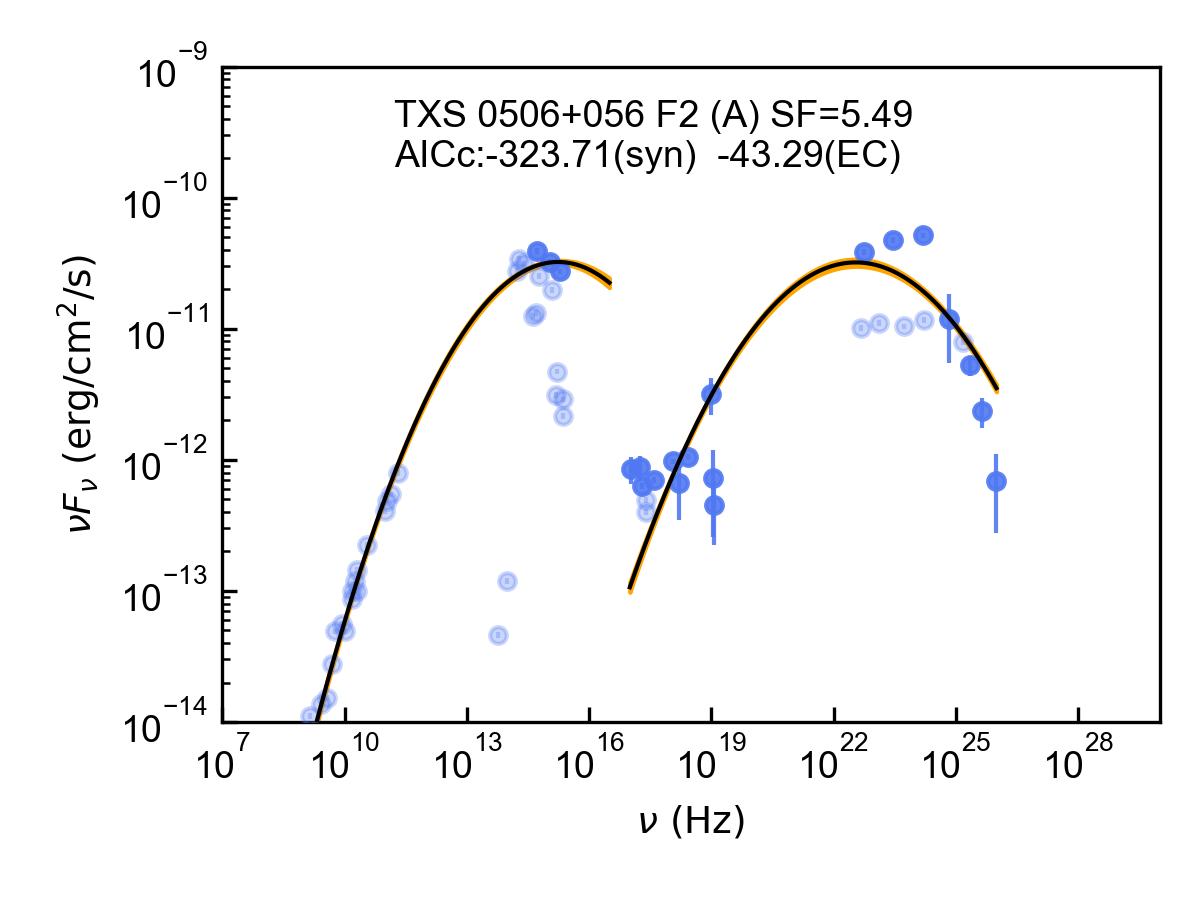}
    \end{subfigure}%
    \begin{subfigure}{.25\textwidth}
        \includegraphics[width = \linewidth]{ 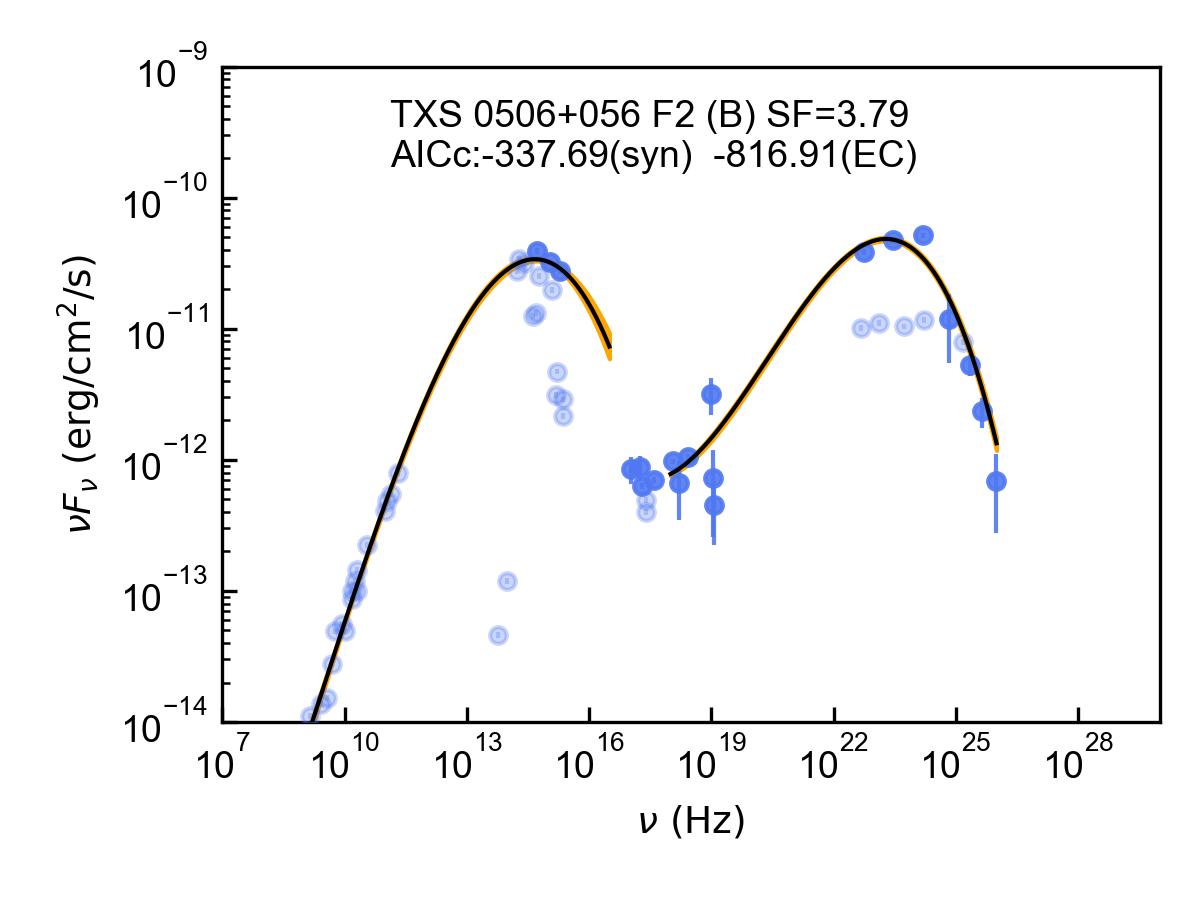}
    \end{subfigure}

    \raggedright
    \begin{subfigure}{.25\textwidth}
        \includegraphics[width = \linewidth]{ 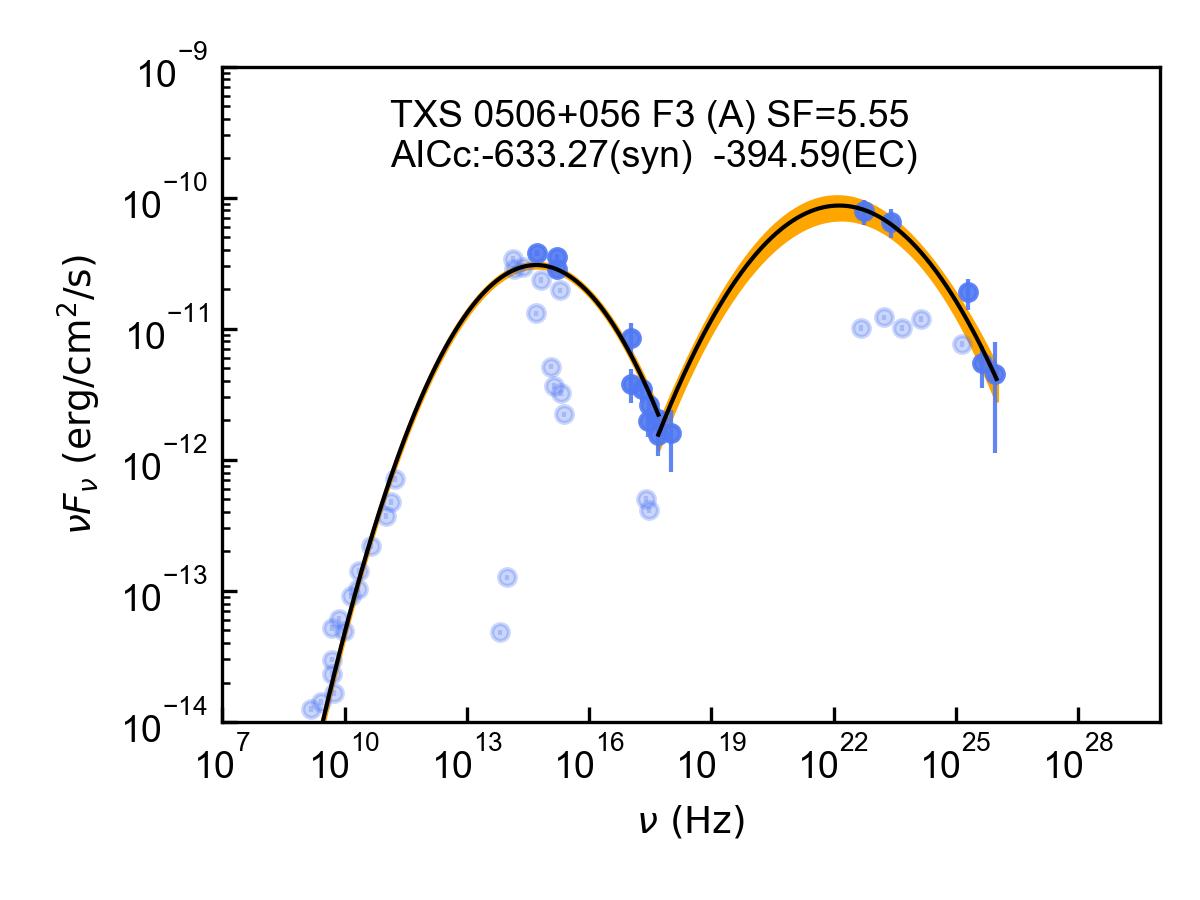}
    \end{subfigure}%
    \begin{subfigure}{.25\textwidth}
        \includegraphics[width = \linewidth]{ 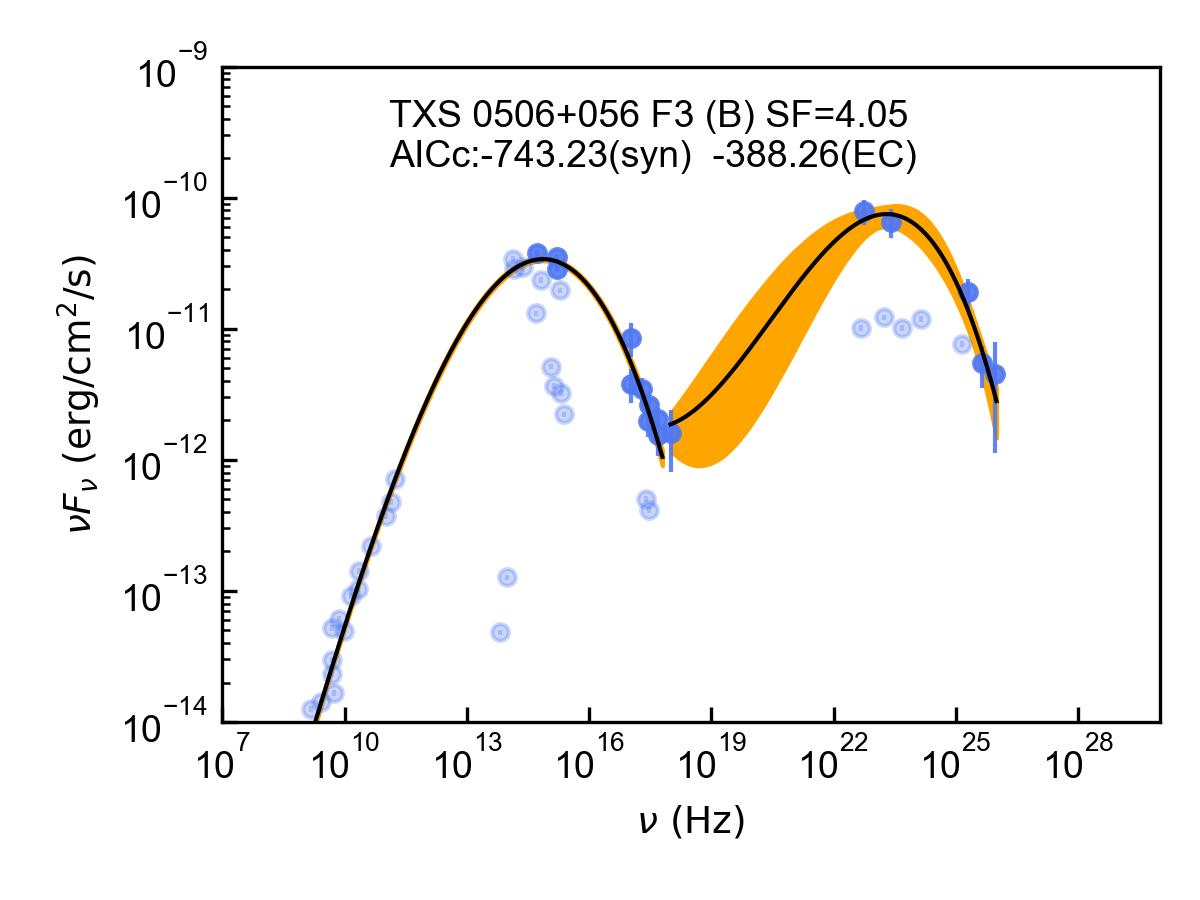}
    \end{subfigure}

    \label{SED TXS 0506+056}
\end{figure*}

\begin{figure*}
    \caption{SED fitting results of OJ 287.}
    \begin{subfigure}{.25\textwidth}
        \includegraphics[width = \linewidth]{ 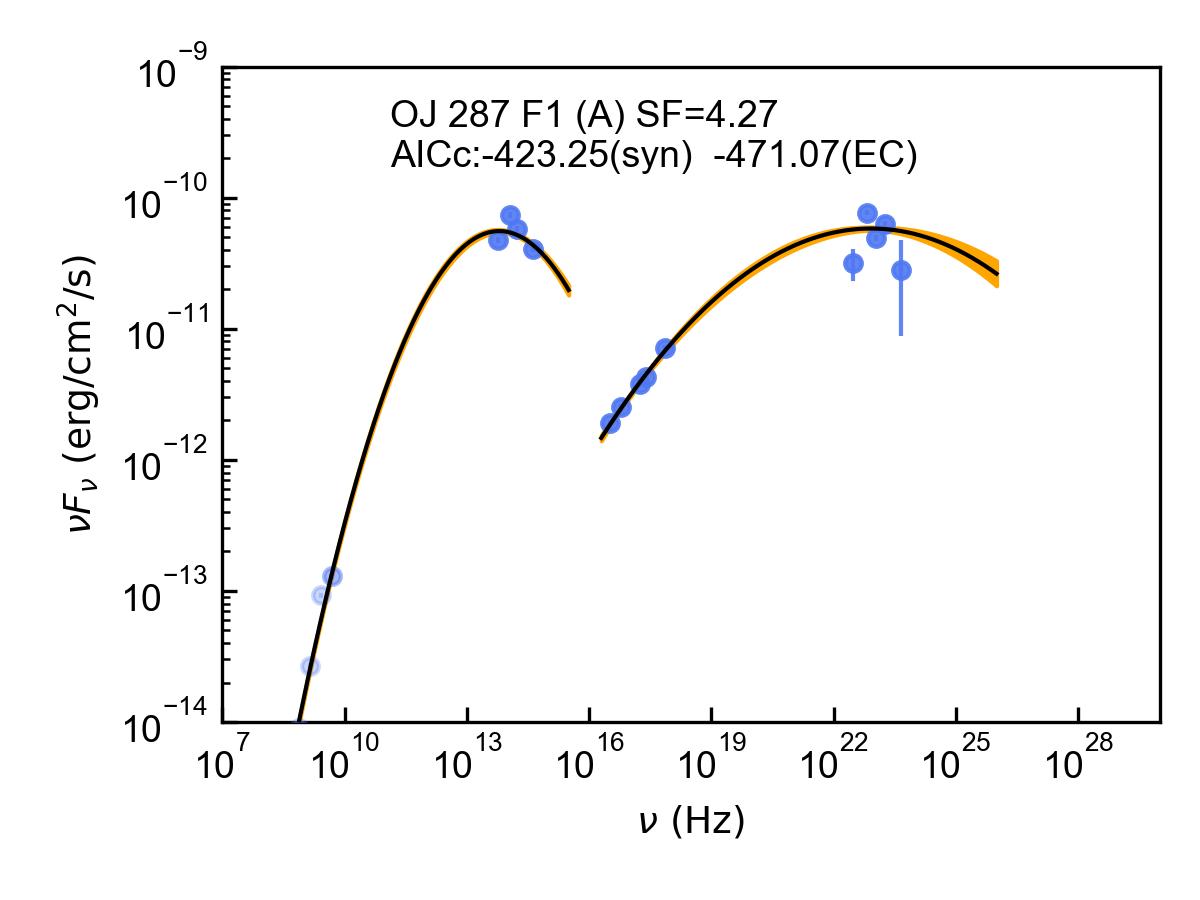}
    \end{subfigure}%
    \begin{subfigure}{.25\textwidth}
        \includegraphics[width = \linewidth]{ 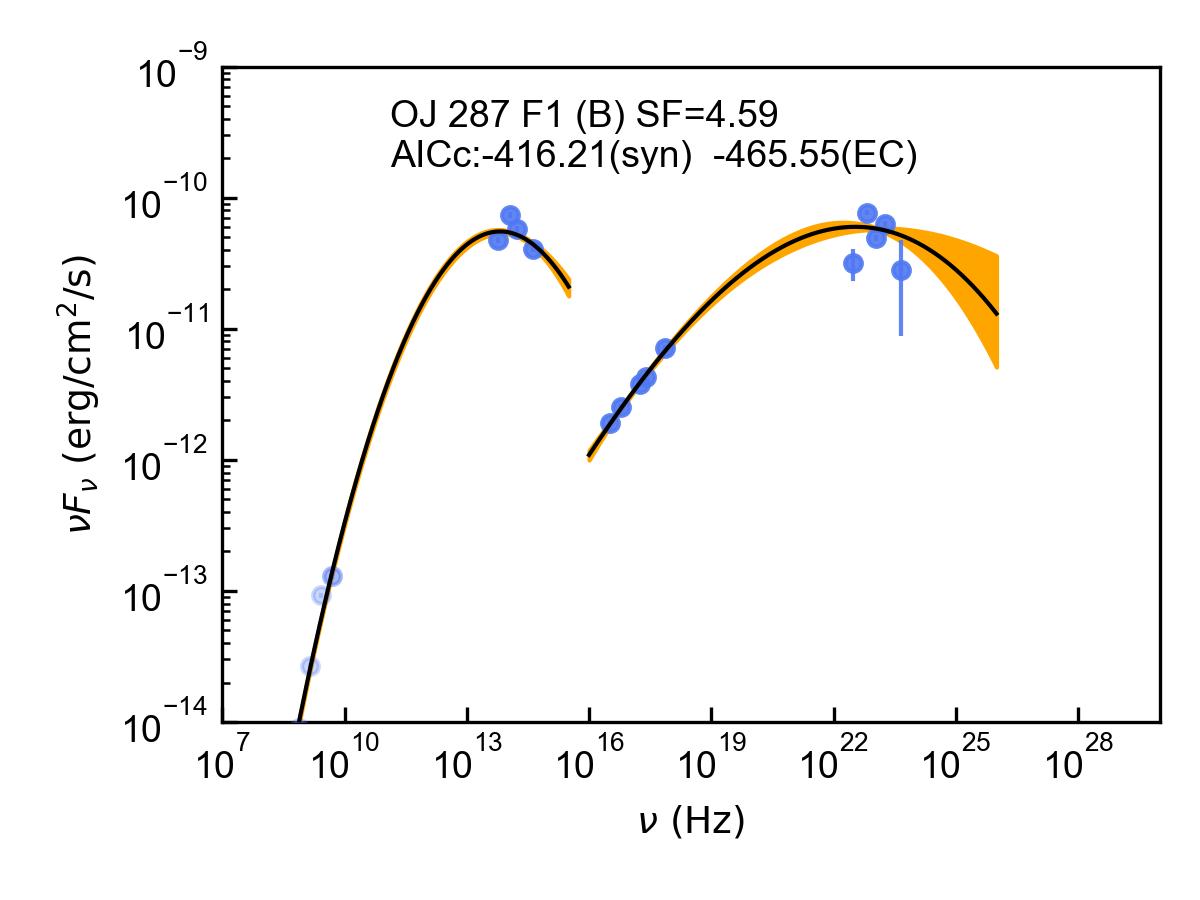}
    \end{subfigure}%
    \begin{subfigure}{.25\textwidth}
        \includegraphics[width = \linewidth]{ 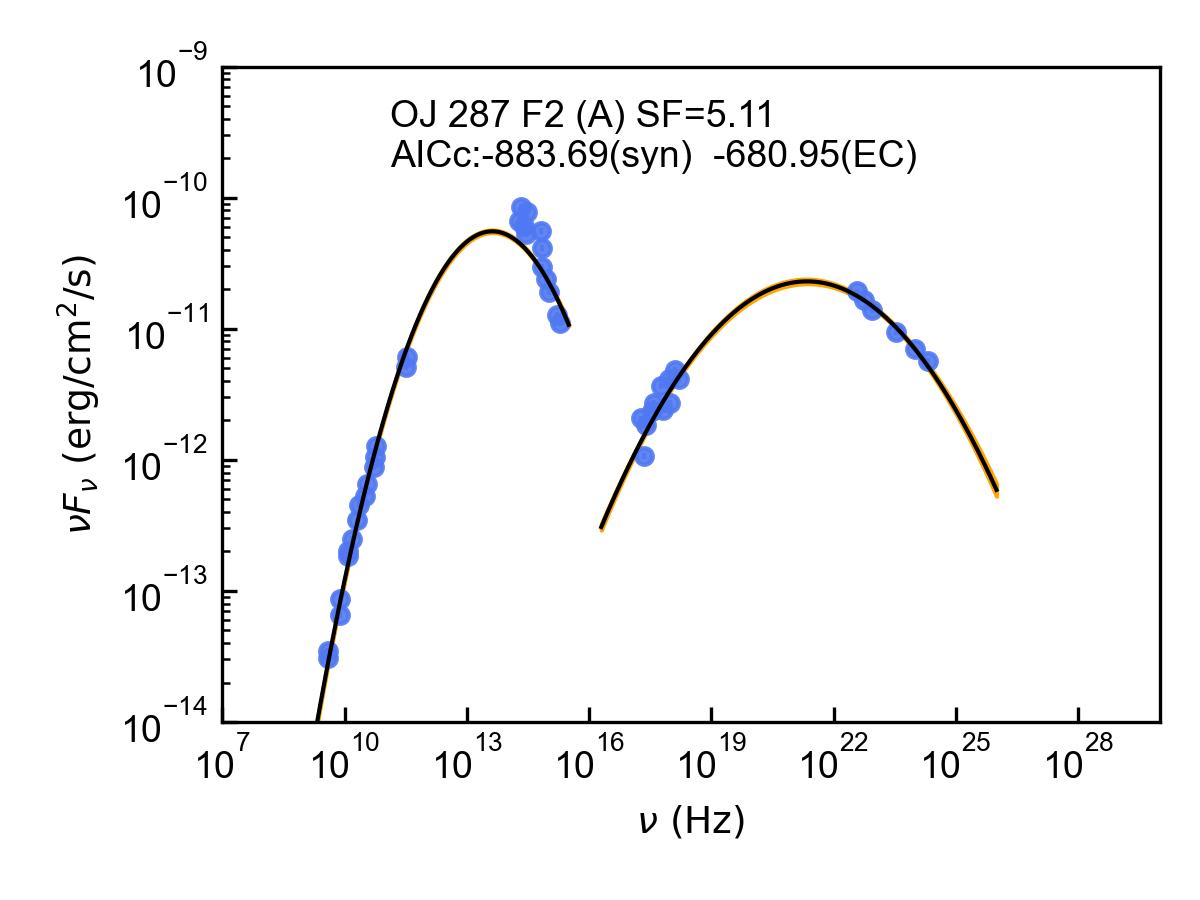}
    \end{subfigure}%
    \begin{subfigure}{.25\textwidth}
        \includegraphics[width = \linewidth]{ 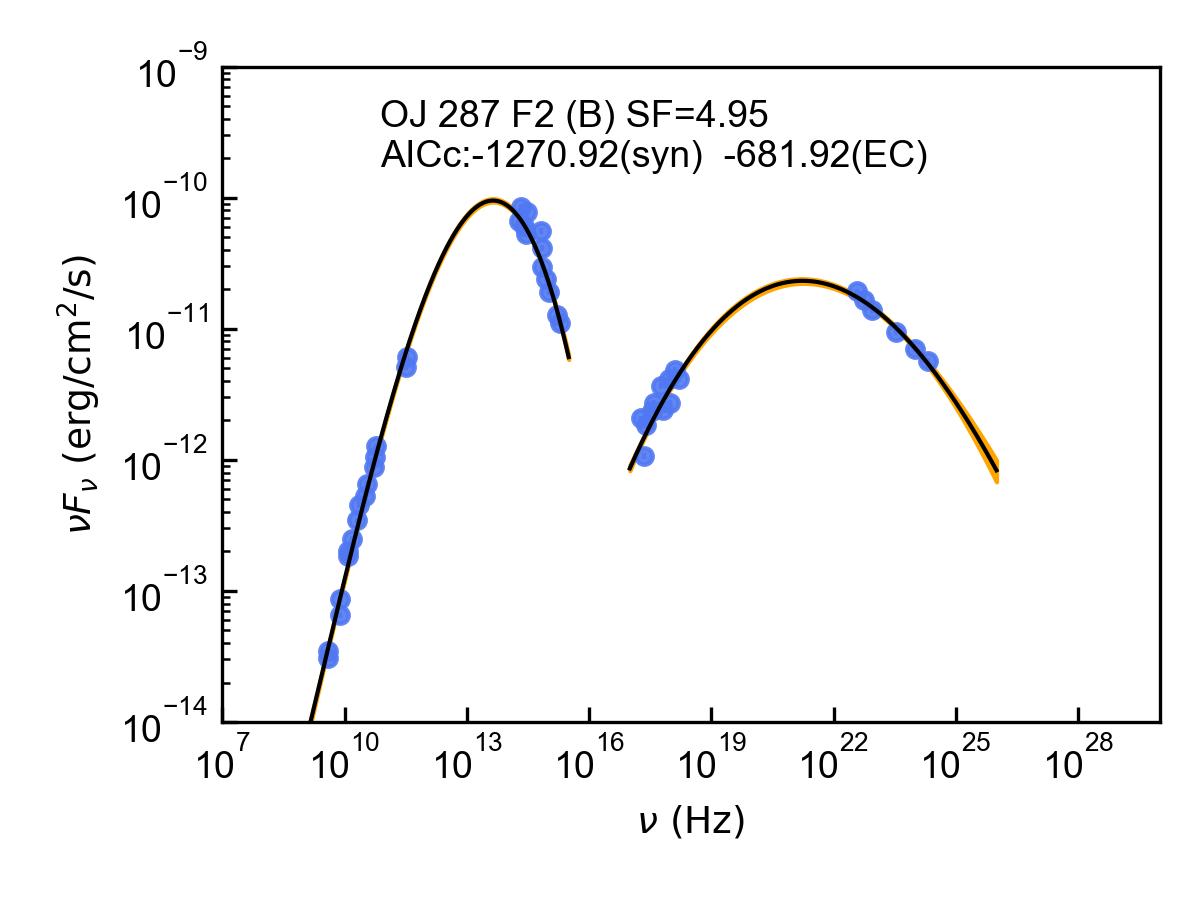}
    \end{subfigure}

    \begin{subfigure}{.25\textwidth}
        \includegraphics[width = \linewidth]{ 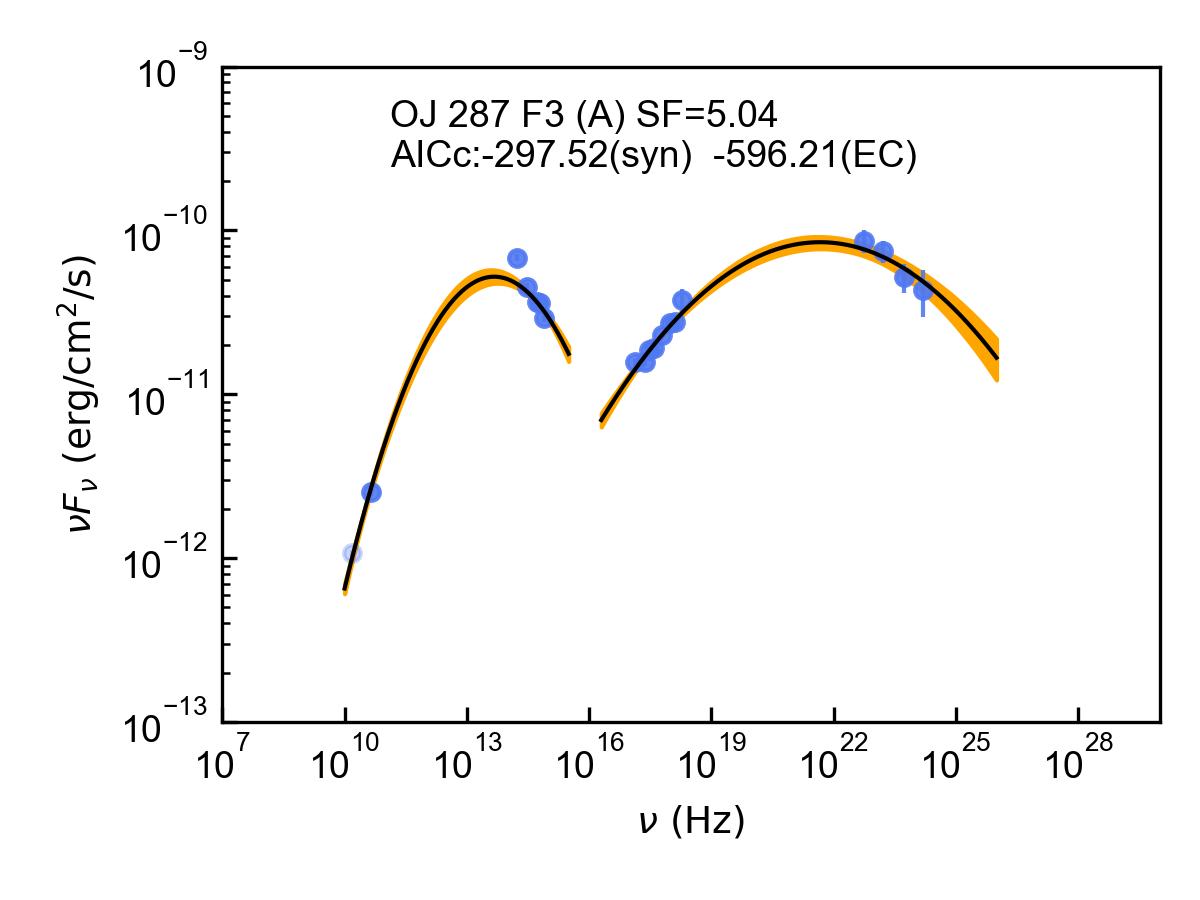}
    \end{subfigure}%
    \begin{subfigure}{.25\textwidth}
        \includegraphics[width = \linewidth]{ 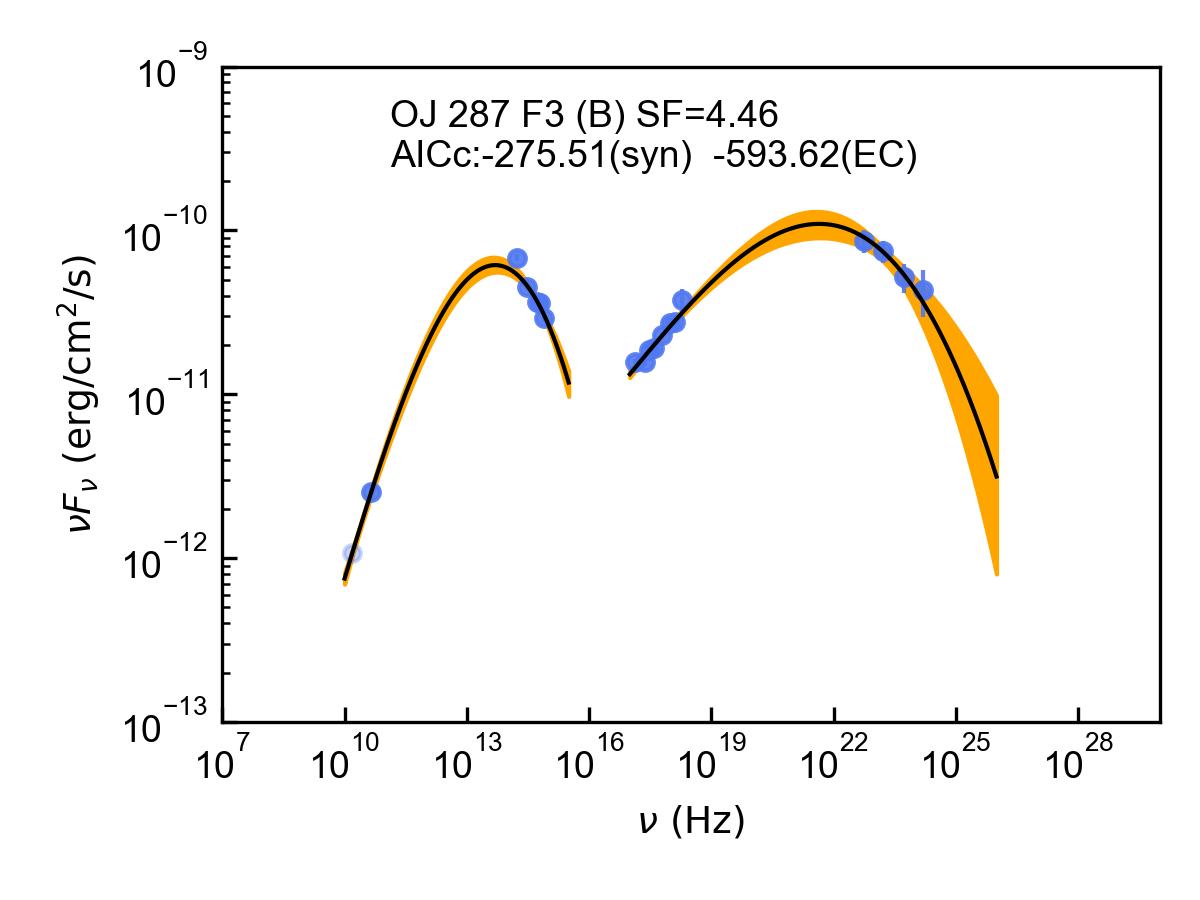}
    \end{subfigure}%
    \begin{subfigure}{.25\textwidth}
        \includegraphics[width = \linewidth]{ 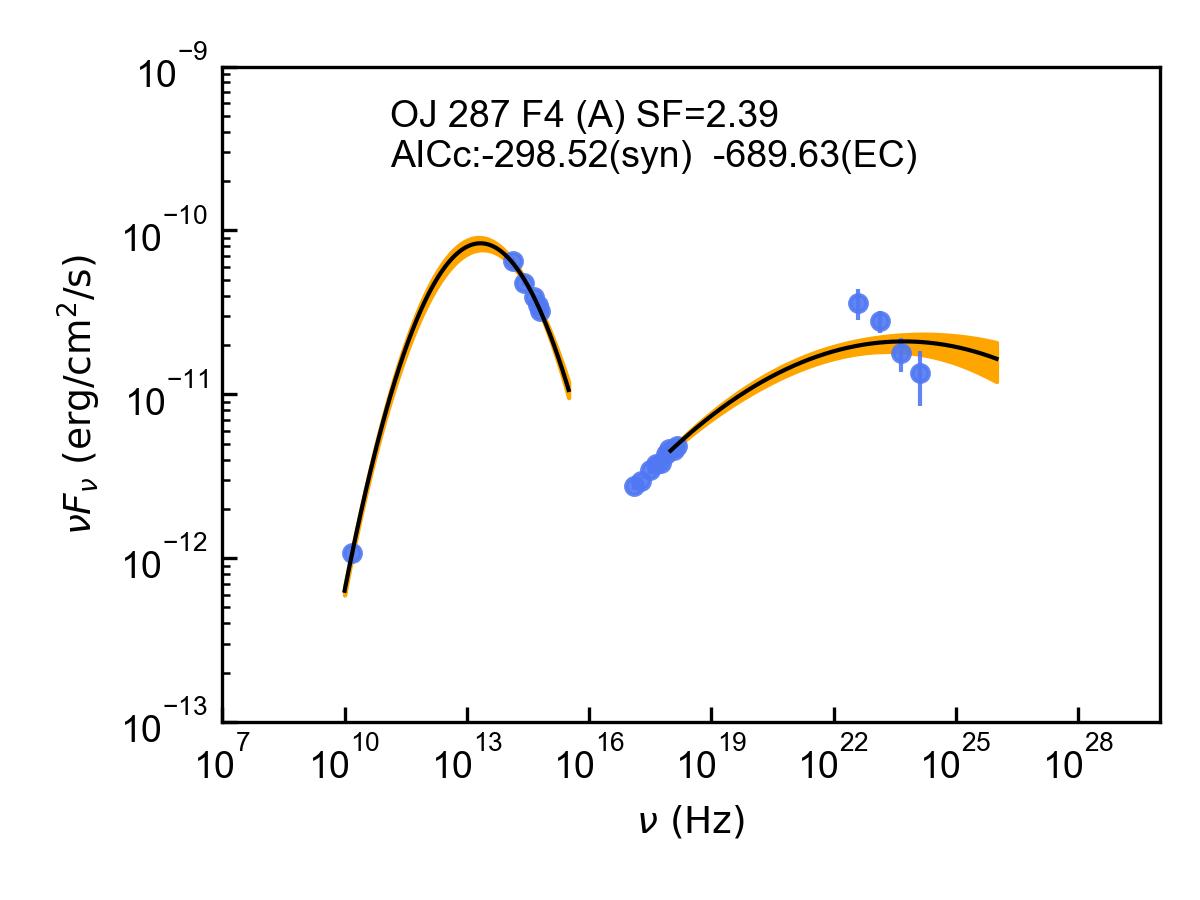}
    \end{subfigure}%
    \begin{subfigure}{.25\textwidth}
        \includegraphics[width = \linewidth]{ 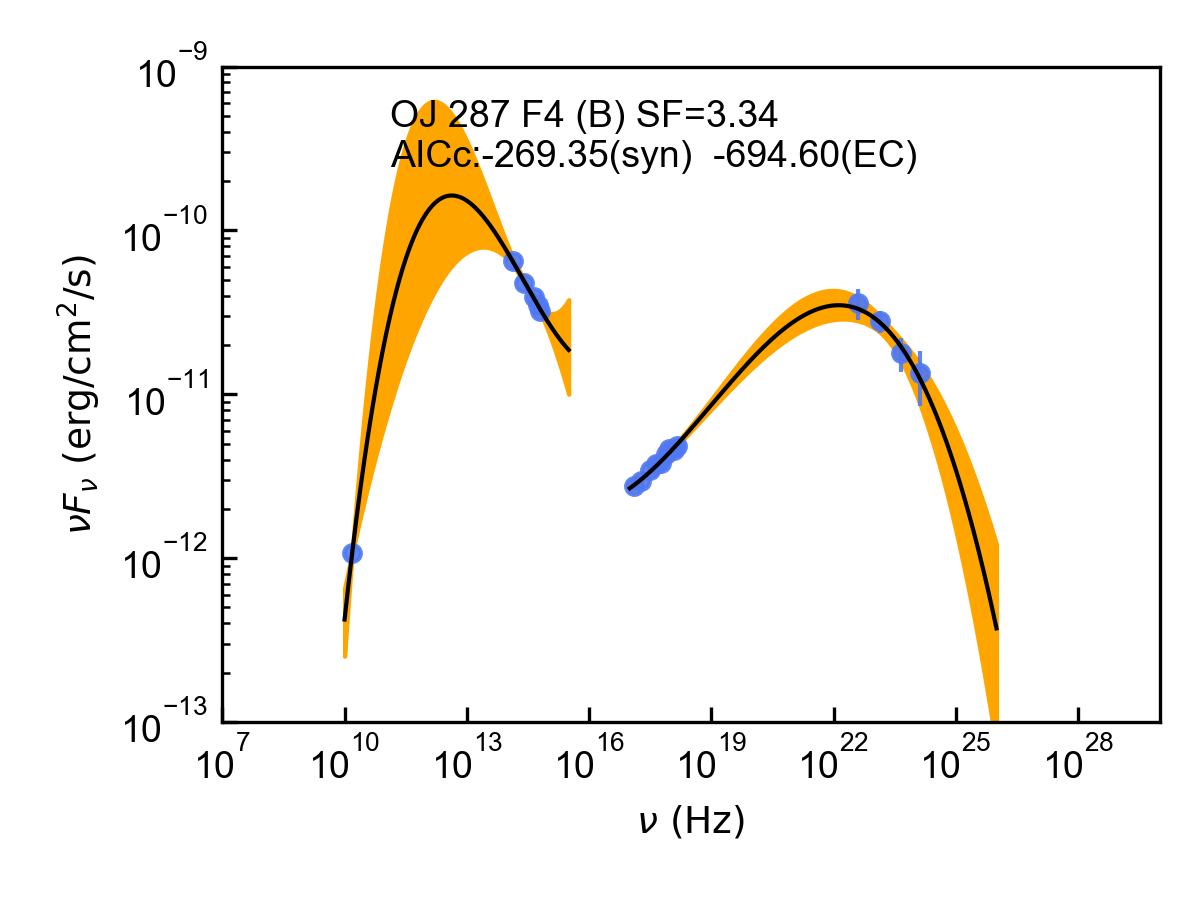}
    \end{subfigure}

    \raggedright
    \begin{subfigure}{.25\textwidth}
        \includegraphics[width = \linewidth]{ 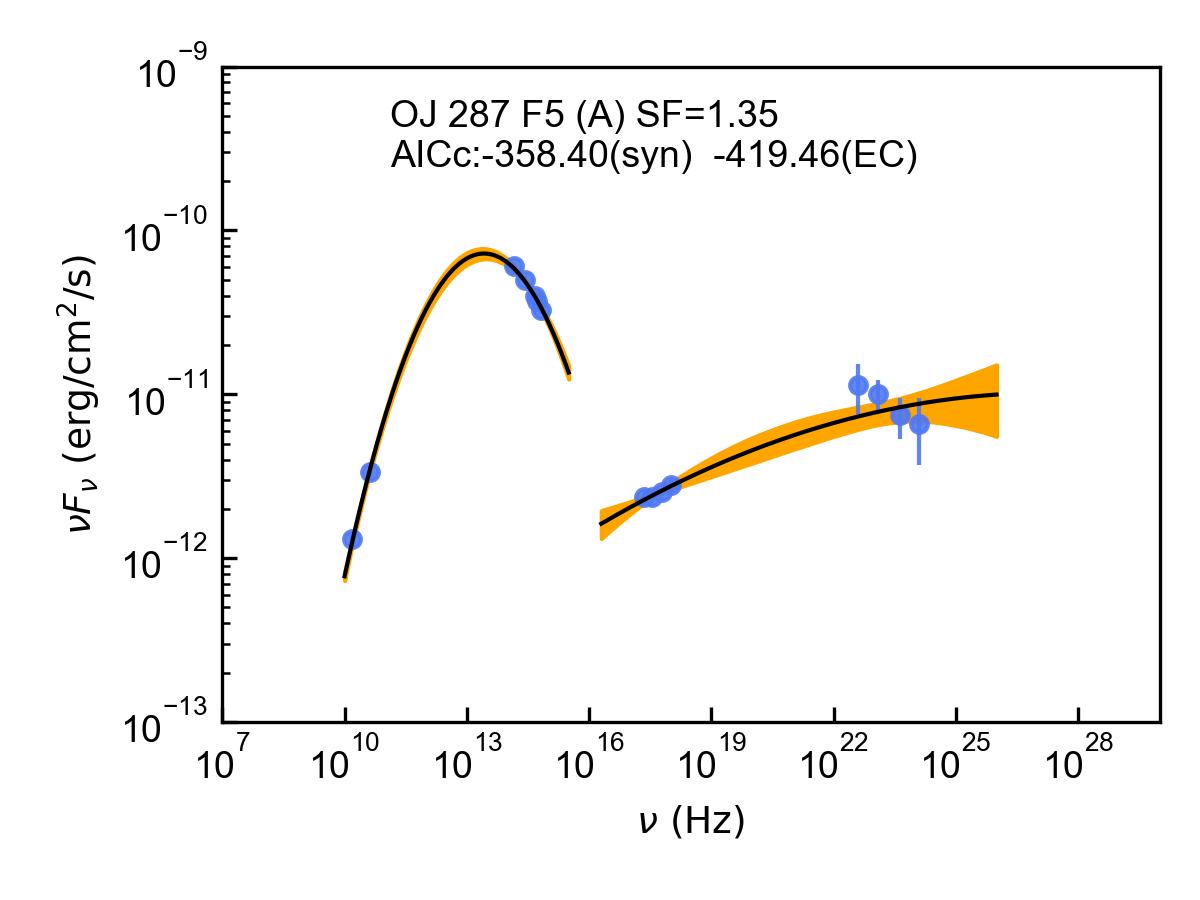}
    \end{subfigure}%
    \begin{subfigure}{.25\textwidth}
        \includegraphics[width = \linewidth]{ 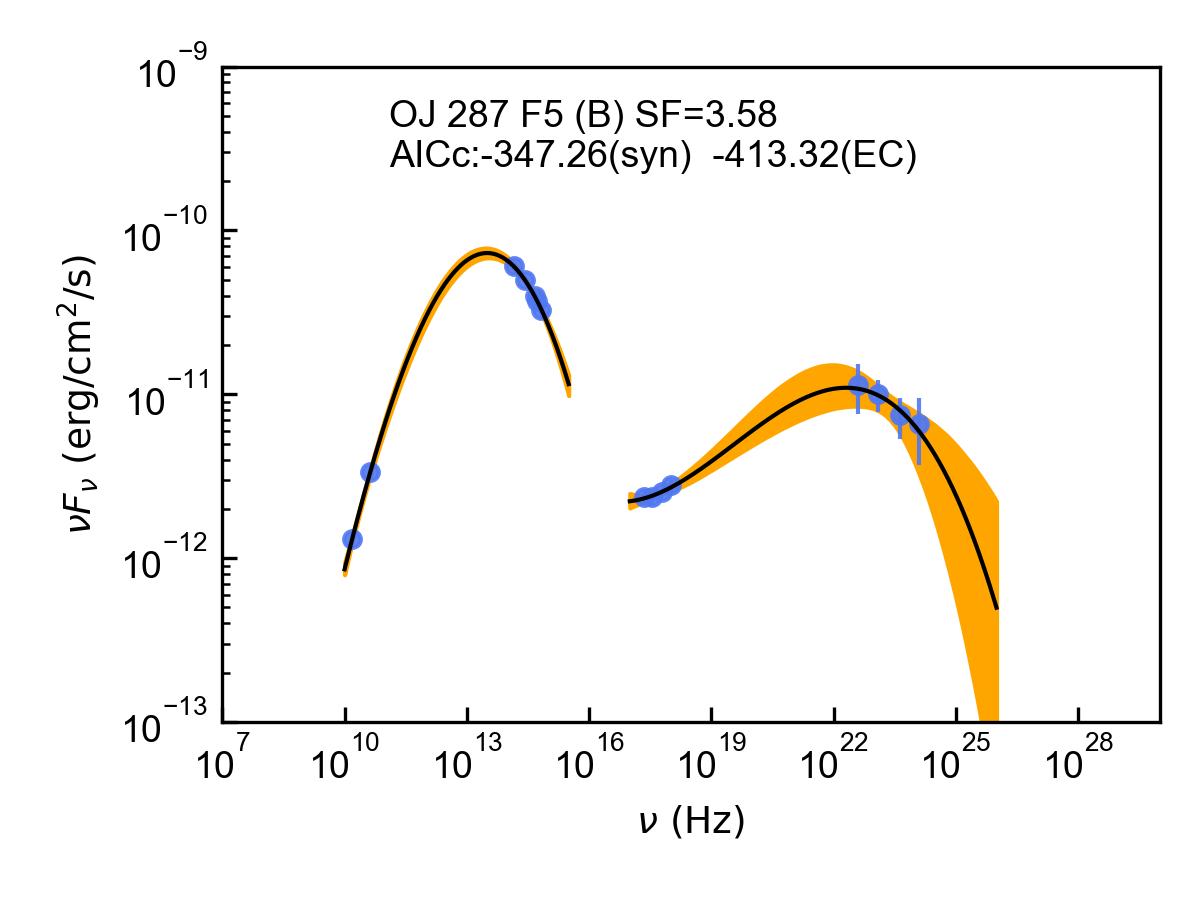}
    \end{subfigure}

    \label{SED OJ 287}
\end{figure*}

Figures \ref{SED DT Doppler} to \ref{SED OJ 287} present all the 45 SEDs fitted by both quadratic and cubic functions. A and B denote the SEDs fitted by quadratic and cubic functions, respectively. Blue and light blue scatters represent the simultaneous and archival data, respectively. Black lines are plotted using the maximum posterior values. Yellow areas denote the 1-$\sigma$ uncertainties under 16-50-84 rule. The AICc of each hump and observed seed factors are displayed on the figures.

\begin{figure*}
    \caption{SED fitting results of the blazars dominated by dusty torus with variability timescales.}

    \begin{subfigure}{.25\textwidth}
        \includegraphics[width = \linewidth]{ 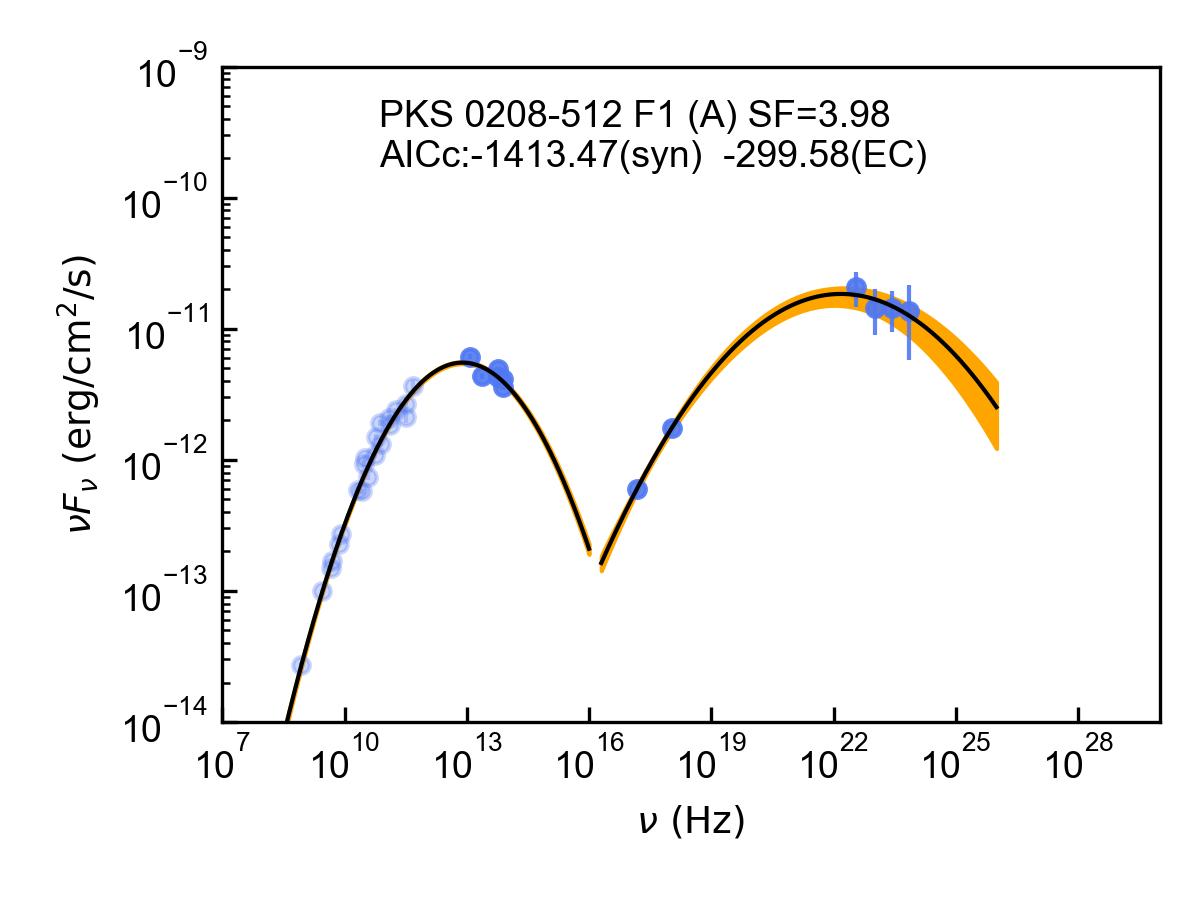}
    \end{subfigure}%
    \begin{subfigure}{.25\textwidth}
        \includegraphics[width = \linewidth]{ 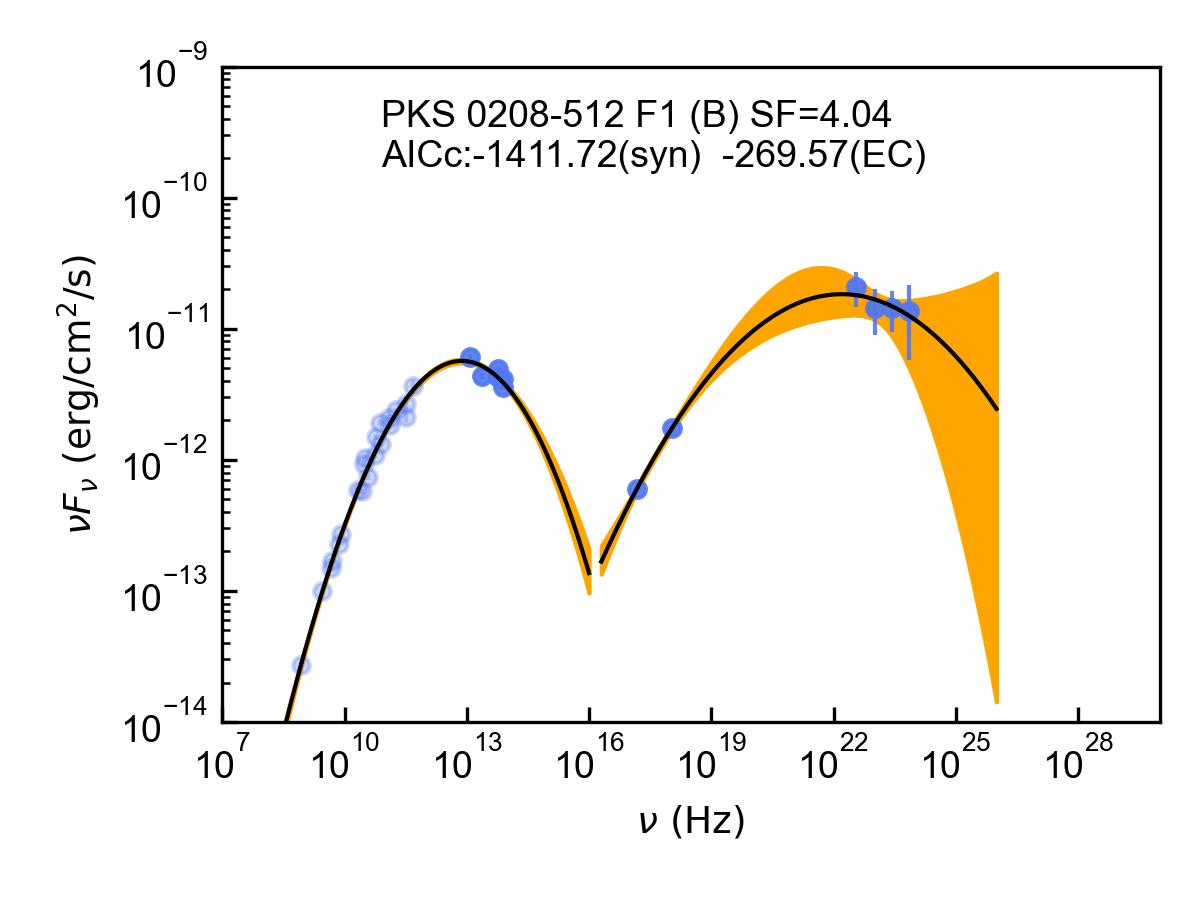}
    \end{subfigure}%
    \begin{subfigure}{.25\textwidth}
        \includegraphics[width = \linewidth]{ 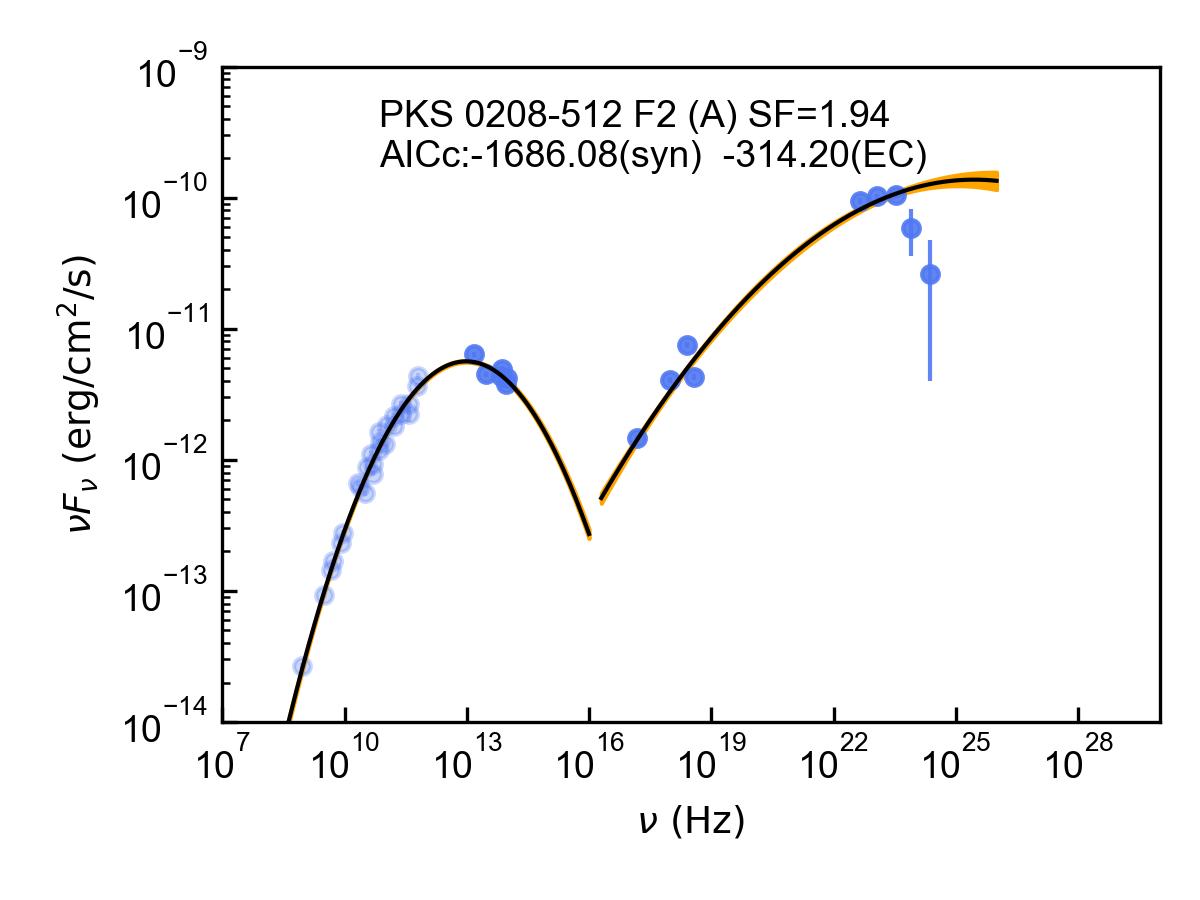}
    \end{subfigure}%
    \begin{subfigure}{.25\textwidth}
        \includegraphics[width = \linewidth]{ 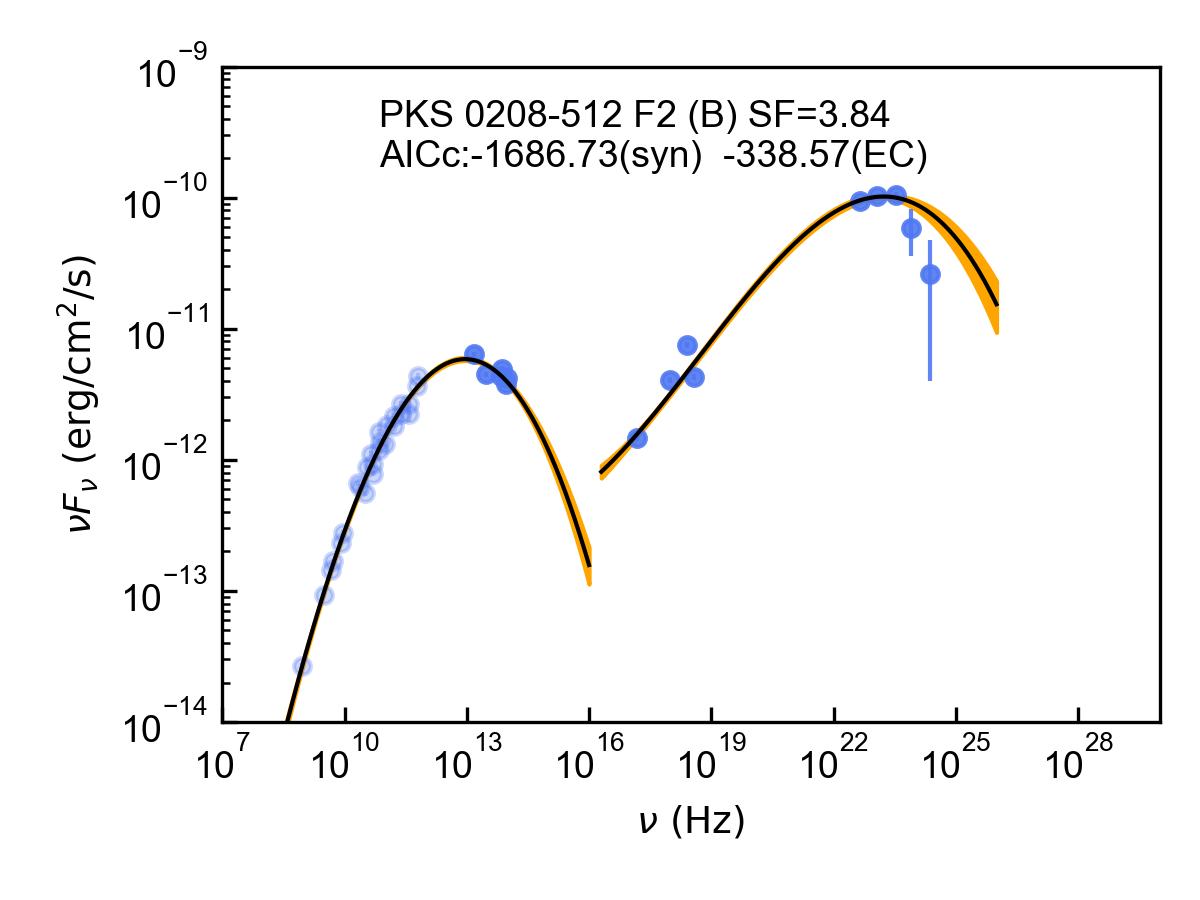}
    \end{subfigure}

    \begin{subfigure}{.25\textwidth}
        \includegraphics[width = \linewidth]{ 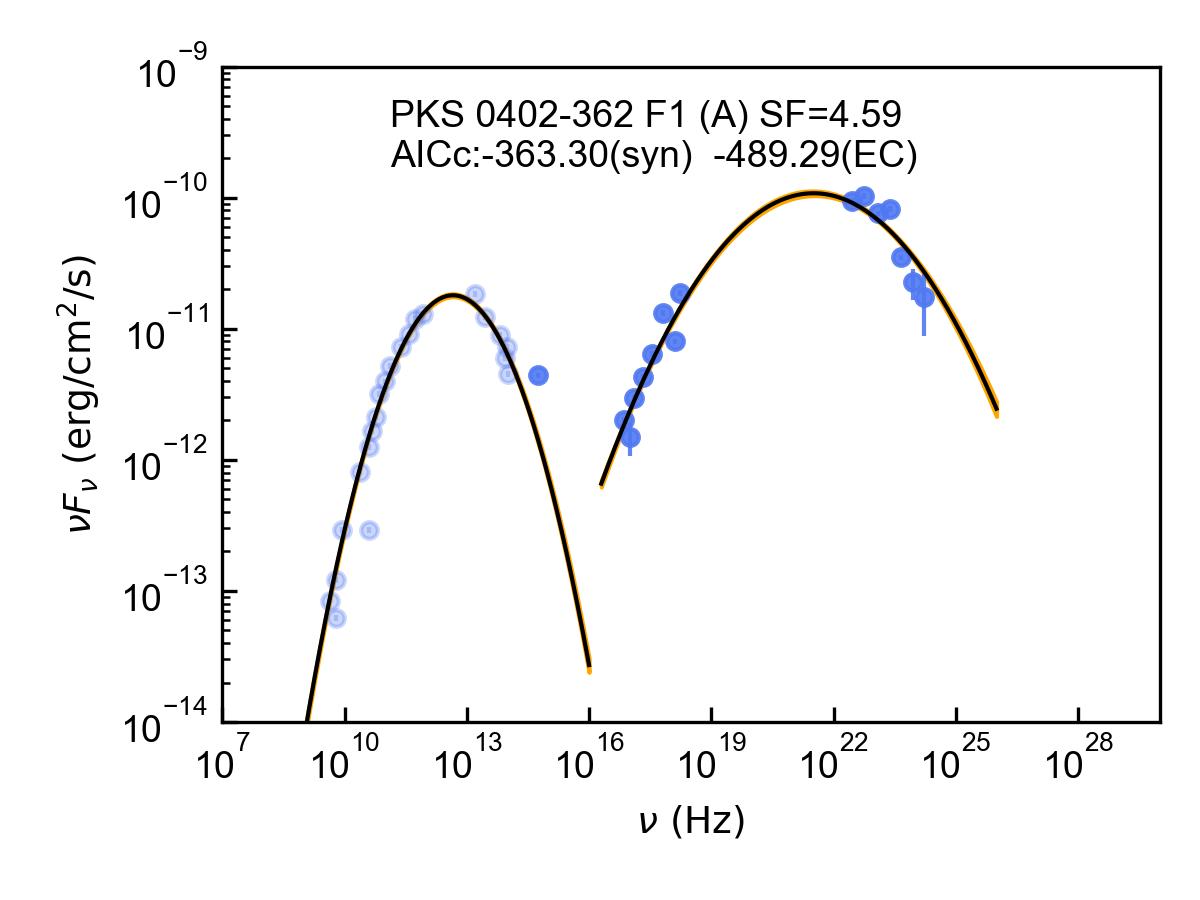}
    \end{subfigure}%
    \begin{subfigure}{.25\textwidth}
        \includegraphics[width = \linewidth]{ 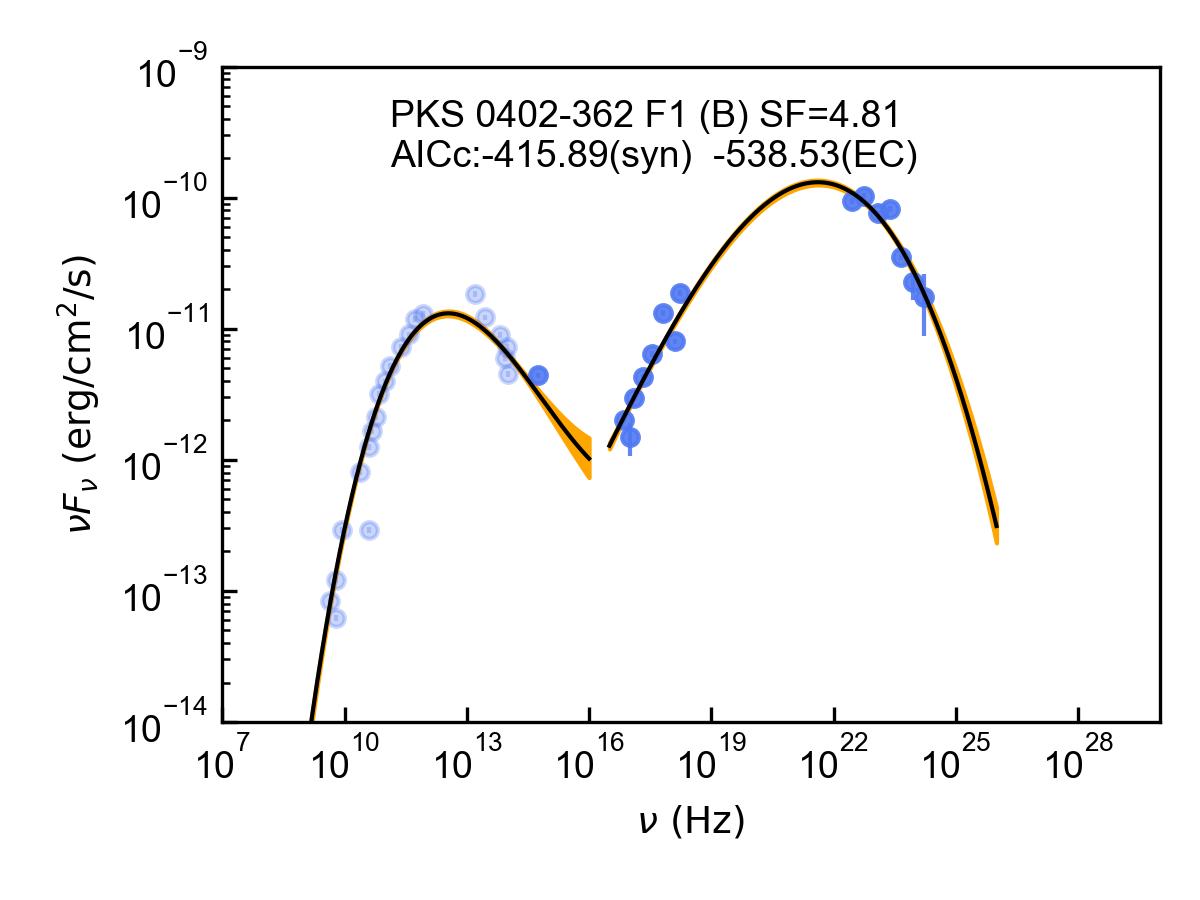}
    \end{subfigure}%
    \begin{subfigure}{.25\textwidth}
        \includegraphics[width = \linewidth]{ 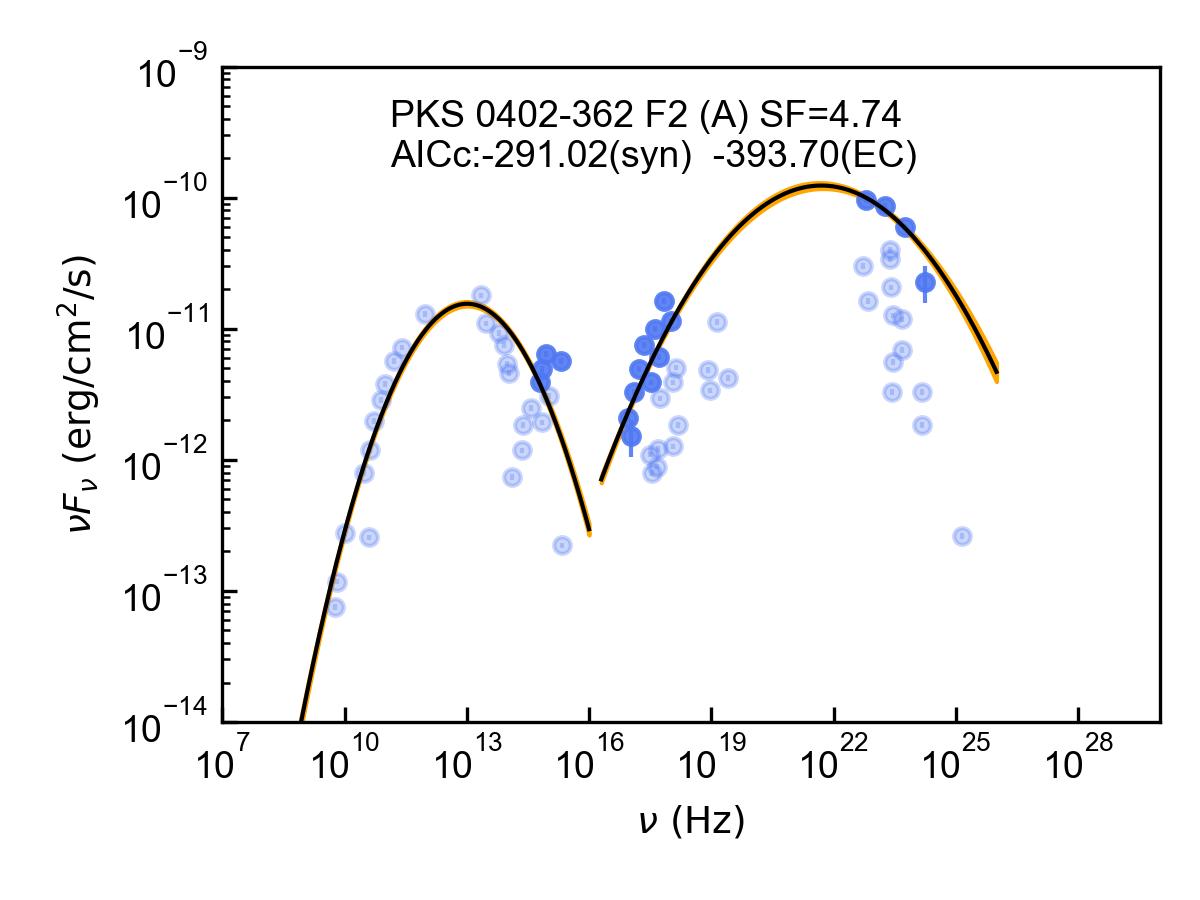}
    \end{subfigure}%
    \begin{subfigure}{.25\textwidth}
        \includegraphics[width = \linewidth]{ 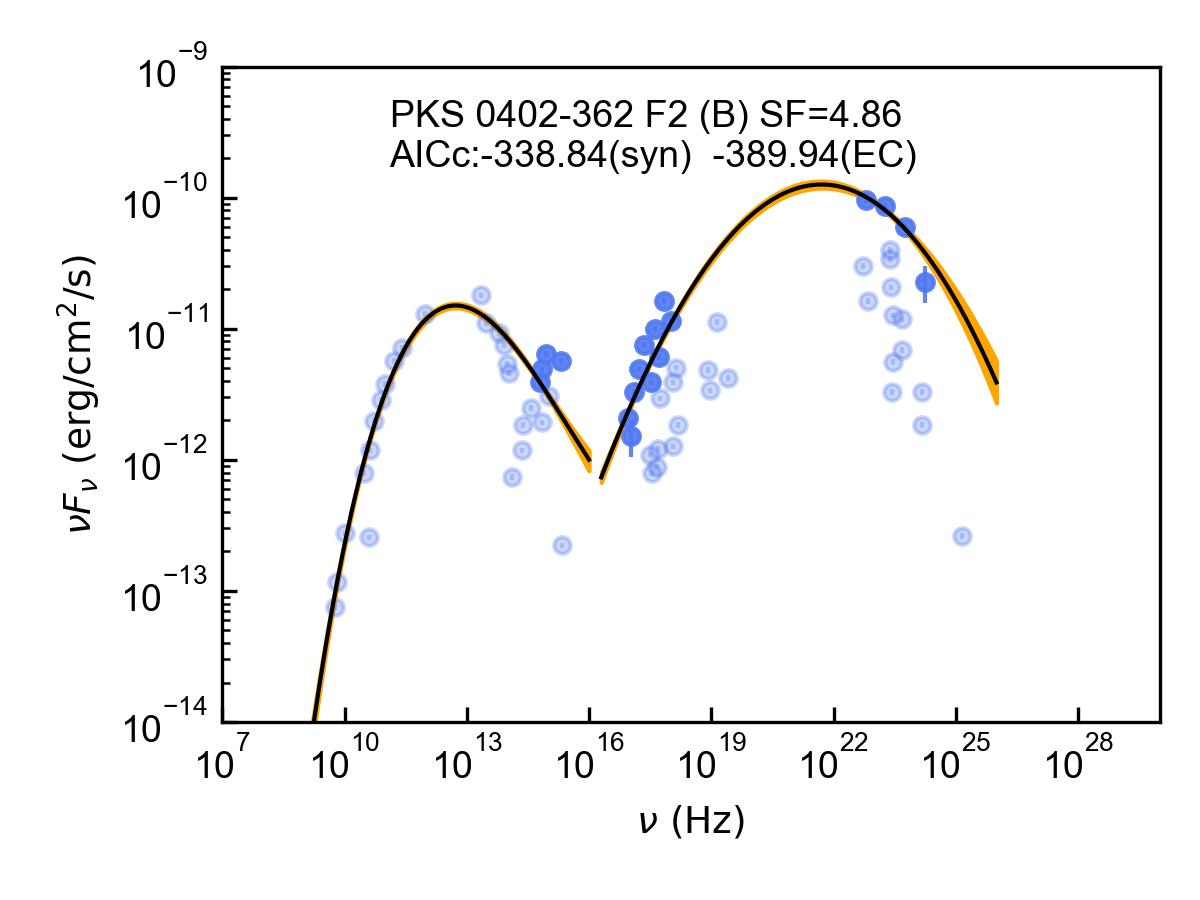}
    \end{subfigure}

    \begin{subfigure}{.25\textwidth}
        \includegraphics[width = \linewidth]{ 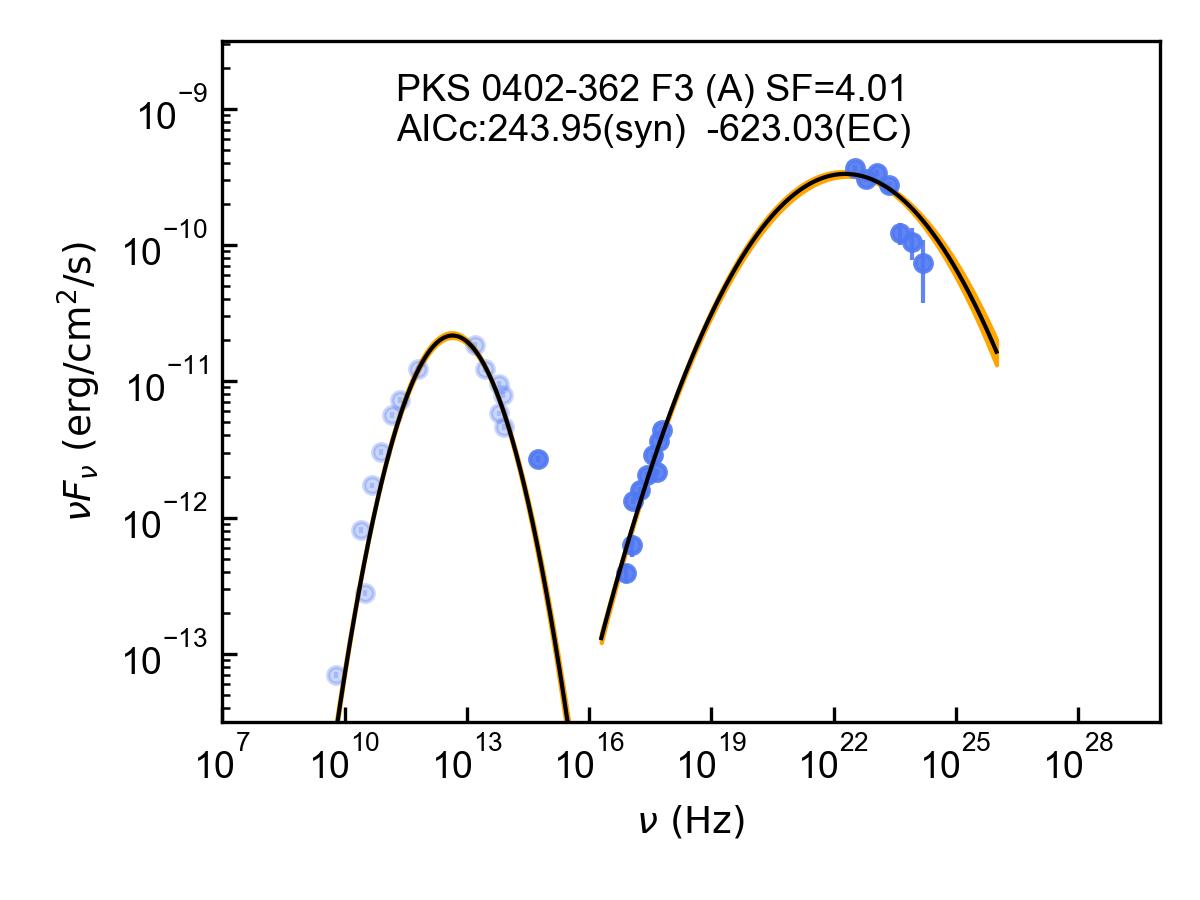}
    \end{subfigure}%
    \begin{subfigure}{.25\textwidth}
        \includegraphics[width = \linewidth]{ 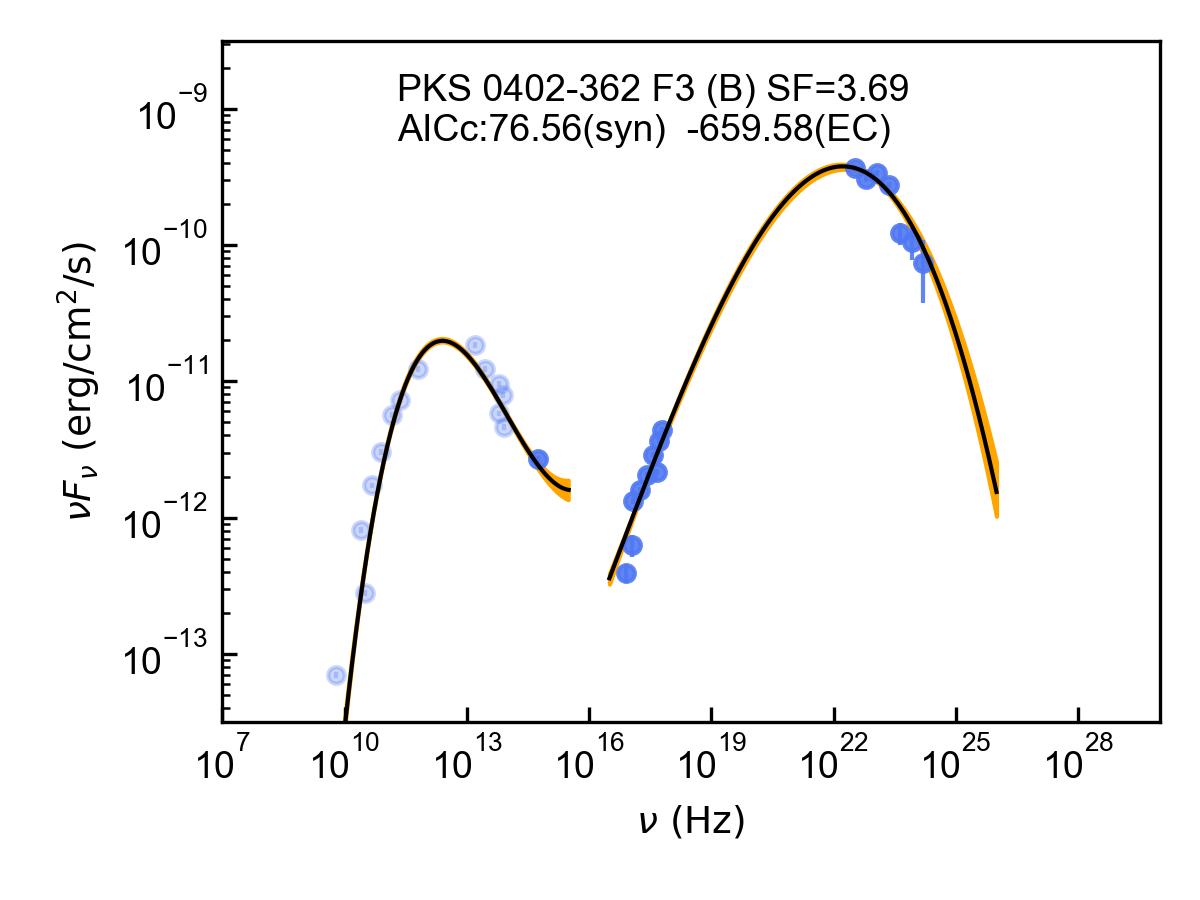}
    \end{subfigure}%
    \begin{subfigure}{.25\textwidth}
        \includegraphics[width = \linewidth]{ 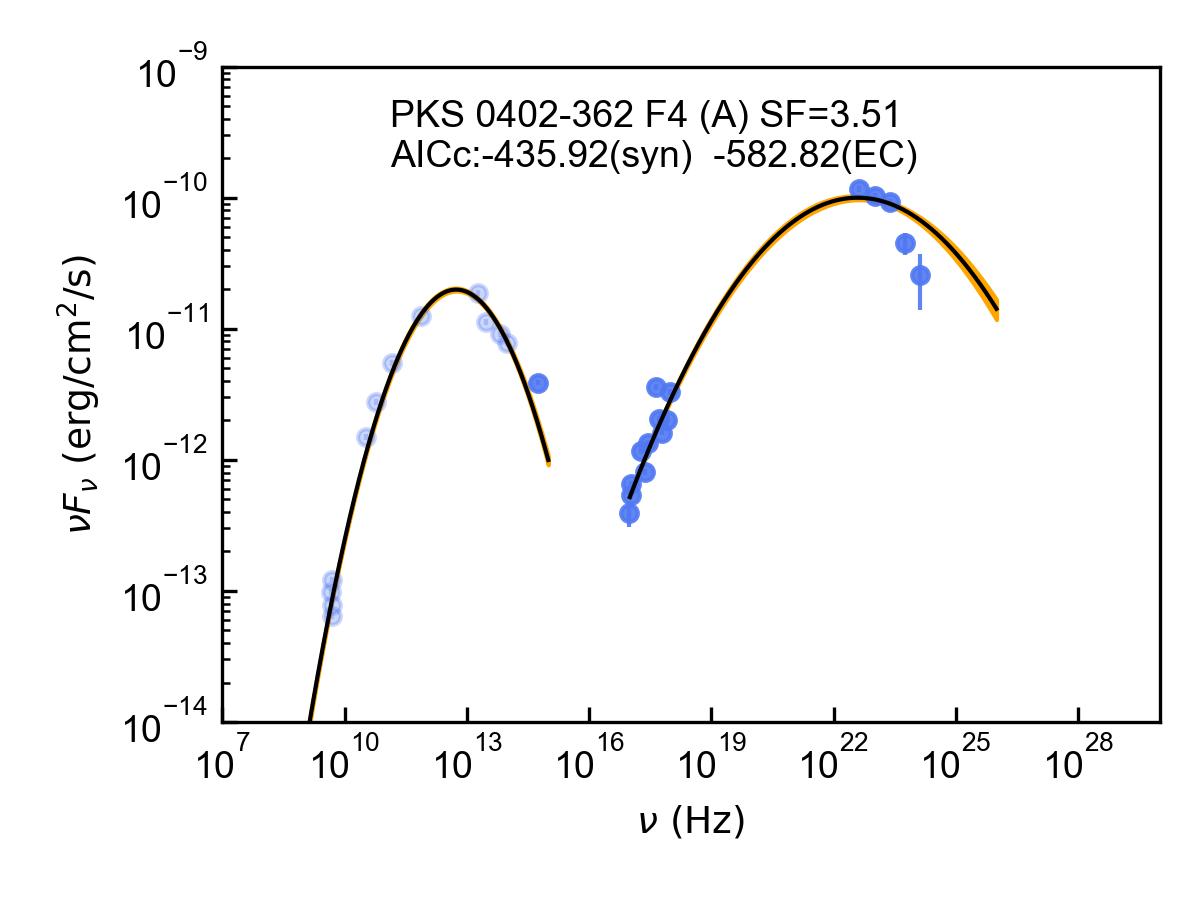}
    \end{subfigure}%
    \begin{subfigure}{.25\textwidth}
        \includegraphics[width = \linewidth]{ 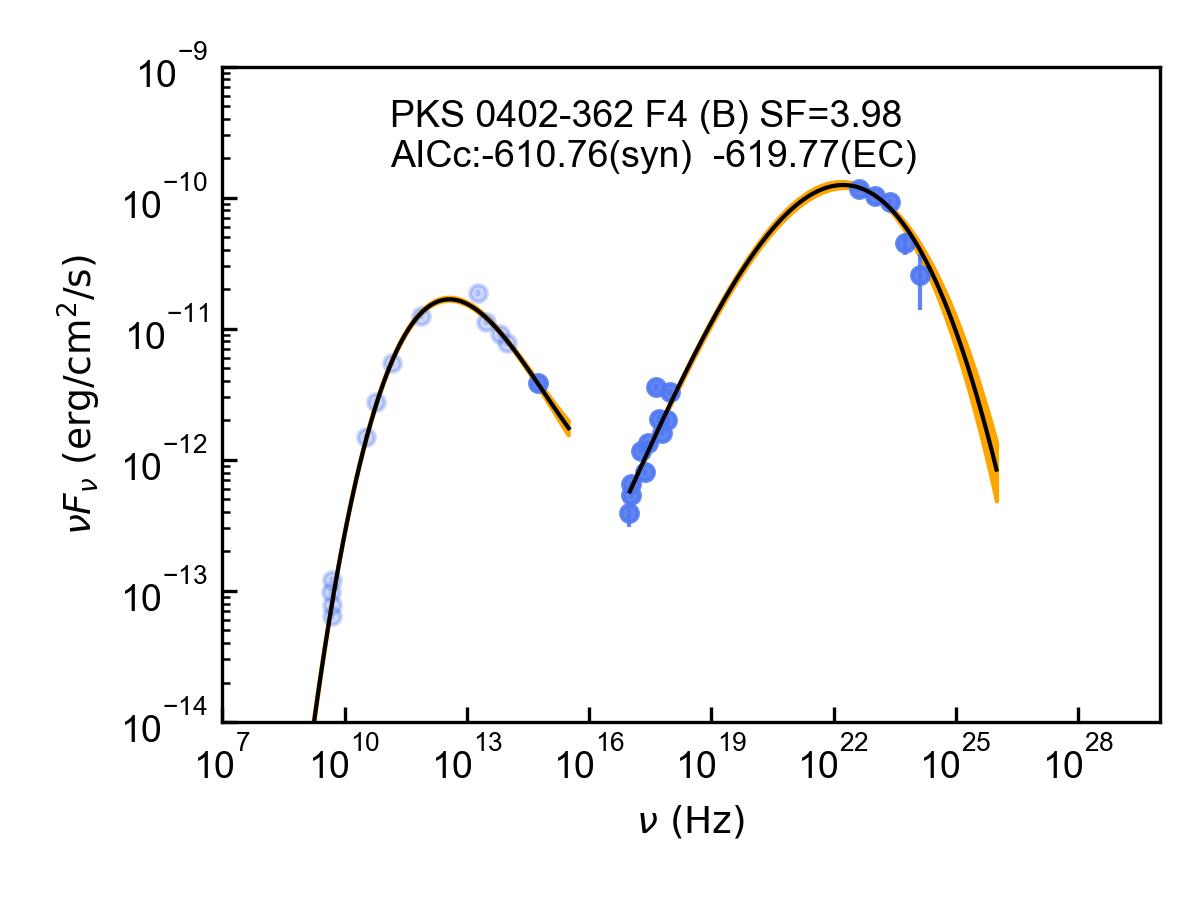}
    \end{subfigure}
    
    \begin{subfigure}{.25\textwidth}
        \includegraphics[width = \linewidth]{ 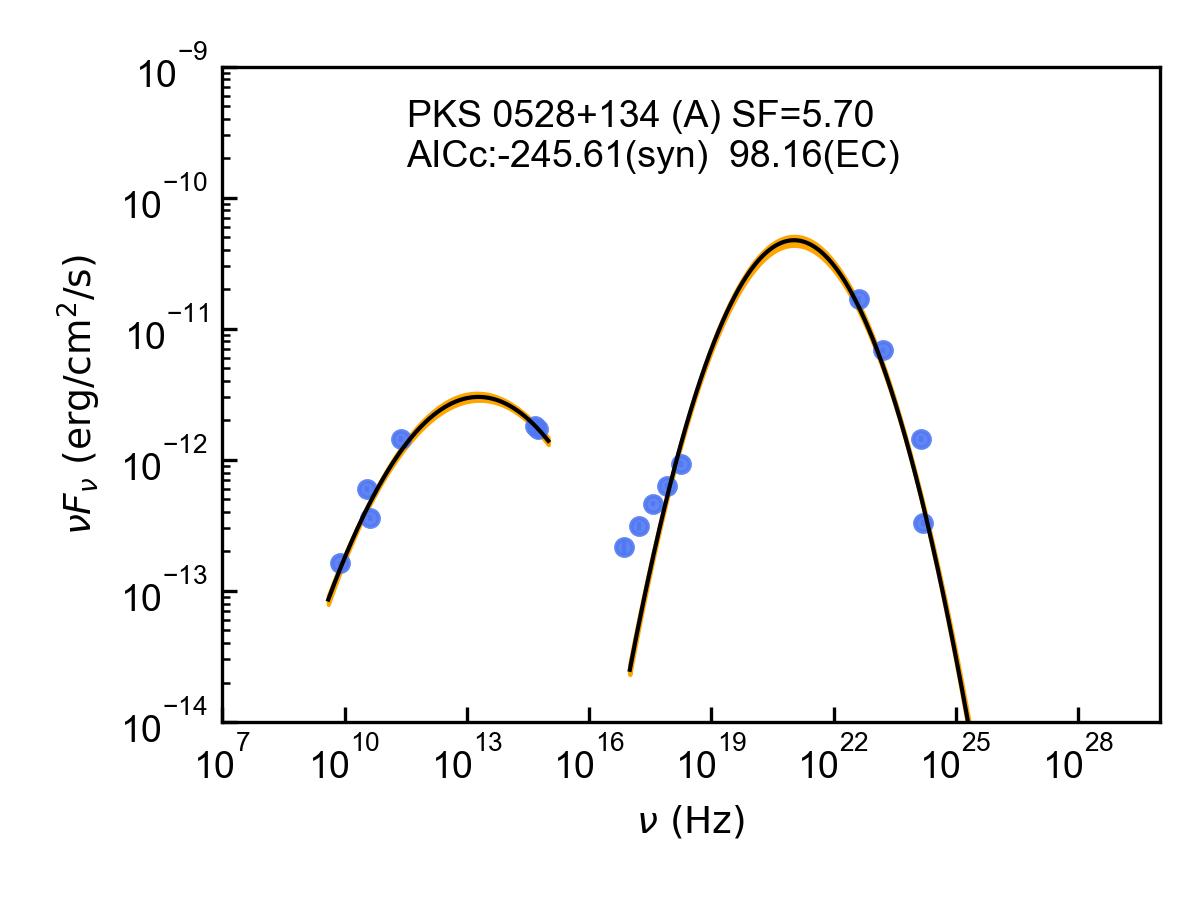}
    \end{subfigure}%
    \begin{subfigure}{.25\textwidth}
        \includegraphics[width = \linewidth]{ 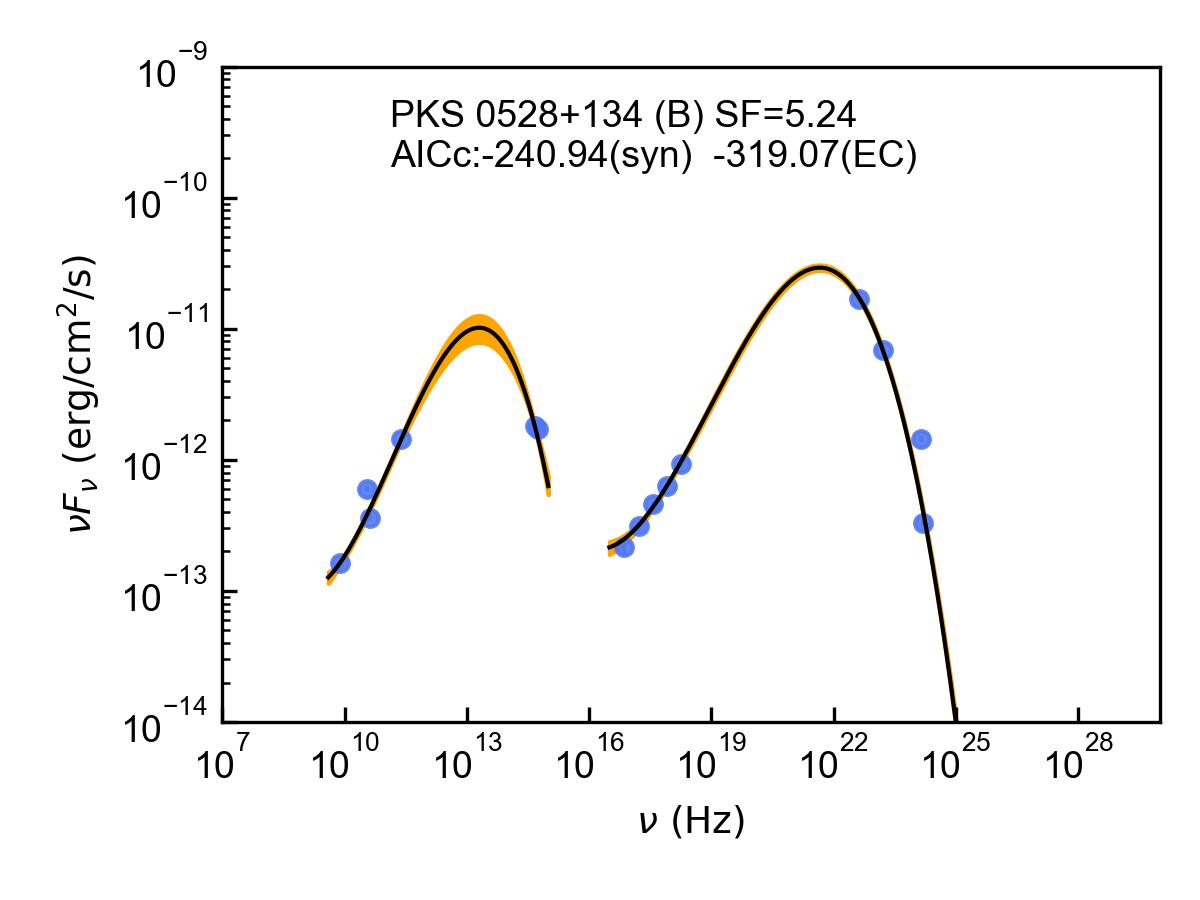}
    \end{subfigure}%
    \begin{subfigure}{.25\textwidth}
        \includegraphics[width = \linewidth]{ 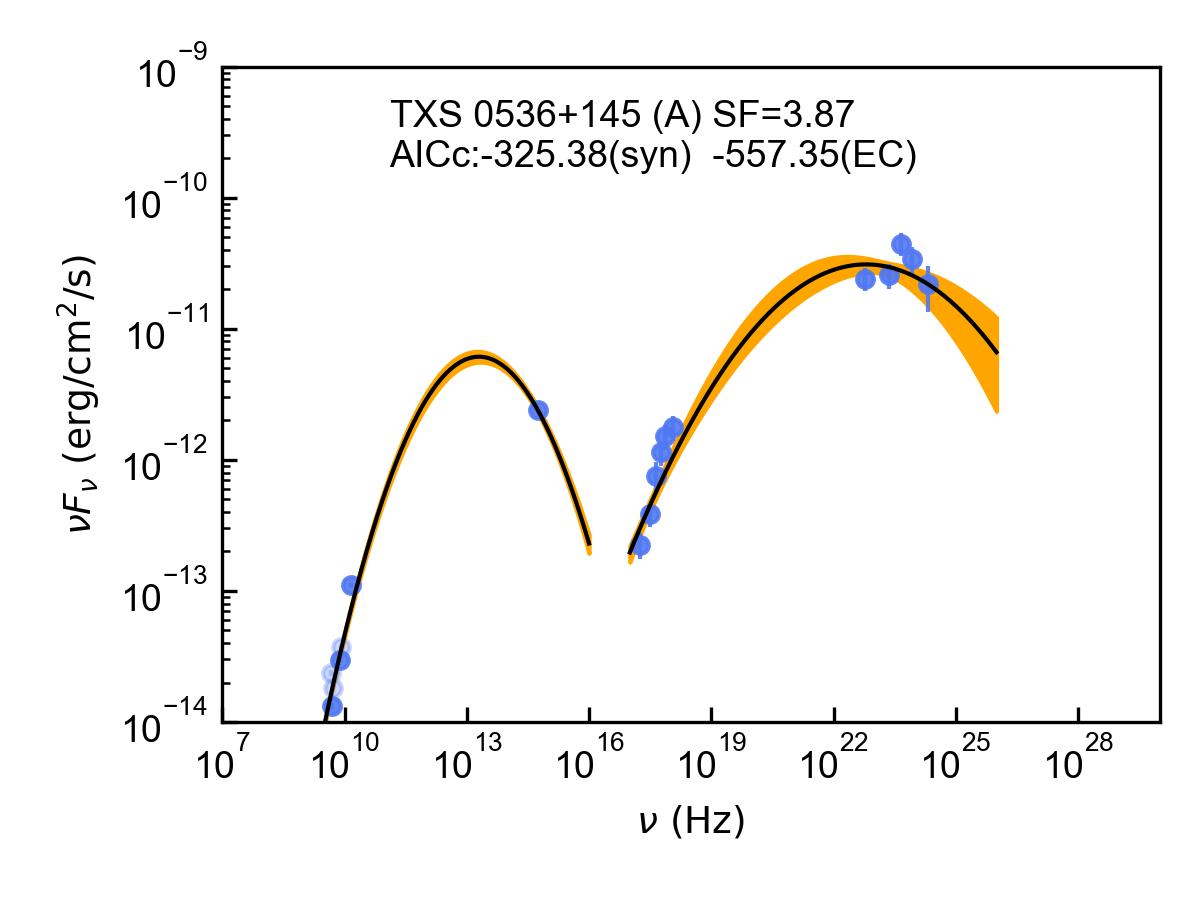}
    \end{subfigure}%
    \begin{subfigure}{.25\textwidth}
        \includegraphics[width = \linewidth]{ 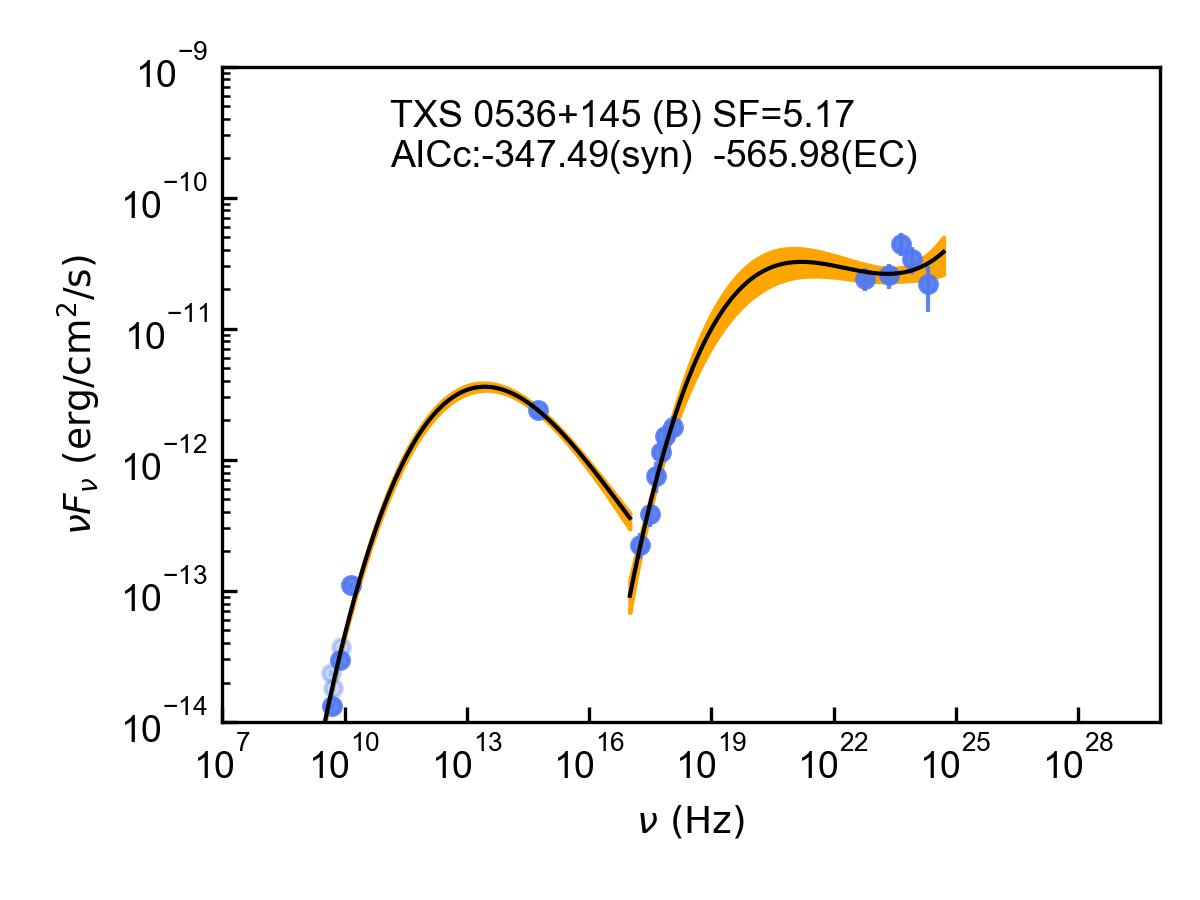}
    \end{subfigure}

    \begin{subfigure}{.25\textwidth}
        \includegraphics[width = \linewidth]{ 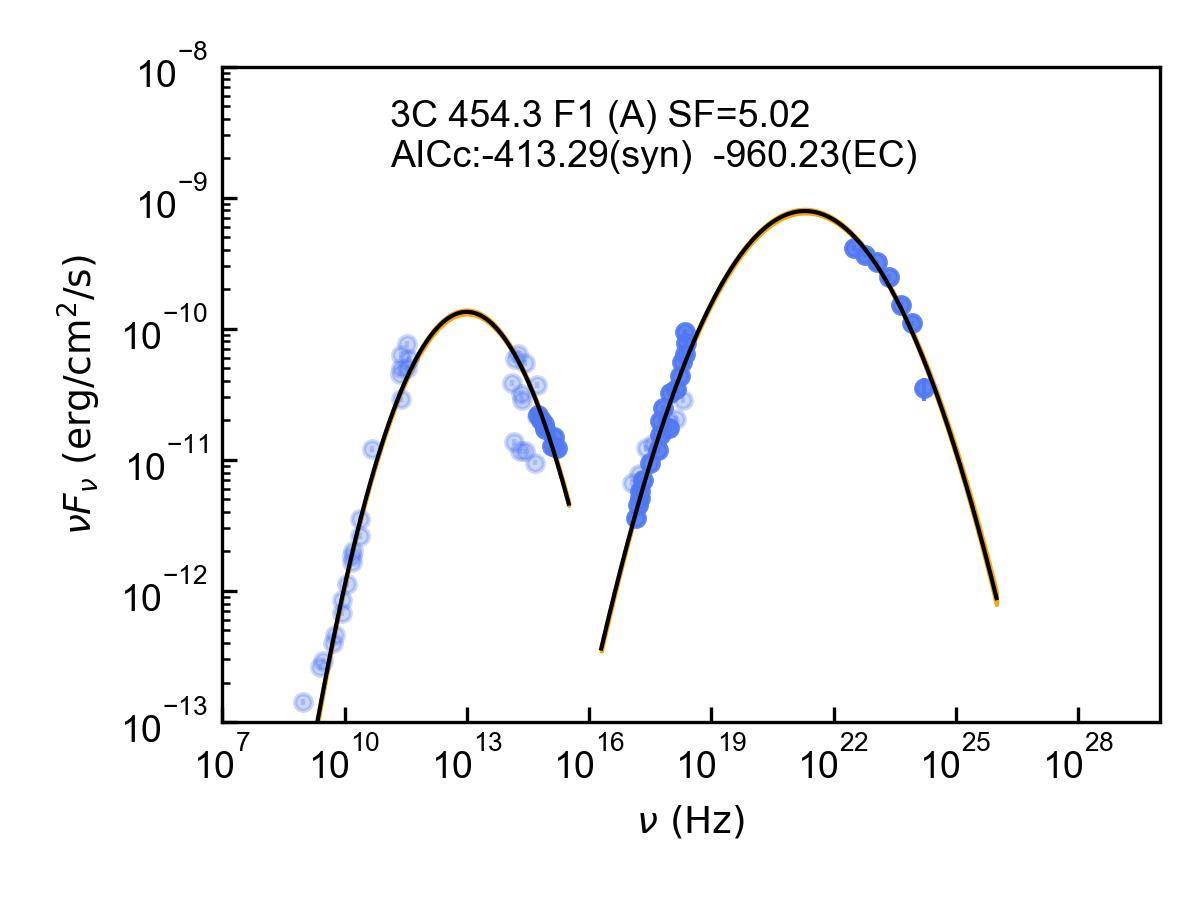}
    \end{subfigure}%
    \begin{subfigure}{.25\textwidth}
        \includegraphics[width = \linewidth]{ 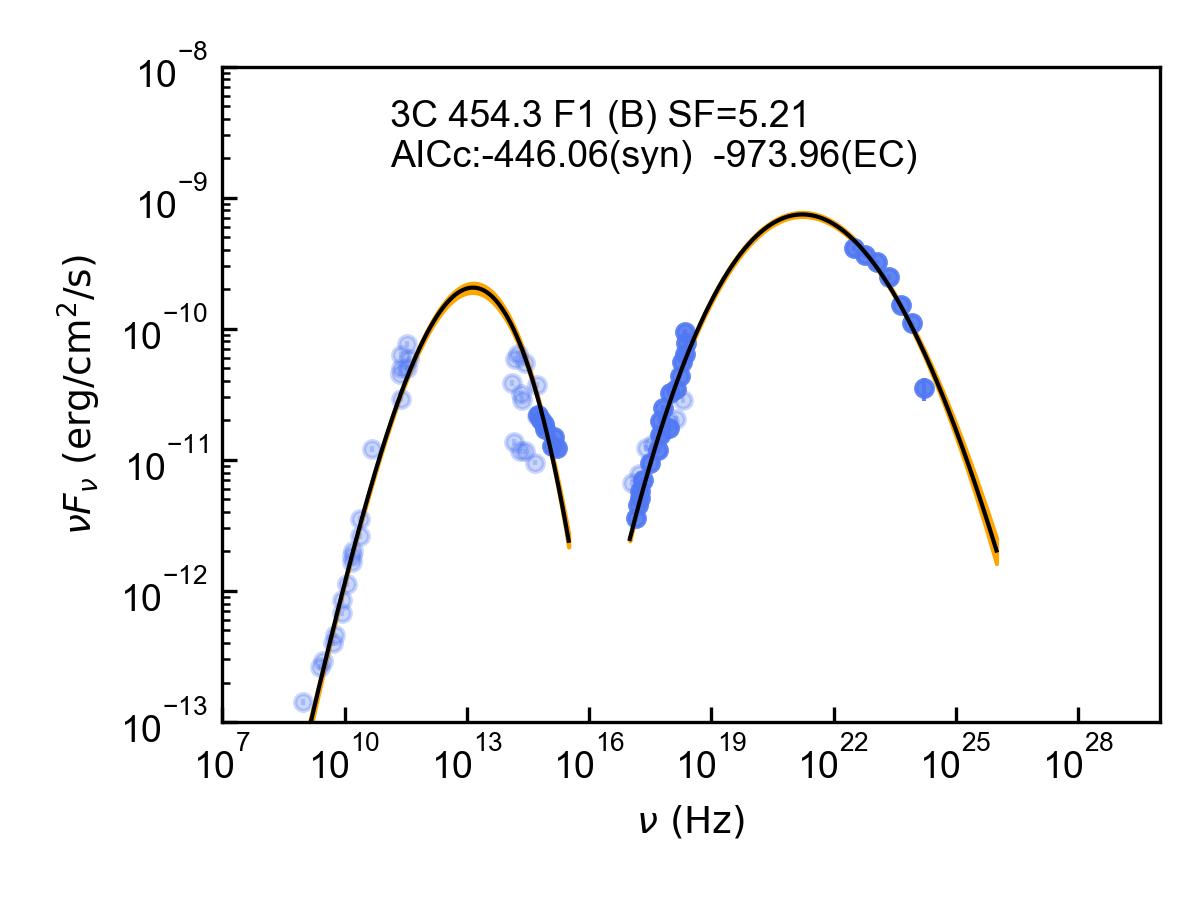}
    \end{subfigure}%
    \begin{subfigure}{.25\textwidth}
        \includegraphics[width = \linewidth]{ 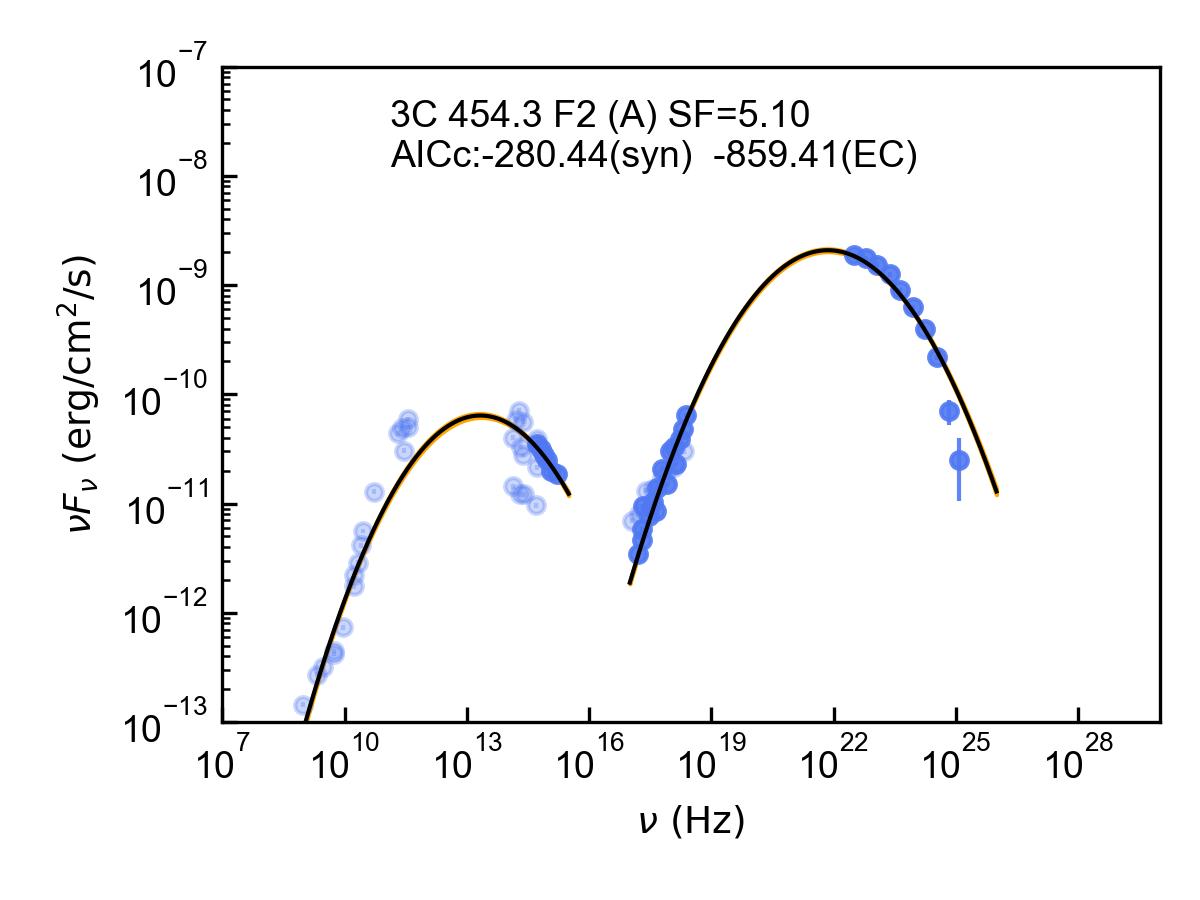}
    \end{subfigure}%
    \begin{subfigure}{.25\textwidth}
        \includegraphics[width = \linewidth]{ 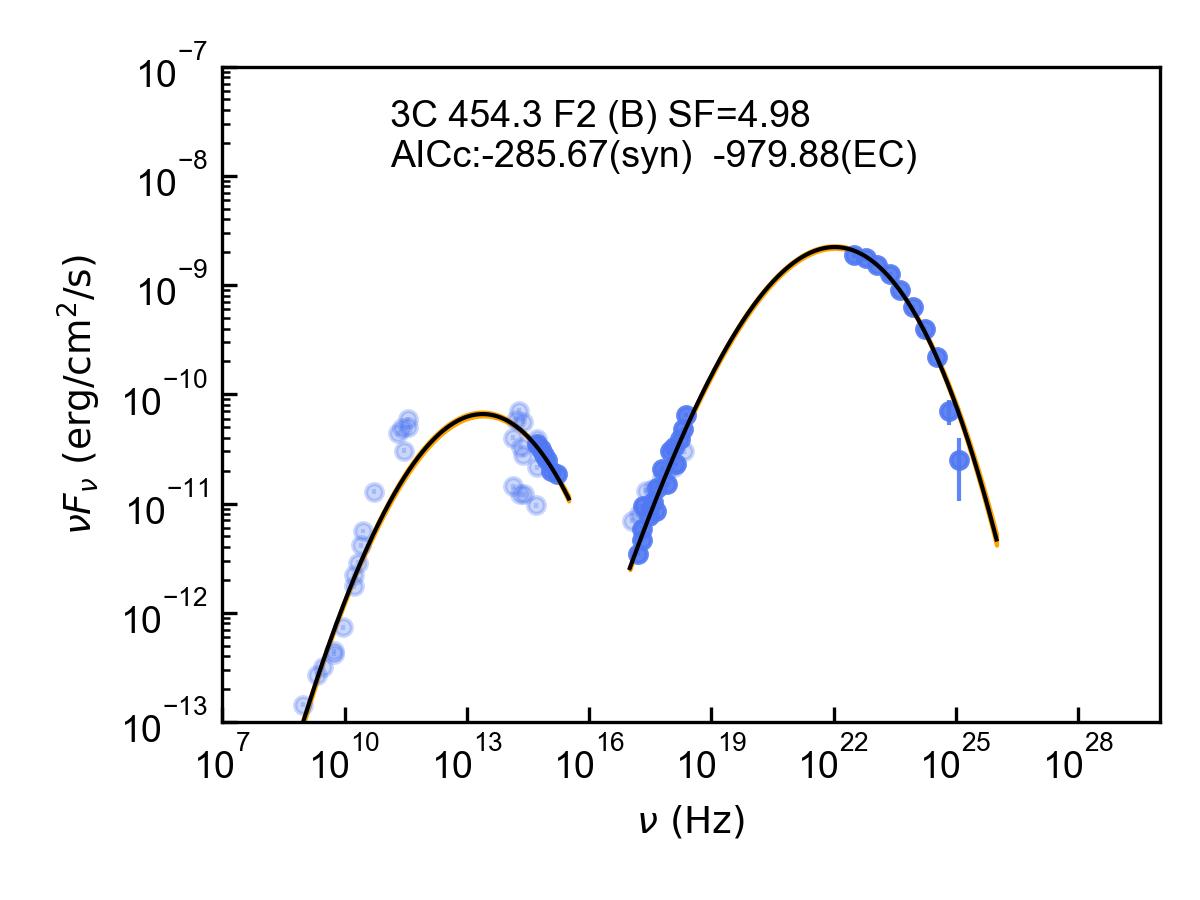}
    \end{subfigure}

    \begin{subfigure}{.25\textwidth}
        \includegraphics[width = \linewidth]{ 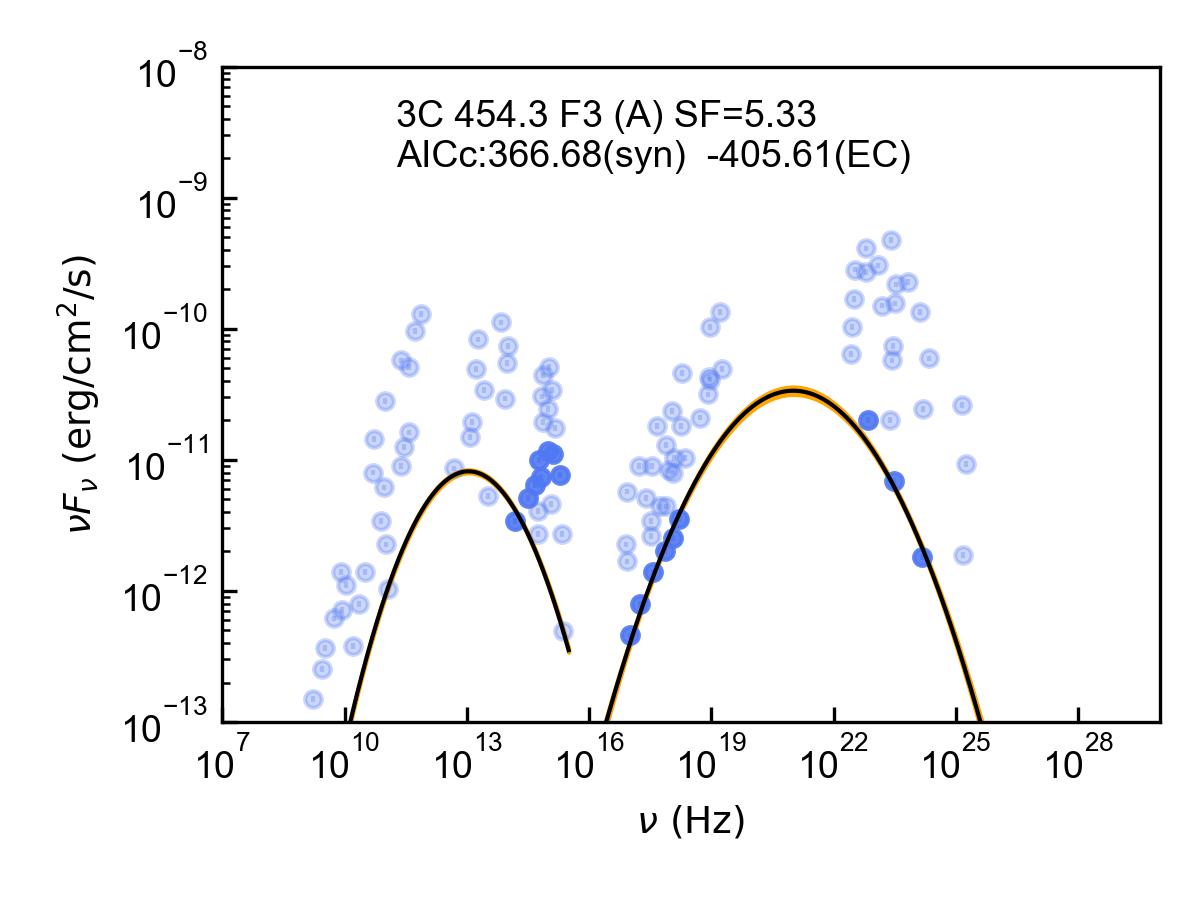}
    \end{subfigure}%
    \begin{subfigure}{.25\textwidth}
        \includegraphics[width = \linewidth]{ 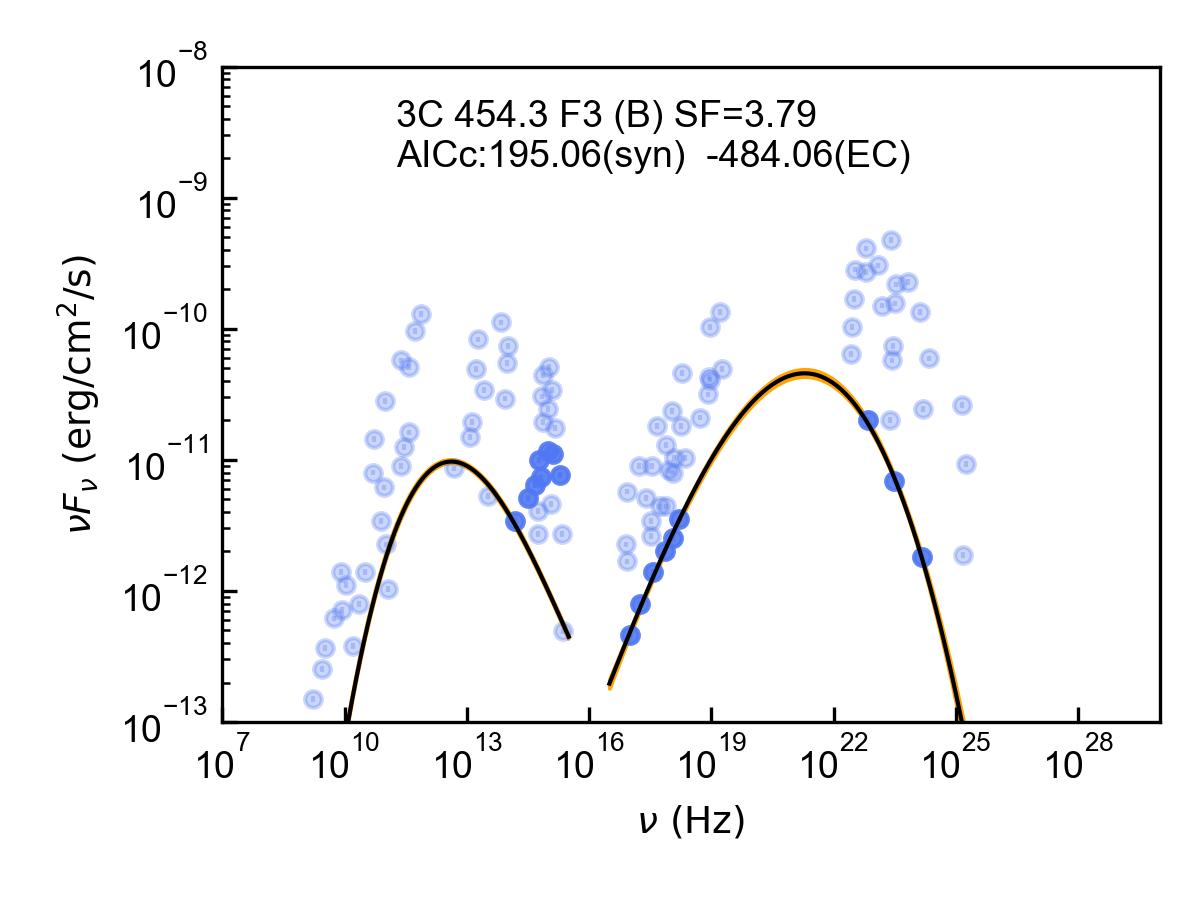}
    \end{subfigure}%
    \begin{subfigure}{.25\textwidth}
        \includegraphics[width = \linewidth]{ 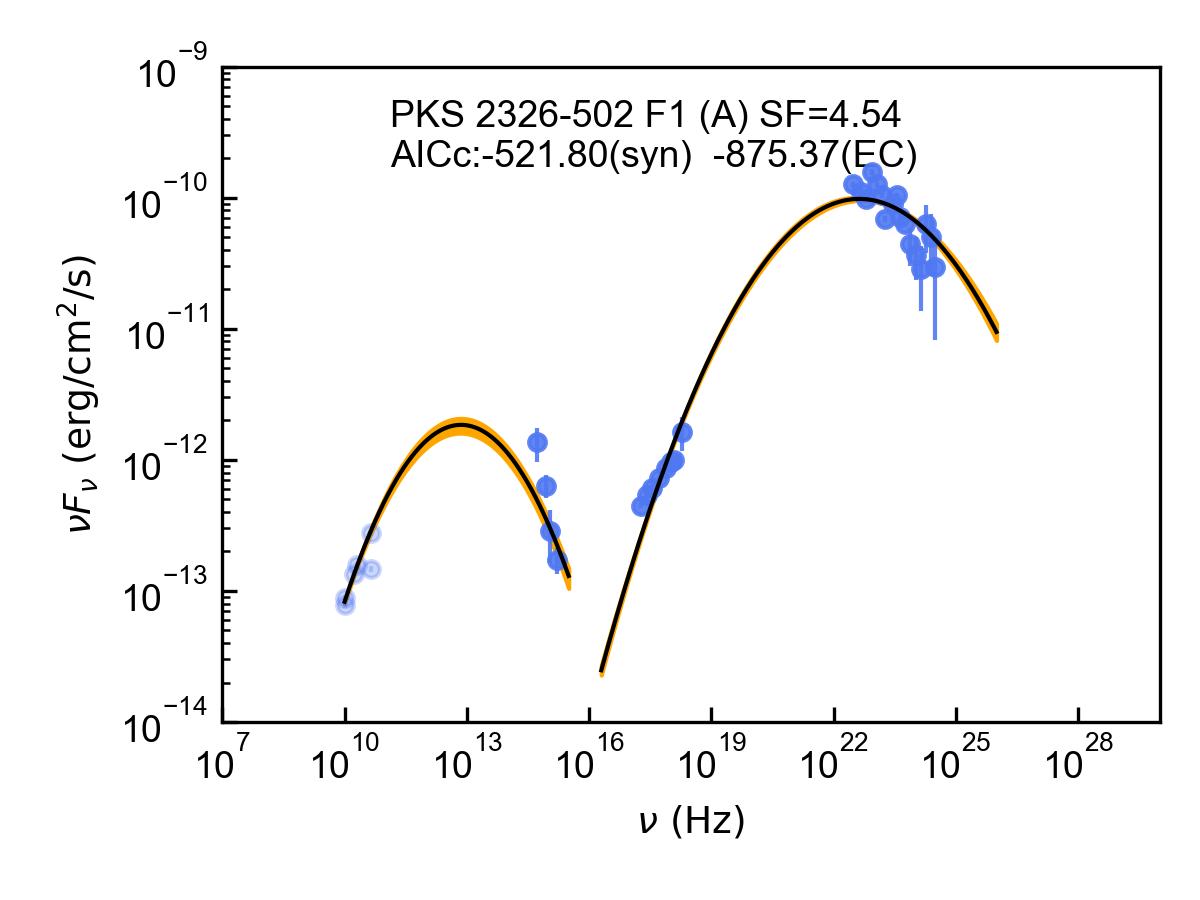}
    \end{subfigure}%
    \begin{subfigure}{.25\textwidth}
        \includegraphics[width = \linewidth]{ 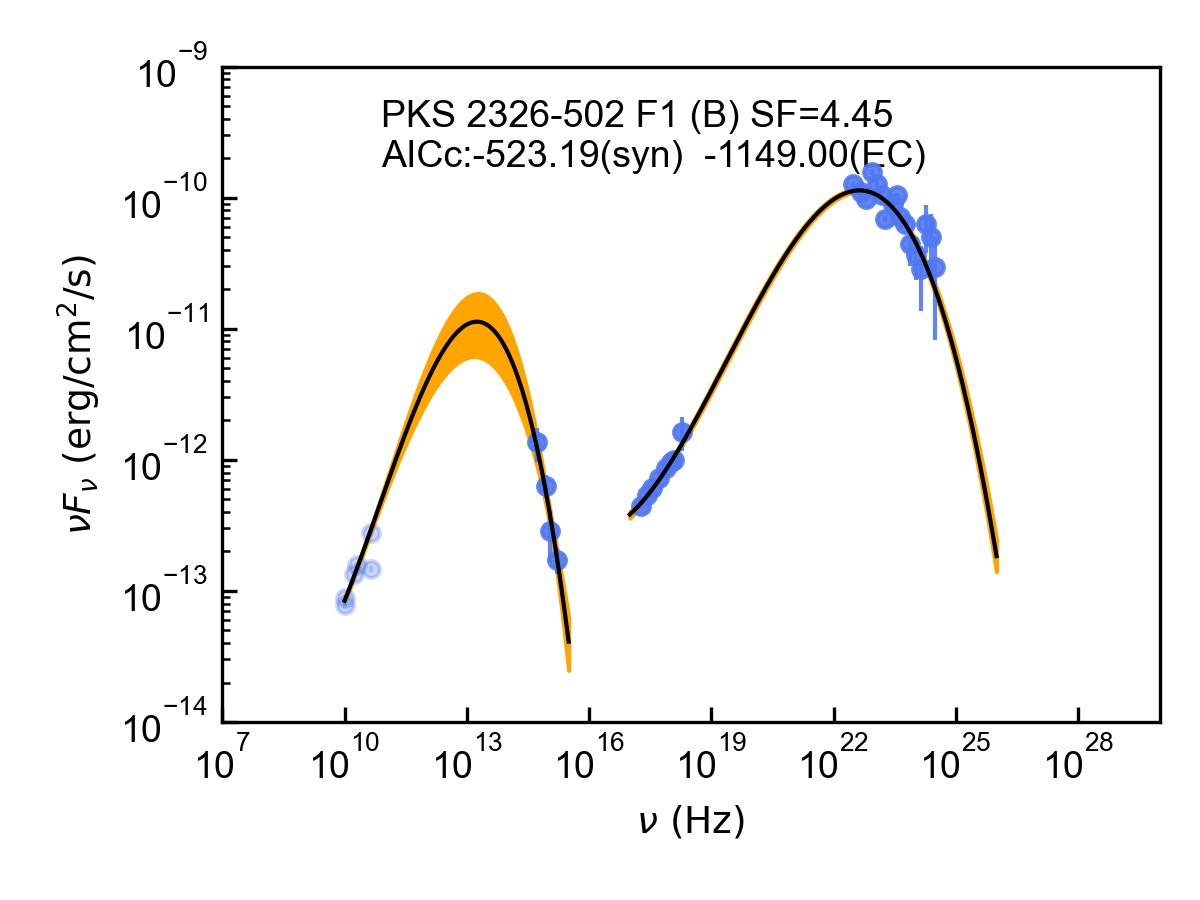}
    \end{subfigure}
    \label{SED DT var}
\end{figure*}

\begin{figure*}
    \ContinuedFloat
    \caption{SED fitting results of the blazars dominated by dusty torus with variability timescales.}
    \begin{subfigure}{.25\textwidth}
        \includegraphics[width = \linewidth]{ 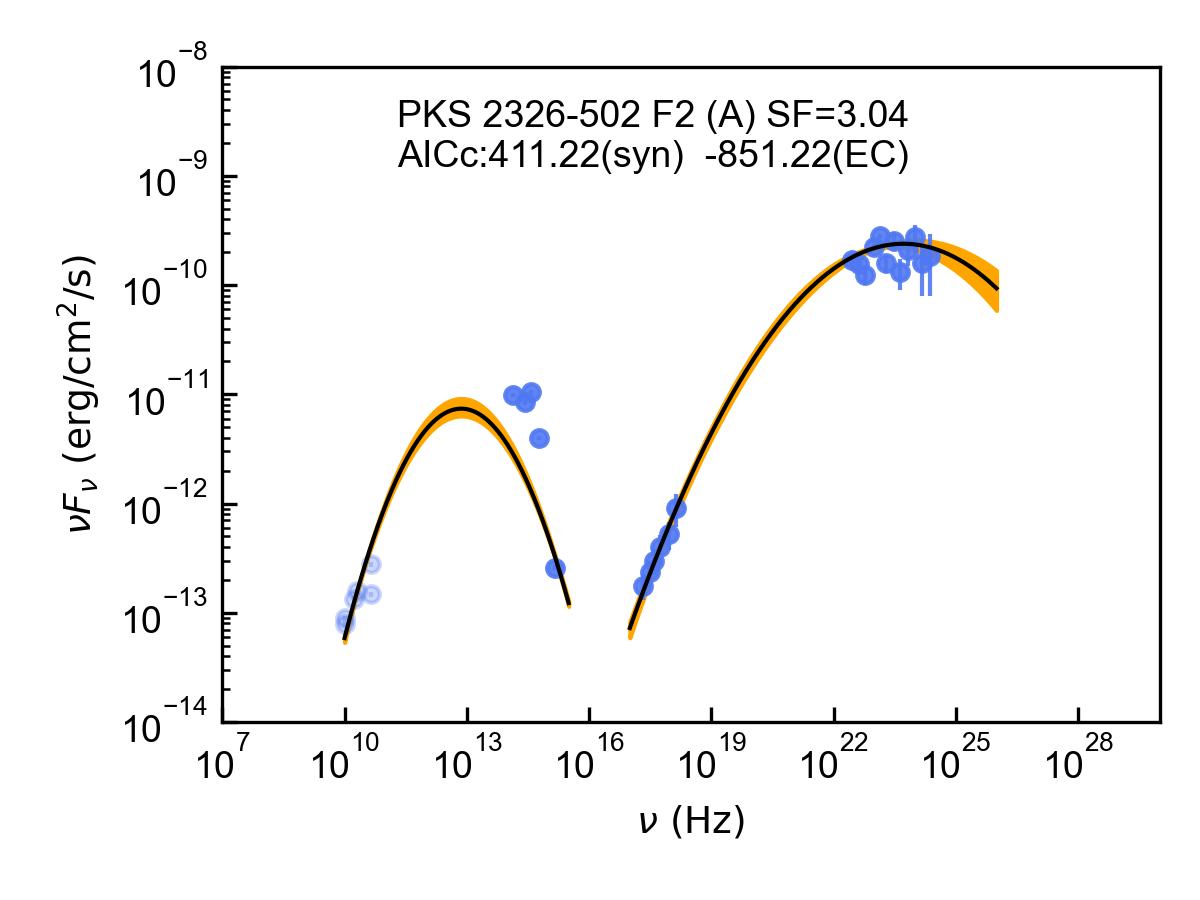}
    \end{subfigure}%
    \begin{subfigure}{.25\textwidth}
        \includegraphics[width = \linewidth]{ 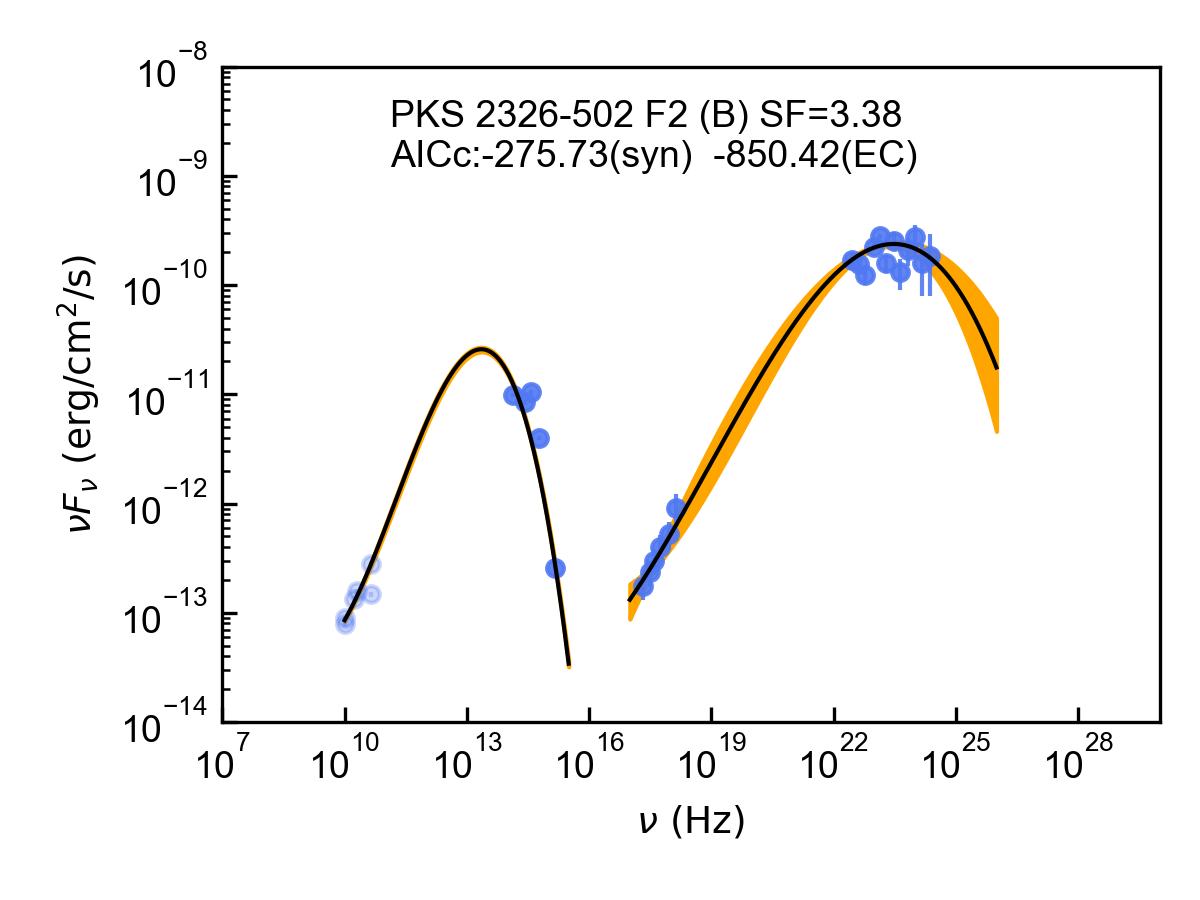}
    \end{subfigure}%
    \begin{subfigure}{.25\textwidth}
        \includegraphics[width = \linewidth]{ 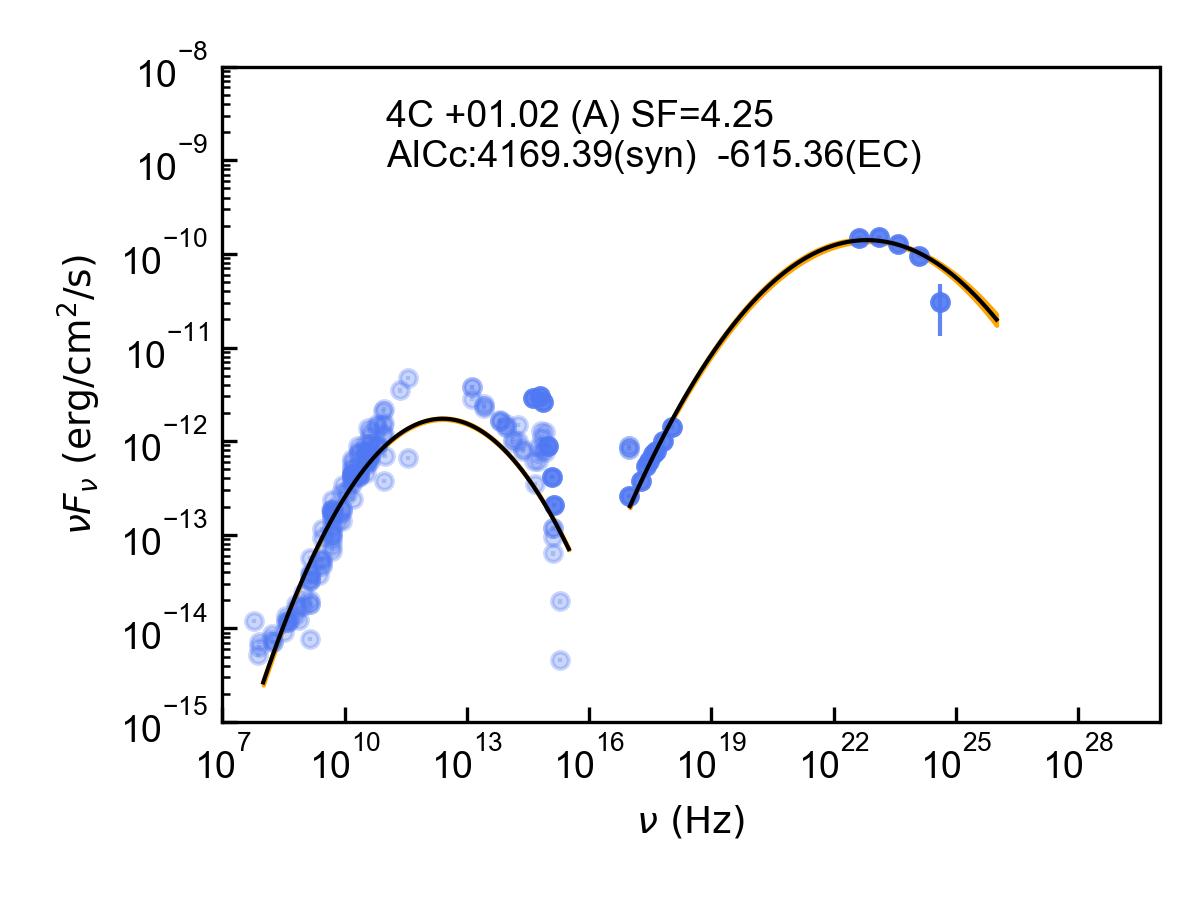}
    \end{subfigure}%
    \begin{subfigure}{.25\textwidth}
        \includegraphics[width = \linewidth]{ 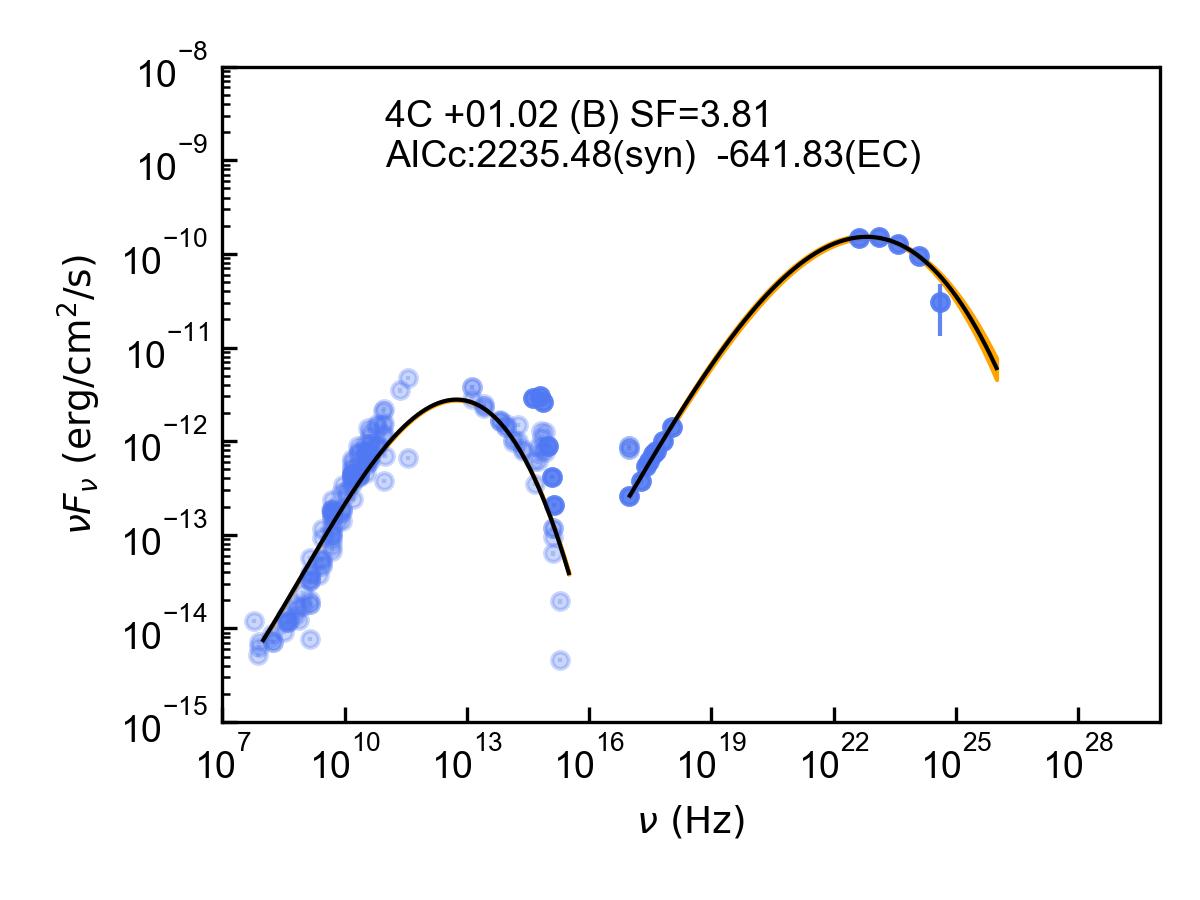}
    \end{subfigure}

    \raggedright
    \begin{subfigure}{.25\textwidth}
        \includegraphics[width = \linewidth]{ 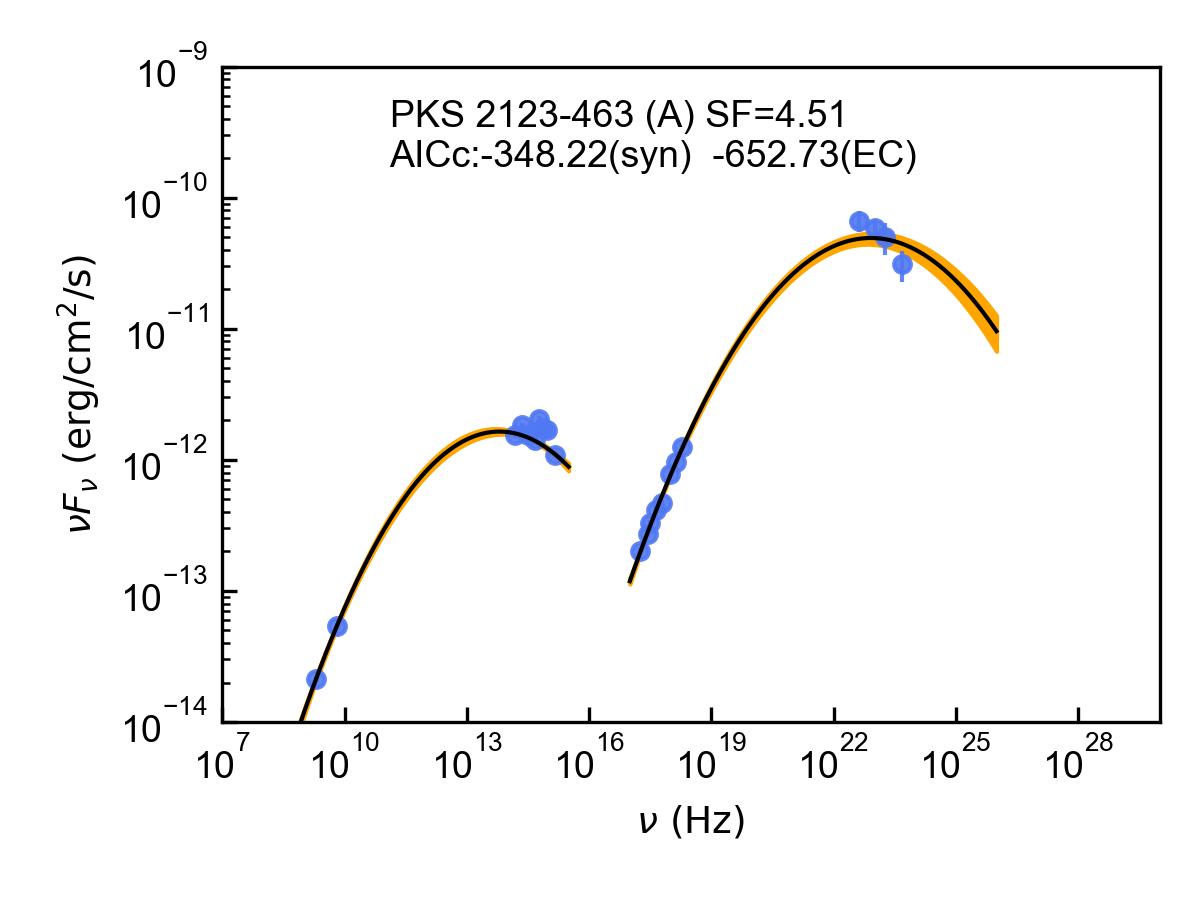}
    \end{subfigure}%
    \begin{subfigure}{.25\textwidth}
        \includegraphics[width = \linewidth]{ 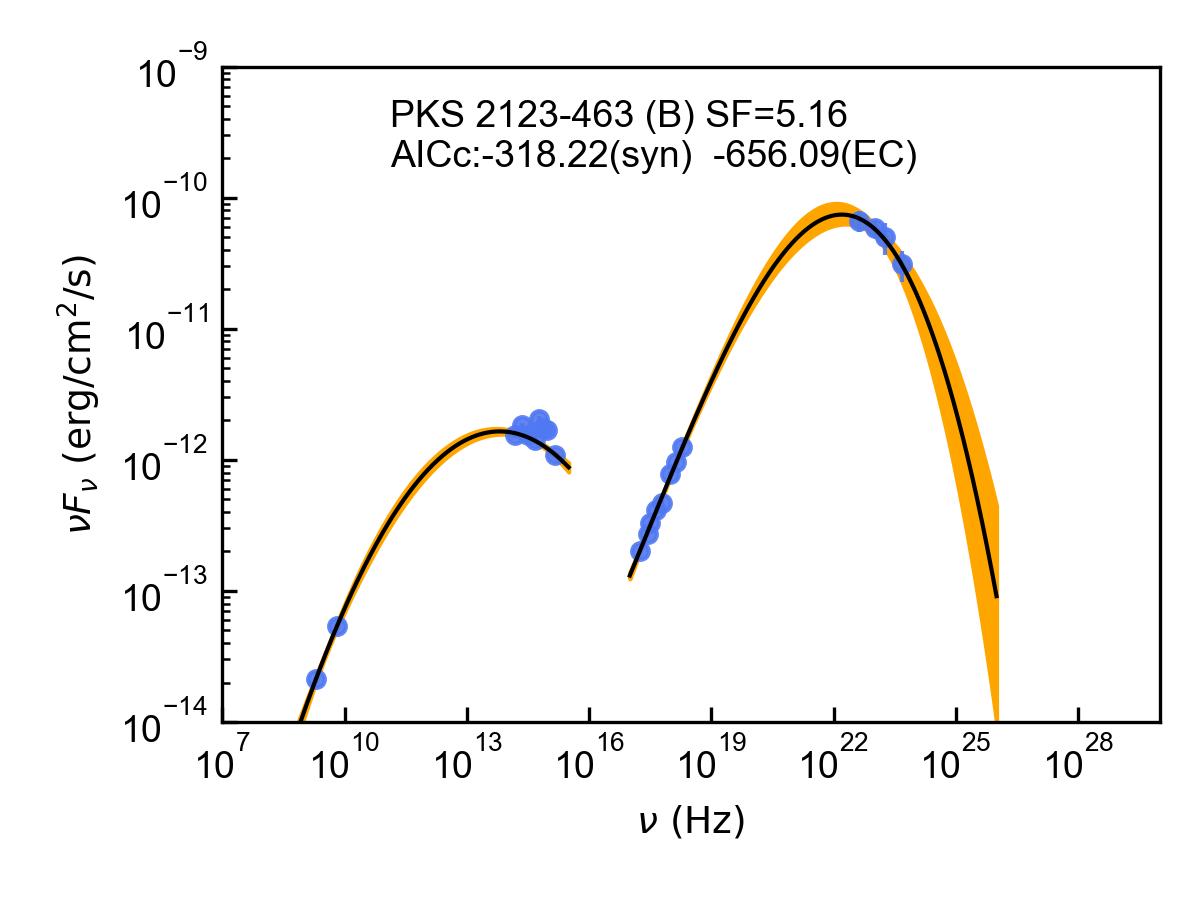}
    \end{subfigure}
    
\end{figure*}

\end{document}